\documentclass[twocolumn]{aastex631}

\DeclareUnicodeCharacter{2212}{-}

\usepackage[T1]{fontenc}
\usepackage[utf8]{inputenc} 

\begin{document}
	
\title{JWST Exoplanetary Worlds and Elemental Survey (JEWELS) II: Condensation Temperature Trends and Galactic Chemical Evolution in JWST Planet-Hosting Stars}
	
\correspondingauthor{Qinghui Sun}
\email{qinghuisun@sjtu.edu.cn}
	
\author[0000-0003-3281-6461]{Qinghui Sun}

\author[0000-0003-2278-6932]{Xianyu Tan}
\affiliation{Tsung-Dao Lee Institute, School of Physics and Astronomy, \& State Key Laboratory of Dark Matter Physics, Shanghai Jiao Tong University, Shanghai 201210, China}

\author[0000-0002-9616-1524]{Gordon (Kai Hou) Yip}
\affiliation{Department of Physics, King’s College London, Strand, London WC2R 2LS, United Kingdom}

\author[0000-0001-5695-8734]{Zitao Lin}
\affiliation{Department of Astronomy, Tsinghua University, Beijing 100084, China}

\author[0000-0003-4794-6074]{Fan Liu}
\affiliation{National Astronomical Observatories, Chinese Academy of Sciences, Beijing 100101, China}

\author[0000-0002-6937-9034]{Sharon Xuesong Wang}
\affiliation{Department of Astronomy, Tsinghua University, Beijing 100084, China}

\author{Zhengduo Li}
\affiliation{Tsung-Dao Lee Institute, School of Physics and Astronomy, \& State Key Laboratory of Dark Matter Physics, Shanghai Jiao Tong University, Shanghai 201210, China}

\begin{abstract}

We present high-precision chemical abundances for 25 FGK-type stars hosting exoplanets observed in JWST Cycle 3 programs and all GTO and DDT programs from Cycles 1–3, based on high-resolution, high signal-to-noise ratio optical spectra from ground-based telescopes. Using a strictly differential, line-by-line analysis relative to the Sun, we derive homogeneous stellar parameters and abundances for 19 elements with atomic number Z $\le$ 30. The sample spans a wide range of stellar properties, with [Fe/H] = −0.6 to +0.4 dex and effective temperatures between 4700 and 6600 K, and includes hosts of terrestrial and giant planets as well as multi-planet systems. We refine carbon and sulfur abundances in cool dwarfs using spectral synthesis, mitigating systematics from line blending. Several chemically interesting systems are identified, including mildly $\alpha$-enhanced metal-poor stars and multi-planet hosts with elevated [C/O]. Using isochrone ages, we derive empirical Galactic chemical evolution (GCE) relations and examine condensation temperature ($T_{\rm cond}$) trends before and after GCE correction. The $T_{\rm cond}$ slopes show no dependence on stellar or planetary properties, indicating that they reflect a mixture of multiple mechanisms, with planet-related signatures entangled in GCE and stellar evolution effects. Thus, $T_{\rm cond}$ trends require careful interpretation. Several systems with significantly positive or negative $T_{\rm cond}$ slopes are identified. Together with forthcoming JWST atmospheric measurements, this homogeneous stellar abundance catalog provides a basis for probing star–planet chemical connections and planet formation pathways.

\end{abstract}
	
\section{Introduction}

The chemical composition of a planet’s atmosphere provides key constraints on its formation environment and evolutionary history. In the core accretion framework \citep{1996Icar..124...62P}, planets form through the accretion of solids followed by gas capture, leading to a mass–metallicity trend in which lower-mass planets are more enriched in heavy elements \citep[e.g.,][]{2018ARA&A..56..175D, 2019ApJ...887L..20W}. Atmospheric compositions also encode information on formation location relative to major ice lines (e.g., H$_2$O, CO$_2$, CO), as well as subsequent migration and gas accretion history \citep[e.g.,][]{2017A&A...608A..92D, 2016SciA....2E1658J, 2022ApJ...929...52K}. Establishing the planet mass–metallicity relation across a wide range of planetary masses is therefore essential for understanding the chemical evolution of protoplanetary disks and the assembly of planetary systems.

Despite extensive efforts, the planet mass–metallicity relation remains poorly constrained. Pre-JWST studies were limited to a few species (typically H$_2$O, Na, and K) and suffered from heterogeneous analyses on both stellar and planetary sides. Even homogeneous \textit{HST} surveys were inconclusive due to limited precision and sample size \citep[e.g.,][]{2023ApJS..269...31E, 2024AJ....167..167S}.

JWST has improved this situation with high-precision spectroscopy from 0.6–28~$\mu$m \citep{2023PASP..135f8001G}, enabling detections of multiple molecule species (H$_2$O, CO$_2$, SO$_2$, CH$_4$) and more precise abundance measurements \citep[e.g.,][]{2023Natur.614..649J,2023ApJ...946L...6M,2024Natur.625...51D}. These data allow comparative studies of atmospheric metallicities and C/O ratios across diverse systems, but progress is now limited by the lack of equally homogeneous and precise stellar abundances.

Recent JWST results show that planetary atmospheres can both match and differ from their host stars (e.g., \citealt{2025MNRAS.540.2535A, 2023Natur.614..653A}), reflecting diverse formation and migration pathways. Because stellar abundances encode the chemical properties of the natal environment, combining them with planetary measurements provides a direct link between planet formation and Galactic chemical evolution \citep[e.g.,][]{2015A&A...581L...2A}, which requires homogeneous, high-precision stellar abundances analyzed in parallel with JWST data.

To address this need, many recent studies have focused on deriving homogeneous stellar parameters and abundances for planet-hosting stars. The GAPS (Global Architecture of Planetary Systems) project, for example, has obtained high-resolution spectra for dozens of exoplanet hosts and analyzed them in a consistent way (e.g., \citealt{2022A&A...664A.161B, 2024A&A...690A.370F}). A primary goal of GAPS is to explore the link between stellar composition and planetary system properties, including possible chemical signatures of planet formation and migration. In parallel, the Ariel group has compiled a homogeneous set of stellar parameters and abundances for about 200 planet hosts \citep{2022A&A...663A.161M, 2024A&A...688A.193D, 2025A&A...697A.102T}, with a focus on population-level questions such as the planet mass–metallicity relation and the Galactic chemical context of exoplanet systems.

Stellar abundance patterns can themselves encode signatures of planet formation, with the dependence of elemental abundances on condensation temperature ($T_{\rm cond}$) proposed as a potential diagnostic of planet formation or engulfment (e.g. \citealt{2009ApJ...704L..66M, 2024Natur.627..501L}). For example, the Sun has been reported to be relatively depleted in refractory elements compared to solar twins, potentially reflecting the formation of planets (e.g., \citealt{2018ApJ...865...68B, 2025ApJ...980..179S, 2025A&A...701A.107S}). While such subtle $T_{\rm cond}$ trend and planet relations can be robustly detected when stellar parameters are tightly controlled, their interpretation becomes more challenging in broader samples. In stars spanning a wide range of stellar parameters, multiple processes, including Galactic chemical evolution (GCE), stellar structure and mixing, and stochastic events such as stellar merger, act simultaneously, and the planet-related signature is often secondary to these effects. As a result, $T_{\rm cond}$ slopes remain a complex and model-dependent diagnostic of planet formation in diverse exoplanet systems.

The JWST Exoplanetary Worlds and Elemental Survey (JEWELS; \citealt{2026ApJS..282...37S}) is designed to provide homogeneous stellar and planetary atmospheric abundances within a consistent framework for interpreting planet formation environments and atmospheric compositions. Compared to efforts such as Ariel and GAPS, JEWELS adopts a complementary approach, focusing on high-precision, homogeneous analyses tailored to stars with current or upcoming JWST observations. The sample is designed around benchmark systems, enabling a direct comparison between stellar abundances and planetary atmospheric measurements. In this way, JEWELS complements larger surveys by providing a consistent reference for detailed star–planet connection studies in the JWST era.

In JEWELS I (\citealt{2026ApJS..282...37S}), we presented abundance measurements for 20 FGK planet-hosting stars observed in JWST Cycle 2 General Observer (GO) programs. In this paper, we extend the survey by reporting abundances for 25 additional JWST host stars. Our sample primarily includes transiting systems spanning hot Jupiters, warm Neptunes, and smaller planets orbiting FGK stars, along with a smaller number of directly imaged, young, wide-orbit giant planets observed with high-contrast imaging and integral-field spectroscopy. In addition to expanding the sample, we compare our derived stellar parameters with literature values, assess the applicability of our differential line-by-line analysis across a broader range of stellar parameters, and examine abundance trends with {$T_{\rm cond}$}, providing the basis for investigating potential signatures of planet formation and stellar evolution.

\section{Target Selection and Observations}

Our initial target list includes all planetary systems observed in JWST Cycle 3 General Observer (GO) programs, as well as those observed in Cycles 1–3 under Guaranteed Time Observations (GTO) and Director’s Discretionary Time (DDT) programs. The full target list is provided in Tables~\ref{tab:jwst_host1} and \ref{tab:jwst_host2}, with program information retrieved from \url{https://www.stsci.edu/jwst/science-execution/approved-programs}. Target selection follows the same overall strategy described in \citet{2026ApJS..282...37S}, which we briefly summarize here.

High-resolution optical spectra are collected from the ESO \citep{2022SPIE12186E..0DR} and Keck \citep{2014SPIE.9152E..2IT} public archives. We initially restrict the sample to FGK-type stars with effective temperatures between 4000 and 7600 K, based on values from the NASA Exoplanet Archive (\citealt{NEA12,NEA13}). For the strictly line-by-line differential analysis relative to the Sun, a narrower range of $T_{\rm eff}=4700$–6600 K is adopted. Each target is visually inspected to assess data quality and suitability for analysis. Systems showing strong rotational broadening ($> 15$ km~s$^{-1}$) or evidence of spectroscopic binarity (e.g., HD~57167, AF~Lep, and V1298~Tau) are excluded. Several additional systems analyzed in \citet{2024AJ....167..167S}, included in the Cycle 2 GO sample of \citet{2026ApJS..282...37S}, or lacking suitable archival spectra are listed in the tables but are not reanalyzed here. After these selections, we finally report abundances for 25 FGK stars.

\section{Stellar and Planetary Parameters}

We derive stellar atmospheric parameters for 20 FGK-type planet-hosting stars using a strictly differential, line-by-line analysis relative to the Sun (\citealt{2025ApJ...980..179S, 2025A&A...701A.107S, 2026ApJS..282...37S}). For five additional stars (TOI-4010, TOI-1807, HAT-P-18, HR~2562, and HD~20689), the available spectra do not allow reliable Fe I and Fe II measurements due to limited wavelength coverage or severe line blending. For these targets, we adopt stellar atmospheric parameters from the literature, as listed in Table~\ref{tab:parameters}.

Stellar parameters are derived following the methods described in \citet{2026ApJS..282...37S}. Briefly, equivalent widths (EWs) for elements with atomic number $Z \leq 30$ are measured using the {\it splot} task in IRAF\footnote{IRAF is distributed by the National Optical Astronomy Observatories, operated by the Association of Universities for Research in Astronomy Inc., under a cooperative agreement with the National Science Foundation.}. Solar spectra obtained with the same instruments are analyzed in parallel to perform a line-by-line differential comparison; for example, for spectra obtained with ESPRESSO, we adopt EWs measured from a solar ESPRESSO spectrum. We retain only lines with EWs between 10 and 150~m\AA. Stellar atmospheric parameters ($T_{\rm eff}$, $\log g$, $V_t$, and [Fe/H]) are then determined with the \texttt{q$^2$} Python package \citep{2014A&A...572A..48R} using MARCS model atmospheres \citep{2008A&A...486..951G}, by enforcing excitation and ionization balance of Fe I and Fe II lines.

Table~\ref{tab:parameters} summarizes the stellar atmospheric parameters and associated uncertainties (with [Fe/H] listed separately in Table~\ref{tab:abundance}), together with basic planetary properties. The first group lists the 20 stars homogeneously analyzed using the strictly differential, line-by-line method, while the second group includes the five stars for which stellar parameters are adopted from the literature. Planetary parameters are compiled from the literature and the NASA Exoplanet Archive, as described in the table notes. EW measurements for these stars are provided in Table~\ref{tab:ew}.

\begin{longrotatetable}[htbp!]
	\begin{deluxetable}{lccccccccccc}
		\tablecaption{Stellar and planetary parameters\label{tab:parameters}}
		\tabletypesize{\scriptsize}
		\tablewidth{0pt}
		\tablehead{
			\colhead{Star$^a$} & 
			\colhead{$T_{\rm eff}^a$ (K)} & 
			\colhead{$\sigma_{T_{\rm eff}^a}$ (K)} & 
			\colhead{$\log g^a$ (dex)} & 
			\colhead{$\sigma_{\log g}^a$} & 
			\colhead{$V_t^a$ (km s$^{-1}$)} & 
			\colhead{$\sigma_{V_t}^a$ (km s$^{-1}$)} &
			\colhead{Planet$^b$} &
			\colhead{$P$ (days)} &
			\colhead{$R_{\rm p}$} &
			\colhead{$M_{\rm p}$}
		}
		\startdata
		\hline
		\multicolumn{11}{c}{20 stars with stellar parameters and abundances derived from a strict line-by-line differential analysis in this work} \\
		\hline\hline
		WASP-77 A & 5615 & 9   & 4.42 & 0.03 & 0.95 & 0.03 & WASP-77 A b & 1.36 & 1.21$R_{\rm J}$ & 1.76$M_{\rm J}$ \\
		WASP-17   & 6577 & 153 & 4.44 & 0.35 & 1.90 & 0.90 & WASP-17 b & 3.74 & 1.93$R_{\rm J}$ & 0.496$M_{\rm J}$ \\
		HAT-P-26  & 4983 & 92  & 4.21 & 0.21 & 0.80 & 0.54 & HAT-P-26 b & 4.23 & 0.565$R_{\rm J}$ & 0.0596$M_{\rm J}$ \\
		WASP-6 & 5697 & 52 & 4.22 & 0.12 & 1.23 & 0.11 & WASP-6 b & 3.36 & 1.12$R_{\rm J}$ & 0.467$M_{\rm J}$ \\
		WASP-76   & 6350 & 20  & 4.25 & 0.05 & 1.56 & 0.03 & WASP-76 b & 1.81 & 1.83$R_{\rm J}$ & 0.926$M_{\rm J}$ \\
		WASP-63   & 5697 & 52  & 4.22 & 0.12 & 1.23 & 0.11 & WASP-63 b & 4.38 & 1.43$R_{\rm J}$ & 0.386$M_{\rm J}$ \\
		WASP-12   & 6184 & 69  & 4.09 & 0.14 & 1.50 & 0.10 & WASP-12 b & 1.09 & 1.79$R_{\rm J}$ & 1.416$M_{\rm J}$  \\
		TOI-1416  & 4977 & 123 & 4.12 & 0.19 & 0.80 & 0.56 & TOI-1416 b & 1.07 & 1.62$R_{\oplus}$ & 3.48$M_{\oplus}$ \\
		TOI-849   & 5778 & 9   & 4.45 & 0.02 & 0.92 & 0.02 & TOI-849b & 0.77 & 3.44$R_{\oplus}$ & 39.1$M_{\oplus}$ \\
		TOI-500   & 4977 & 91  & 4.76 & 0.22 & 1.13 & 0.25 & TOI-500 b & 0.55 & 1.17$R_{\oplus}$ & 1.42$M_{\oplus}$ \\
		&  &   &  &  &  &  & TOI-500 c & 6.64 & 2.09$R_{\oplus}$ & 5.03$M_{\oplus}$ \\
		&  &   &  &  &  &  & TOI-500 d & 26.2 & 6.36$R_{\oplus}$ & 33.12$M_{\oplus}$ \\
		&  &   &  &  &  &  & TOI-500 e & 61.3 & 3.99$R_{\oplus}$ & 15.05$M_{\oplus}$ \\
		TOI-451   & 5550 & 36  & 4.65 & 0.07 & 1.40 & 0.07 & TOI-451 b & 1.86 & 1.91$R_{\oplus}$ & 4.31$M_{\oplus}$ \\
		&  &   &  &  &  &  & TOI-451 c & 9.19 & 3.10$R_{\oplus}$ & 9.8$M_{\oplus}$ \\
		&  &   &  &  &  &  & TOI-451 d & 16.36 & 4.07$R_{\oplus}$ & 15.6$M_{\oplus}$ \\
		TOI-431   & 5021 & 103 & 4.64 & 0.18 & 1.33 & 0.23 & TOI-431 b & 0.49 & 1.28$R_{\oplus}$ & 3.07$M_{\oplus}$ \\
		&  &  &  &  & &  & TOI-431 c & 4.8 & 1.49$R_{\oplus}$ & 2.83$M_{\oplus}$ \\
		&  & &  & &  & & TOI-431 d & 12.5 & 3.29$R_{\oplus}$ & 9.9$M_{\oplus}$ \\
		TOI-199   & 5089 & 59  & 4.20 & 0.10 & 0.80 & 0.29 & TOI-199 b & 104.85 & 9.08$R_{\oplus}$ & 54$M_{\oplus}$ \\
		&  &   & &  & 0 &  & TOI-199 c & 273.69 & 1.01$R_{\rm J}$ & 0.28$M_{\rm J}$ \\
		KELT-8    & 5783 & 35  & 4.38 & 0.09 & 1.05 & 0.08 & KELT-8 b & 3.24 & 1.62$R_{\rm J}$ & 0.66$M_J$ \\
		HD 209100 & 4822 & 76  & 4.31 & 0.18 & 0.87 & 0.22 & $\epsilon$ Indi A b & 45.20 yrs & 3.25$R_{\rm J}$ & 6.31$M_{\rm J}$ \\
		HD 207496 & 4827 & 117 & 4.20 & 0.20 & 0.80 & 0.57 & HD 207496 b & 6.44 & 2.25$R_{\oplus}$ & 6.1$M_{\oplus}$ \\
		HD 20329  & 5612 & 8   & 4.37 & 0.02 & 0.92 & 0.02 & HD 20329 b & 0.93 & 1.72$R_{\oplus}$ & 7.42$M_{\oplus}$ \\
		HD 3167   & 5269 & 21  & 4.38 & 0.05 & 0.80 & 0.08 & HD 3167 b & 0.96 & 1.70$R_{\oplus}$ & 5.02$M_{\oplus}$ \\
		&  &  &  &  &    &  & HD 3167 c & 29.85 & 3.01$R_{\oplus}$ & 9.80$M_{oplus}$ \\
		&  &  &  &  &    &  & HD 3167 d & 8.51 & 1.92$R_{\oplus}$ & 6.90$M_{\oplus}$ \\
		&  &  &  &  &    &  & HD 3167e & 96.63 & 2.83$R_{\oplus}$ & 8.41$M_{\oplus}$ \\
		HATS-72   & 4701 & 146 & 4.10 & 0.24 & 0.80 & 0.48 & HATS-72 b & 7.33 & 8.10$R_{\oplus}$ & 39.86$M_{\oplus}$ \\
		eps Eri & 5085 & 24 & 4.48 & 0.08 & 0.84 & 0.10 & eps Eri b & 2671 & 1.25$R_{\rm J}$ & 0.66$M_{\rm J}$ \\
		\hline\hline
		\multicolumn{11}{c}{Five stars with stellar parameters adopted from the literature} \\
		\hline\hline
		HR 2562$^c$ & 6597 & 81 & 4.30 & 0.20 & 1.99 & 0.27 & HR 2562 b & 120 yrs & 1.11$R_{\rm J}$ & 30.06$M_{\rm J}$  \\
		HD 206893$^c$ & 6500 & 100 & 4.45 & 0.15 & 1.71 & 0.30 & HD 206893 b & 9350 & 1.25$R_{\rm J}$ & 28$M_{\rm J}$ \\
		&  &  &  &  &  &  & HD 206893 c & 2090 & 1.46$R_{\rm J}$ & 12.7$M_{\rm J}$ \\
		TOI-4010$^c$ & 4960 & 36 & 4.54 & 0.02 & 0.80 & 0.04 & TOI-4010 b & 1.3 & 3.02$R_{\oplus}$ & 11.00$M_{\oplus}$ \\
		&  &  &  &  &  &  & TOI-4010 c & 5.4 & 5.93$R_{\oplus}$ & 20.31$M_{\oplus}$ \\
		&  &  &  &  &    &  & TOI-4010 d & 14.7 & 6.18$R_{\oplus}$ & 38.15$M_{\oplus}$ \\
		TOI-1807$^c$ & 4914 & 59 & 4.60 & 0.02 & 0.80 & 0.05 & TOI-1807 b & 0.55 & 1.37$R_{\oplus}$ & 2.57$M_{\oplus}$ \\
		HAT-P-18$^c$ & 4803 & 80 & 4.57 & 0.04 & 0.80 & 0.08 & HAT-P-18 b & 5.51 & 0.995$R_{\rm J}$ & 0.197$R_{\rm J}$ \\
		\enddata
		\tablecomments{a. The star name, derived effective temperature ($T_{\rm eff}$), surface gravity ($\log g$), microturbulence ($V_t$), and their associated uncertainties. \\
			b. Planet names, orbital periods, radii, and masses compiled from the literature. Values for WASP-77 A b are from \citet{2013PASP..125...48M}, WASP-17 b from \citet{2010ApJ...709..159A}, HR 2562 b from \citet{2018AA...612A..92M}, HAT-P-26 b from \citet{2011ApJ...728..138H}, WASP-63 b from \citet{2012MNRAS.426..739H}, WASP-12 b from \citet{2009ApJ...693.1920H}, WASP-6 b from \citet{2023ApJ...944L..56M}, HD 206893 b, c from \citet{2023AA...671L...5H},TOI-4010 b, c, d from \citet{2023AJ....166....7K}, TOI-1807 b from \citet{2022AA...664A.163N}, TOI-1416 b from \citet{2023AA...677A..12D}, TOI-500 b, c, d, e from \citet{2022NatAs...6..736S}, TOI-451 b, c, d from \citet{2021AJ....161...65N}, TOI-431 b, c, d from \citet{2021MNRAS.507.2782O}, TOI-199 b, c from \citet{2023AJ....166..201H}, KELT-8 b from \citet{2017AJ....153..136S}, $\epsilon$ Indi Ab from \citet{2019MNRAS.490.5002F}, HD 207496 b from \citet{2023AA...673A...4B}, HD 20329 b from \citet{2022AA...668A.158M}, HD 3167 b, c, d, e from \citet{2025ApJS..278...52H}, HATS-72 b from \citet{2020AJ....159..173H}, HAT-P-18 b from \citet{2011ApJ...726...52H}, $\epsilon$ Eri b from \citet{2021AJ....162..181L}, and CT Cha b from \citet{2008AA...491..311S}. When either the planetary radius or mass is unavailable in the reference, the missing parameter is adopted from the NASA Exoplanet Archive. \\
			c. The $T_{\rm eff}$, $\log g$ of HR 2562 are adopted from Gaia DR3 (\citealt{2024AA...689A.185G}), HD 206893 from \citet{2021AJ....161....5W}, TOI-4010 from \citet{2023AJ....166....7K}, TOI-1807 from \citet{2023AJ....166...33M}, and HAT-P-18 from \citet{2017AA...602A.107B}. The microturbulence for these stars are estimated from the relation of \citet{1993AA...275..101E}, adopting 0.8 km s$^{-1}$ as the lower limit. Planetary parameters and references are compiled from the NASA Exoplanet Archive (accessed 2025 October).}
	\end{deluxetable}
\end{longrotatetable}

Among the five stars with stellar parameters from the literature, two exhibit strong line blending. For HD 206893, the high $T_{\rm eff}$ results in severe blending across most of the spectrum, allowing reliable EW measurements only for a small number of oxygen (EW = 145.9, 128.5, and 92.6~m\AA) and potassium (EW = 134.5~m\AA) lines. The spectrum of HR 2562 also shows significant blending, but the potassium line at 129.7 m\AA\ remains measurable.

This strict line-by-line differential analysis relative to the Sun is primarily designed for solar analogs with similar atmospheric properties. In the JEWELS series, we find that the method remains applicable over a moderately extended parameter space, covering $T_{\rm eff} = 4700$–6600 K and [Fe/H] = $-0.6$ to $0.5$ dex. Within this range, the atmospheric parameters derived from the excitation and ionization balance of Fe I and Fe II lines converge reliably.

\begin{figure}
	\centering
	\includegraphics[width=0.46\textwidth]{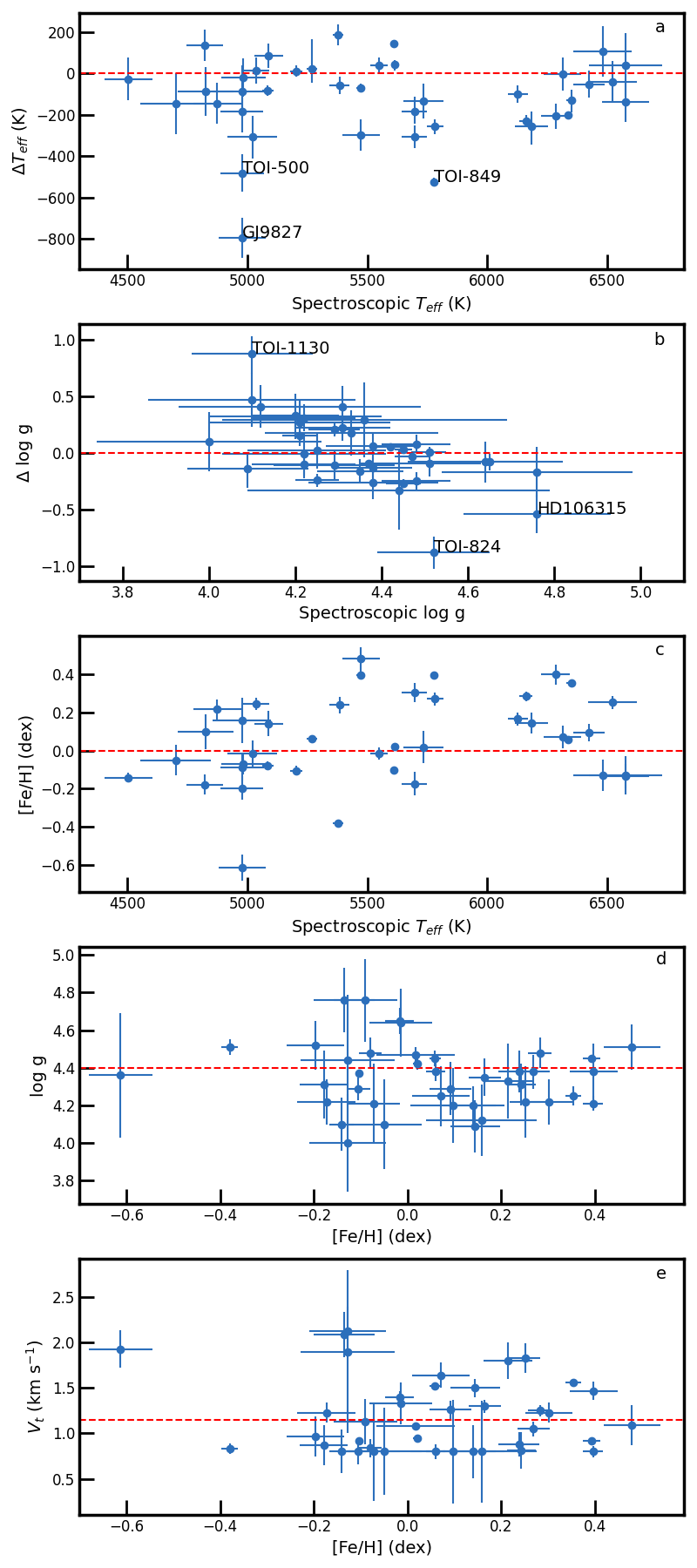}
	\caption{From top to bottom: (a) Gaia DR3 $T_{\rm eff}$ minus spectroscopic $T_{\rm eff}$; (b) Gaia DR3 $\log g$ minus spectroscopic $\log g$; (c) [Fe/H] versus spectroscopic $T_{\rm eff}$; (d) spectroscopic $\log g$ versus [Fe/H]; and (e) $V_t$ versus [Fe/H]. All JEWELS I \& II stars are shown. No trends are seen in these relations, except for a systematic trend in Gaia versus spectroscopic $\log g$.}
	\label{fig:diff_teff}
\end{figure}

In Figure~\ref{fig:diff_teff}a, we compare our spectroscopic $T_{\rm eff}$ of 39 stars with values from Gaia DR3 ($T_{\rm eff, \mathrm{GSP\text{-}Phot}}$), which are based on photometry. Our spectroscopic $T_{\rm eff}$ are listed in Table~\ref{tab:parameters} and Table 1 of JEWELS I, while the Gaia values are taken directly from the DR3 release (\citealt{2023A&A...674A...1G}) and shown in Table \ref{tab:age}. For most stars, the two measurements agree within $\sim$200 K, and we do not find any trend of $\Delta T_{\rm eff}$ with $T_{\rm eff}$. While small offsets between photometric and spectroscopic $T_{\rm eff}$ are expected, the absence of a systematic trend toward either hotter or cooler stars suggests that our temperature scale is stable across this parameter space. Three stars (TOI-849, TOI-500, and GJ~9827) show larger differences exceeding 400 K, with Gaia $T_{\rm eff}$ significantly lower than our spectroscopic values. These outliers are probably related to degeneracies in photometry-based $T_{\rm eff}$ estimates.

While the comparison in $T_{\rm eff}$ shows only a systematic offset, the Gaia DR3 $\log g$ values exhibit a slight trend relative to our spectroscopic results (Figure~\ref{fig:diff_teff}b). This likely reflects the indirect and model-dependent nature of Gaia $\log g$, which is inferred from luminosities and stellar evolution models. The observed trend is consistent with a mild compression of the $\log g$ range in the Gaia values.

We also examine the dependence of the derived [Fe/H] on $T_{\rm eff}$, $\log g$, and $V_t$ (Figure~\ref{fig:diff_teff}c–e). No trends are found with any of these parameters across our $T_{\rm eff}$ range, indicating that our analysis does not introduce systematic biases over this parameter space. We do, however, observe a slight increase in the [Fe/H] uncertainties toward cooler ($T_{\rm eff} < 5100$ K) and hotter ($T_{\rm eff} > 6400$ K) stars (up to $\sim$0.1 dex), but without any systematic trends. By adopting a homogeneous spectroscopic $T_{\rm eff}$ scale throughout the JEWELS series, together with a strictly differential line-by-line analysis, we minimize potential systematics arising from the use of different methods.

\begin{figure}
	\centering
	\includegraphics[width=0.46\textwidth]{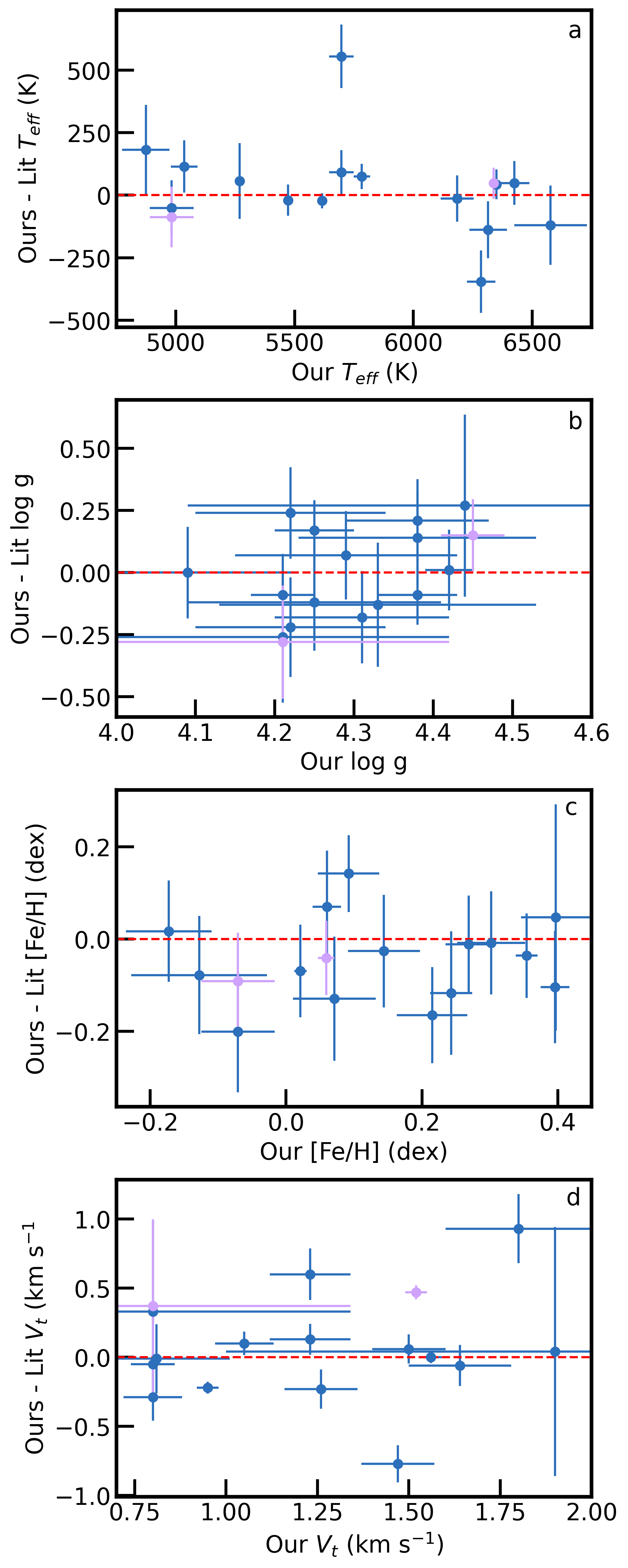}
	\caption{Differences between the stellar parameters derived in this work and those reported in the literature. Our sample includes 15 stars in common with the Ariel reference sample (blue circles; \citealt{2022A&A...663A.161M}) and two stars overlapping with GAPS (lilac circles; \citealt{2022A&A...664A.161B}). The observed scatter likely reflects a combination of instrumental offsets, differences in equivalent-width measurements and adopted atmospheric model grids, and methodological variations in the determination of stellar parameters. No systematic trends or offsets are apparent.}
	\label{fig:comp_Ariel}
\end{figure}

We further compare our derived stellar parameters with literature values from both the GAPS (\citealt{2022A&A...664A.161B}) and Ariel programs. The GAPS analysis adopts similar method as ours, including EW measurements with {\it splot} in IRAF, line-by-line differential analysis relative to the Sun over a comparable $T_{\rm eff}$ range, and abundance calculations with MOOG, but adopts a different grid of model atmospheres (ATLAS9). Two targets are in common, enabling a direct comparison (Figure~\ref{fig:comp_Ariel}); residual differences likely arise from the choice of model atmospheres, instrumental offsets, and variations in EW measurements due to continuum placement. A larger overlap exists with the Ariel sample (\citealt{2022A&A...663A.161M}), with 15 stars in common. Unlike our manual EW measurements, the Ariel analyses use the automated DAOSPEC pipeline, introducing a distinct set of systematics. While some scatter is apparent (Figure~\ref{fig:comp_Ariel}), no significant systematic offsets or trends are identified. We note that these methodological differences, combined with uncertainties propagated from stellar parameters, can affect derived chemical abundances. Here, we discuss results within our own abundance scale, without individual literature comparisons.

\section{Chemical Abundance Analysis}

We derive abundances for 19 elements, from C (Z = 6) to Zn (Z = 30), including Fe, for 23 planet hosting stars using the strictly differential, line by line method established in JEWELS I (\citealt{2026ApJS..282...37S}). Stellar parameters for 20 stars are determined on the same scale, while for the remaining five we adopt literature values (Table~\ref{tab:parameters}). Among these, three stars, TOI-4010, TOI-1807, and HAT-P-18, are analyzed with literature parameters, and we report abundances for most elements. The other two stars, HD~206893 and HR~2562, also rely on literature parameters but exhibit severe line blending, which prevents reliable measurements for most lines; nevertheless, a few lines remain usable upon visual inspection. For HD~206893, we measure the O~I triplet (EW = 145.9, 128.5, and 92.6 m\AA) and the K~I line (EW = 134.5 m\AA), yielding [O/H] = 0.046 $\pm$ 0.081 and [K/H] = 0.167 $\pm$ 0.085 dex; while for HR~2562 only the K~I line (EW = 129.7 m\AA) is measurable, giving [K/H] = 0.111 $\pm$ 0.120 dex.

Absolute abundances, A(X), are computed for each line using the {\it abfind} driver in MOOG and then converted to differential abundances relative to the Sun, [X/H]\footnote{[X/H] = A(X)${\star}$ – A(X)${\odot}$, where A(X) = 12 + $\log(N_{\rm X}/N_{\rm H})$, and $N_{\rm X}$ and $N_{\rm H}$ are the number densities of element X and hydrogen, respectively.}. For each element, the final abundance is derived by averaging the [X/H] values of all lines in linear space after excluding $2\sigma$ outliers. Line-to-line scatter ($\sigma$) and the standard deviation of the mean ($\sigma_\mu = \sigma/\sqrt{N}$) are computed for all elements. For species measured from a single line, such as potassium, $\sigma_\mu$ is estimated from the EW uncertainty, which depends on spectral S/N and instrument characteristics \citep{2025ApJ...980..179S}. Non-local thermodynamic equilibrium (NLTE) corrections from \citet{2007AA...465..271R} are applied to the oxygen triplet, which is most affected by NLTE effects. Table~\ref{tab:abundance} lists the final averaged abundances ([X/H]) for all 23 planet hosts, along with total uncertainties obtained by combining $\sigma_\mu$ and errors propagated from stellar atmospheric parameters in quadrature.

For the synthesis, microturbulence is included in the model atmospheres and consistently adopted in the spectral synthesis. Line broadening due to instrumental resolution, macroturbulence, and rotation is approximated using a single Gaussian profile with a fixed FWHM, as implemented in MOOG. This Gaussian approximation is appropriate given that all targets are relatively slow rotators, for which rotational broadening does not dominate the line profile; a rotational profile would be required for more rapidly rotating stars. Limb darkening is not included.

\begin{deluxetable*}{lccccccccccccccccccccccccc}
		\tabletypesize{\scriptsize}
		\tablecaption{Elemental abundances [X/H] and uncertainty for 23 planet-hosting stars in JWST.\label{tab:abundance}}
		\tablehead{
			\colhead{Atom$^a$} & \colhead{Species$^a$} & \colhead{T$_{\rm cond}^a$} &
			\multicolumn{4}{c}{WASP-77A$^b$} &
			\multicolumn{4}{c}{WASP-17} &
			\multicolumn{4}{c}{HAT-P-26} &
			\colhead{...} \\
			\colhead{} & \colhead{} & \colhead{} &
			\colhead{[X/H]$^b$} & \colhead{$\sigma_\mu^b$} & \colhead{Err$_{\rm atm}^b$} & \colhead{Err$_{\rm tot}^b$} &
			\colhead{[X/H]} & \colhead{$\sigma_\mu$} & \colhead{Err$_{\rm atm}$} & \colhead{Err$_{\rm tot}$} &
			\colhead{[X/H]} & \colhead{$\sigma_\mu$} & \colhead{Err$_{\rm atm}$} & \colhead{Err$_{\rm tot}$} &
			\colhead{...$^c$}
		}
		\startdata
		26 & Fe & 1334 & 0.021 & 0.003 & 0.009 & 0.009 & -0.128 & 0.020 & 0.098 & 0.100 & -0.071 & 0.005 & 0.054 & 0.054 & ... \\
		6  & C  & 40   & 0.086 & 0.060 & 0.008 & 0.061 & -0.135 & 0.033 & 0.108 & 0.113 & 0.090 & 0.163 & 0.129 & 0.208 & ... \\
		8  & O  & 180  & 0.008 & 0.020 & 0.011 & 0.023 & -0.070 & 0.037 & 0.162 & 0.166 & 0.156 & 0.020 & 0.161 & 0.162 & ...    \\
		11 & Na & 958  & 0.010 & 0.007 & 0.005 & 0.009 & -0.180 & 0.027 & 0.056 & 0.062 & 0.027 & 0.043 & 0.072 & 0.084 & ... \\
		12 & Mg & 1336 & 0.062 & 0.051 & 0.008 & 0.052 & -0.244 & 0.098 & 0.080 & 0.127 & 0.062 & 0.049 & 0.084 & 0.097 & ... \\
		13 & Al & 1653 & 0.023 & 0.038 & 0.005 & 0.038 & -0.140 & 0.081 & 0.044 & 0.092 & 0.233 & 0.010 & 0.068 & 0.069 & ... \\
		14 & Si & 1310 & -0.001 & 0.005 & 0.003 & 0.006 & -0.070 & 0.030 & 0.047 & 0.056 & 0.098 & 0.017 & 0.046 & 0.049 & ... \\
		16 & S  & 664  & 0.099 & 0.035 & 0.007 & 0.036 & -0.243 & 0.076 & 0.092 & 0.119 & -0.004 & --    & --    & 0.077 & ...    \\
		19 & K  & 1006 & 0.013 & 0.020 & 0.014 & 0.024 & -0.221 & 0.040 & 0.195 & 0.199 & --     & --    & --    & --    & ...   \\
		20 & Ca & 1517 & 0.051 & 0.009 & 0.010 & 0.013 & -0.177 & 0.025 & 0.088 & 0.091 & -0.006 & 0.014 & 0.120 & 0.121 & ... \\
		21 & Sc & 1659 & 0.019 & 0.014 & 0.010 & 0.017 & -0.176 & 0.067 & 0.118 & 0.136 & 0.098 & 0.045 & 0.088 & 0.099 & ... \\
		22 & Ti & 1582 & 0.046 & 0.006 & 0.011 & 0.013 & -0.008 & 0.054 & 0.112 & 0.124 & 0.057 & 0.012 & 0.107 & 0.108 & ... \\
		23 & V  & 1429 & 0.102 & 0.016 & 0.011 & 0.019 & 0.079 & 0.074 & 0.124 & 0.144 & 0.348 & 0.066 & 0.141 & 0.156 & ... \\
		24 & Cr & 1296 & 0.037 & 0.006 & 0.010 & 0.012 & -0.278 & 0.041 & 0.114 & 0.121 & -0.009 & 0.026 & 0.102 & 0.105 & ... \\
		25 & Mn & 1158 & 0.064 & 0.011 & 0.010 & 0.015 & 0.186 & 0.260 & 0.085 & 0.274 & 0.014 & 0.050 & 0.091 & 0.104 & ... \\
		27 & Co & 1352 & 0.059 & 0.011 & 0.007 & 0.013 & -0.102 & 0.095 & 0.094 & 0.134 & 0.135 & 0.021 & 0.039 & 0.044 & ... \\
		28 & Ni & 1353 & 0.031 & 0.005 & 0.006 & 0.008 & -0.184 & 0.056 & 0.083 & 0.100 & -0.030 & 0.009 & 0.011 & 0.014 & ... \\
		29 & Cu & 1037 & 0.062 & 0.033 & 0.006 & 0.034 & -0.276 & 0.000 & 0.077 & 0.077 & 0.018 & 0.038 & 0.010 & 0.039 & ... \\
		30 & Zn & 726  & -0.008 & 0.013 & 0.007 & 0.015 & -0.154 & 0.262 & 0.074 & 0.272 & 0.084 & 0.053 & 0.035 & 0.064 & ... \\
		\enddata
	\tablecomments{a. Atomic number, species, andcondensation temperature ($T_{cond}$) of each element. \\
		b. Host star name, mean abundance [X/H] for each element, $\sigma_{\mu}$, error propagated from atmosphere, and total uncertainty, which combines $\sigma_{\mu}$ and the propagated uncertainties from stellar atmosphere parameters in quadrature. The Oxygen abundance is after NLTE correction. \\
		c. Only the first three stars are shown here for illustration. The full table for all 23 targets is available in machine-readable format online. Among the 23 stars, TOI-4010, TOI-1807, and HAT-P-18 adopt stellar atmospheric parameters from the literature and are therefore not on the same abundance scale as the other 20 planet hosts. For HR~2562 and HD~206893, most spectral lines are blended; only a small number of important lines are measured, and their derived abundances are reported in the main text rather than in this table.}
\end{deluxetable*}

For cool dwarfs with $T_{\rm eff}$ below 5000 K, reliable carbon abundance determinations from atomic C I lines become increasingly challenging. In this temperature regime, many commonly used atomic transitions are intrinsically weak and/or affected by blending with molecular features, leading to systematically elevated abundances when EWs are used. In our sample, this behavior is evident for eight cool stars, HAT-P-26, TOI-4010, TOI-1807, TOI-1416, TOI-500, HD 209100, HD 207496, and HATS-72, for which [C/H] inferred from atomic-line EWs are anomalously high. A detailed inspection of the adopted C I lines shows that several lines (e.g., 5052.17, 6587.61, 7111.47, and 7113.18 \AA) are too weak to be measured robustly, while others (e.g., 5380.34, 7111.47, and 7113.18 \AA) suffer significant blending with molecular bands. This has also been demonstrated by previous work (\citealt{2020A&A...640A.123B, 2021A&A...655A..99D, 2022A&A...664A.161B}), which also show that different carbon indicators can yield distinct trends as a function of metallicity.

To obtain more reliable carbon abundances for these stars, we instead perform spectral synthesis of the C$_2$ molecular bands near 5165 \AA\ using the {\it synth} driver in MOOG. The adopted abundances, listed in Table \ref{tab:abundance}, correspond to the best-fitting synthetic spectra that simultaneously reproduce the C$_2$ band features, nearby atomic features, and the local continuum. For the spectral synthesis, we adopt an isotopic ratio of 99\% $^{12}$C and 1\% $^{13}$C. We also tested spectral synthesis of the C$2$ bands near 5128,\AA, but found that this region does not yield a satisfactory simultaneous fit to the molecular features and the surrounding spectrum, primarily due to difficulties in defining the local continuum. We therefore rely exclusively on the 5165,\AA\ bands for carbon abundance determinations in cool dwarfs ($T{\rm eff} < 5000$ K). For TOI-1807 and TOI-4010, the available spectra do not cover the C$_2$ band at 5165 \AA; therefore, carbon abundances are not reported for these stars.

We note that sulfur (S) abundances derived from atomic S I lines in cool dwarfs ($T_{\rm eff} < 5000$ K) are systematically high, sometimes exceeding 0.6 dex, primarily due to line blending. Specifically, S I 6046 and 6052.66 \AA\ are strongly blended; S I 6743.54 \AA\ shows moderate to strong blending; S I 6757.15 \AA\ exhibits moderate blending with a moderate NLTE effect; and the redder lines S I 8693.93 and 8694.62 \AA\ are mostly free of blends but are more affected by NLTE effects. Inspection of the NLTE corrections for S from \citet{2017ASPC..510..141K} indicates that, when performing a strict differential line-by-line analysis relative to the Sun, NLTE effects largely cancel out in cool dwarfs, leaving blending as the dominant source of systematic overestimation.

To improve the [S/H] determinations for cool dwarfs, we recompute [S/H] using spectral synthesis of atomic S I lines with the {\it synth} driver in MOOG. In the synthesis, we adopt sulfur isotope fractions of 95\% for $^{32}$S, 4.2\% for $^{34}$S, 0.75\% for $^{33}$S, and 0.02\% for $^{36}$S,. For the eight dwarfs with $T_{\rm eff}$ $<$ 5000 K, the redder S I lines at 8693.93 and 8694.62 \AA are not available. In addition, the spectra of TOI-1807 and TOI-4010 do not cover the 6000–7000 \AA\ region. For the remaining six cool dwarfs, we therefore derive [S/H] from four S I lines in the 6000–6800 \AA\ range and compute the mean [S/H] by averaging the individual line abundances in linear space. Table \ref{tab:abund_S} lists the synthesized abundances for each of the four S I lines for these six cool dwarfs, along with the corresponding measurements for the solar spectrum. The final sulfur abundances reported in Table \ref{tab:abundance} are the mean [S/H] values derived from these synthesized lines.

To ensure consistency with JEWELS I, we perform similar spectral synthesis of the S I lines for four cool dwarfs ($T_{\rm eff}$ $<$ 5000 K) in that sample. For a substantial fraction of the JEWELS I targets, however, the synthetic spectra do not simultaneously reproduce the S I absorption features and the local continuum. We therefore report sulfur abundances only for those JEWELS I stars for which the synthesis yields a satisfactory fit; these results are also listed in Table \ref{tab:abund_S}. We further note that sulfur abundances derived from EWs and from spectral synthesis can exhibit systematic offsets.

For the remaining elements, our strictly line-by-line differential analysis relative to the Sun largely mitigates NLTE effects for most stars. Departures may become noticeable only for stars with parameters significantly different from solar. For carbon, we use molecular features in cool dwarfs ($T_{\rm eff} < 5000$ K), which are insensitive to NLTE effects. In stars with $5000 \lesssim T_{\rm eff} \lesssim 6000$ K, any residual effects are expected to cancel in a differential sense. NLTE corrections for hotter stars are generally small. For sodium, NLTE effects are modest in solar and near-solar metallicity stars (e.g., \citealt{2011A&A...528A.103L}), with corrections below 0.03 dex for the commonly used 6154 and 6160 \AA\ lines. Even smaller effects are expected for the weaker 4751 and 5148 \AA\ lines. Stronger transitions that exhibit larger NLTE effects are not used. For manganese, available studies (e.g., \citealt{2019A&A...631A..80B}) indicate that NLTE effects largely cancel in a differential analysis relative to the Sun, with residuals around 0.05 dex across the parameter range of our sample.

Cobalt is affected by departures from LTE, with typical NLTE corrections of 0.05–0.1 dex in solar-type stars and larger corrections in cooler or metal-poor regimes (e.g., \citealt{2010MNRAS.401.1334B}). These corrections are strongly line-dependent, are not available for all transitions used here, and do not fully remove line-to-line scatter. For the remaining elements, no significant NLTE effects are known, particularly within a strict line-by-line differential analysis, though this may reflect the limited availability of comprehensive NLTE studies. Applying incomplete or heterogeneous NLTE corrections could introduce additional systematics, especially for a sample spanning a wide range of stellar parameters. For this reason, we apply NLTE corrections only to the oxygen triplet, which is both non-negligible and well-documented. Residual NLTE effects may persist, particularly for the coolest and most metal-poor stars, but these are expected to be minor and do not affect the relative abundance trends discussed in this work.

\section{Galactic chemical evolution} \label{sec:GCE}

We estimate stellar ages, masses, and radii by comparing spectroscopically derived parameters with theoretical stellar evolution models. $T_{\rm eff}$), log g, and [Fe/H] are used as inputs to the q$^2$ code (\citealt{Ramirez2014}), which interpolates the Yonsei-Yale isochrones (\citealt{2001ApJS..136..417Y}) over a wide range of ages and metallicities. q$^2$ employs a maximum likelihood approach to derive the most probable stellar age, mass, and radius, along with associated uncertainties. The results are compiled in Table~\ref{tab:age}. All 39 stars in JEWELS I \& II are analyzed using the same strictly differential, line-by-line method, ensuring that the derived ages are on the same scale. Here we note that ages for individual field stars remain challenging to determine, with potential systematic offsets relative to independent methods such as gyrochronology.

\begin{figure*}
	\centering
	\includegraphics[width=1.00\textwidth]{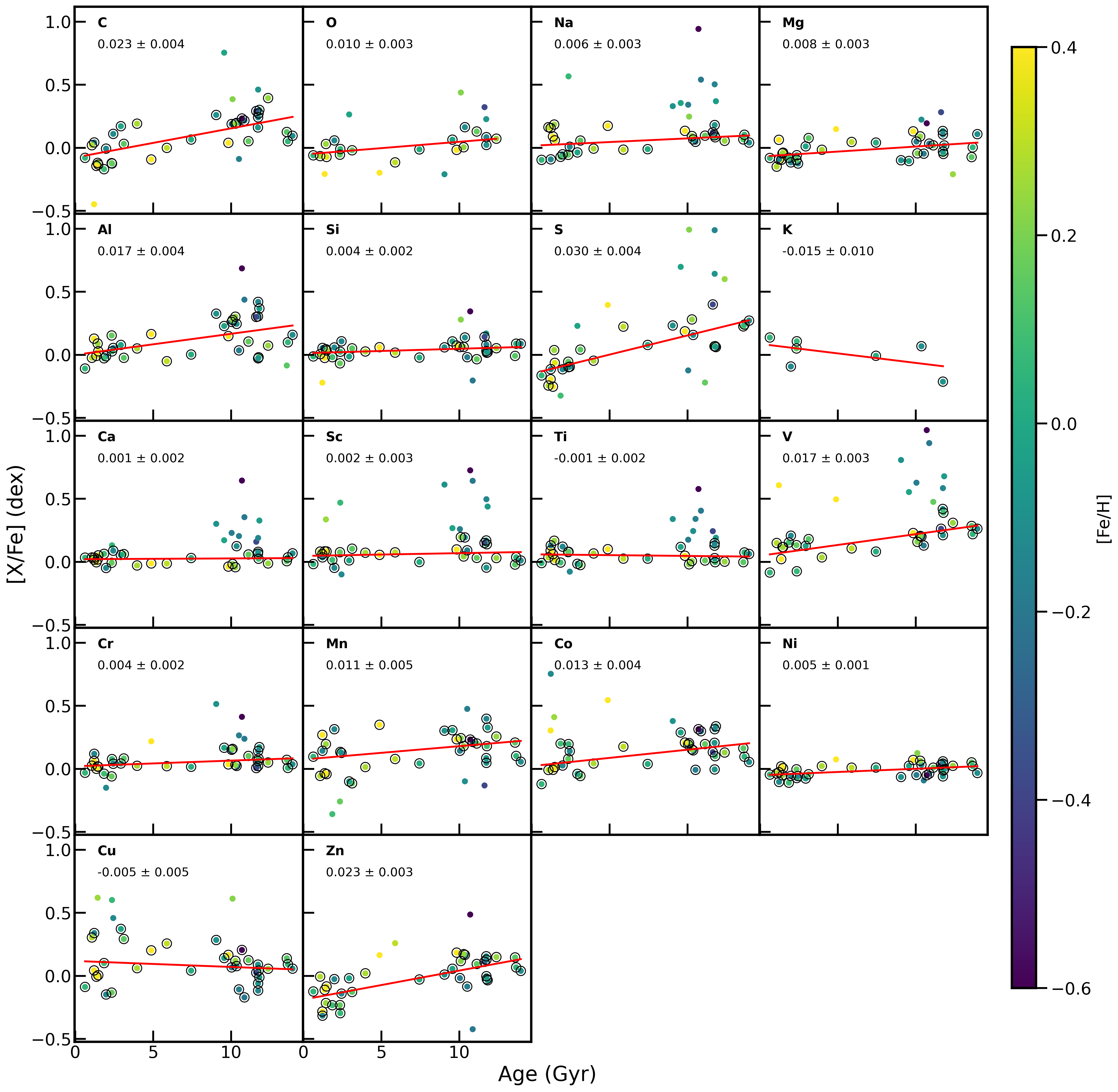}
	\caption{[X/Fe] as a function of isochrone age for 18 elements. The 39 JWST planet-hosting stars are color-coded by [Fe/H], and linear fits are shown after $2\sigma$ clipping. The slope of each linear fit and its uncertainty are indicated in each subplot.}
	\label{fig:GCE}
\end{figure*}

For our empirical study of Galactic chemical evolution, we adopt the isochrone ages derived above. Figure~\ref{fig:GCE} presents [X/Fe] as a function of age for 18 elements, with points colored by [Fe/H] and a linear trend fitted after 2$\sigma$ clipping. We generally observe [X/Fe] increasing with age for several elements (e.g., C, S, V, Zn), this behavior is qualitatively consistent with the increasing contribution of long-timescale nucleosynthetic sources relative to Fe. In contrast, elements such as Ca and Ti show relatively flat trends, which may suggest a dominant contribution from core-collapse supernovae with a roughly steady enrichment pattern. Potassium exhibits a decreasing trend, but the result is based on only a few stars and remains uncertain.

\begin{deluxetable*}{lccccccc}
	\tablecaption{Stellar properties of JEWELS I \& II planet-hosting stars \label{tab:age}}
	\tablehead{
		\colhead{Star} & \colhead{$T_{\rm eff}^a$} & \colhead{log g$^a$} & \colhead{Age$^b$} & \colhead{Mass$^b$} & \colhead{Radius$^b$} & \colhead{$T_{cond}$ slope$^c$} & \colhead{GCE-corrected $T_{cond}$ slope$^c$} \\
		\colhead{} & \colhead{(K)} & \colhead{(dex)} & \colhead{(Gyr)} & \colhead{($M_\odot$)} & \colhead{($R_\odot$)} & \colhead{($\times10^{-5}$ dex Gyr$^{-1}$)} & \colhead{($\times10^{-5}$ dex Gyr$^{-1}$)}
	}
	\startdata
	WASP-77A & 5659$_{-27}^{+19}$ & 4.47$_{-0.02}^{+0.01}$ & $7.44^{+1.21}_{-1.20}$ & $0.950^{+0.007}_{-0.008}$ & $0.995^{+0.033}_{-0.038}$ & $1.95^{+10.27}_{-10.09}$ & $5.29_{-10.08}^{+10.27}$ \\
	WASP-17  & 6618$_{-13}^{+13}$ & 4.11$_{-0.01}^{+0.02}$ & $1.98^{+0.95}_{-0.94}$ & $1.283^{+0.105}_{-0.143}$ & $1.539^{+0.345}_{-0.346}$ & $4.41^{+10.27}_{-10.09}$ & $7.24_{-10.08}^{+10.27}$ \\
	HAT-P-26 & 4962$_{-2}^{+7}$ & 4.48$_{-0.01}^{+0.01}$ & $11.73^{+3.15}_{-1.73}$ & $0.850^{+0.103}_{-0.103}$ & $1.315^{+0.707}_{-0.507}$ & $3.09^{+5.10}_{-6.49}$ & $14.11_{-10.08}^{+10.27}$ \\
	WASP-6   & 5391$_{-14}^{+13}$ & 4.53$_{-0.01}^{+0.01}$ & $10.51^{+1.69}_{-1.69}$ & $0.938^{+0.036}_{-0.036}$ & $1.186^{+0.185}_{-0.185}$ & $15.45^{+10.27}_{-10.09}$ & $21.36_{-10.08}^{+10.27}$ \\
	WASP-76  & 6220$_{-54}^{+47}$ & 4.01$_{-0.04}^{+0.03}$ & $1.48^{+0.46}_{-0.47}$ & $1.361^{+0.024}_{-0.024}$ & $1.427^{+0.108}_{-0.108}$ & $9.80^{+10.27}_{-10.09}$ & $5.79_{-1.45}^{+2.15}$ \\
	WASP-63  & 5513$_{-27}^{+52}$ & 4.12$_{-0.01}^{+0.01}$ & $5.88^{+1.18}_{-1.59}$ & $1.115^{+0.052}_{-0.053}$ & $1.348^{+0.179}_{-0.202}$ & $-1.02^{+2.49}_{-3.44}$ & $-1.21_{-3.55}^{+4.85}$ \\
	WASP-12  & 5930$_{-56}^{+18}$ & 3.95$_{-0.09}^{+0.03}$ & $3.12^{+0.65}_{-0.64}$ & $1.308^{+0.109}_{-0.110}$ & $1.844^{+0.457}_{-0.513}$ & $-2.25^{+2.95}_{-3.92}$ & $-0.16_{-10.08}^{+10.27}$ \\
	TOI-1416 & 4892$_{-1}^{+1}$ & 4.53$_{-0.01}^{+0.01}$ & $11.12^{+3.77}_{-1.77}$ & $0.960^{+0.126}_{-0.126}$ & $1.380^{+0.624}_{-0.554}$ & $13.94^{+10.27}_{-10.09}$ & $22.04_{-10.08}^{+10.27}$ \\
	TOI-849  & 5254$_{-4}^{+19}$ & 4.48$_{-0.02}^{+0.01}$ & $1.21^{+0.75}_{-0.54}$ & $1.120^{+0.008}_{-0.008}$ & $1.058^{+0.023}_{-0.026}$ & $35.65^{+10.27}_{-10.09}$ & $18.13_{-12.56}^{+17.44}$ \\
	TOI-500  & 4495$_{-12}^{+3}$ & 4.59$_{-0.01}^{+0.01}$ & $9.05^{+4.64}_{-4.64}$ & $0.771^{+0.028}_{-0.028}$ & $0.725^{+0.037}_{-0.037}$ & $20.88^{+10.27}_{-10.09}$ & $35.49_{-10.08}^{+10.27}$ \\
	TOI-451  & 5591$_{-4}^{+4}$ & 4.57$_{-0.01}^{+0.01}$ & $2.93^{+2.03}_{-2.03}$ & $0.945^{+0.017}_{-0.021}$ & $0.877^{+0.031}_{-0.026}$ & $-10.43^{+5.13}_{-5.47}$ & $-10.31_{-10.08}^{+10.27}$ \\
	TOI-431  & 4713$_{-18}^{+3}$ & 4.56$_{-0.01}^{+0.01}$ & $9.56^{+5.81}_{-3.30}$ & $0.793^{+0.030}_{-0.031}$ & $0.755^{+0.041}_{-0.035}$ & $-16.96^{+12.37}_{-12.46}$ & $-3.97_{-10.08}^{+10.27}$ \\
	TOI-199  & 5175$_{-9}^{+8}$ & 4.53$_{-0.01}^{+0.01}$ & $13.58^{+1.69}_{-1.34}$ & $0.916^{+0.078}_{-0.112}$ & $1.269^{+0.823}_{-0.445}$ & $-17.06^{+7.43}_{-7.36}$ & $-13.19_{-9.52}^{+10.19}$ \\
	KELT-8   & 5526$_{-9}^{+11}$ & 4.25$_{-0.01}^{+0.01}$ & $3.97^{+1.39}_{-1.40}$ & $1.104^{+0.024}_{-0.023}$ & $1.104^{+0.089}_{-0.120}$ & $4.20^{+10.27}_{-10.09}$ & $5.10_{-10.08}^{+10.27}$ \\
	HD 209100 & 4959$_{-12}^{+12}$ & 4.72$_{-0.02}^{+0.02}$ & $10.04^{+4.84}_{-4.84}$ & $0.722^{+0.024}_{-0.024}$ & $0.675^{+0.025}_{-0.025}$ & $3.61^{+6.07}_{-8.08}$ & $29.90_{-10.08}^{+10.27}$ \\
	HD 207496 &  4741$_{-1}^{+1}$ & 4.52$_{-0.01}^{+0.01}$ & $11.77^{+3.15}_{-2.15}$ & $0.852^{+0.116}_{-0.114}$ & $1.297^{+0.606}_{-0.472}$ & $-10.06^{+7.41}_{-8.28}$ & $0.30_{-10.08}^{+10.27}$ \\
	HD 20329 & 5755$_{-3}^{+3}$ & 4.28$_{-0.01}^{+0.01}$ & $10.36^{+0.64}_{-0.65}$ & $0.911^{+0.004}_{-0.004}$ & $1.033^{+0.030}_{-0.023}$ & $1.10^{+4.79}_{-5.19}$ & $8.50_{-10.08}^{+10.27}$ \\
	HD 3167$^a$  & 5292$_{-63}^{+141}$ & -- & $13.65^{+1.57}_{-1.28}$ & $0.850^{+0.010}_{-0.009}$ & $0.926^{+0.026}_{-0.034}$ & $-1.46^{+2.99}_{-3.34}$ & $5.07_{-4.57}^{+5.78}$ \\
	HATS-72  & 4556$_{-26}^{+27}$ & 4.57$_{-0.01}^{+0.01}$ & $11.82^{+3.10}_{-3.10}$ & $0.833^{+0.148}_{-0.149}$ & $1.467^{+1.131}_{-0.835}$ & $13.92^{+10.27}_{-10.09}$ & $54.15_{-10.08}^{+10.27}$ \\
	eps Eri  & 5002$_{-5}^{+5}$ & 4.56$_{-0.01}^{+0.01}$ & $11.74^{+3.14}_{-3.14}$ & $0.793^{+0.015}_{-0.016}$ & $0.782^{+0.026}_{-0.022}$ & $-18.55^{+9.14}_{-9.60}$ & $-5.56_{-10.08}^{+10.27}$ \\
	WASP-121 & 6483$_{-11}^{+11}$ & 4.21$_{-0.01}^{+0.01}$ & $1.44^{+0.66}_{-0.59}$ & $1.403^{+0.112}_{-0.106}$ & $1.512^{+0.701}_{-0.201}$ & $11.22^{+10.27}_{-10.09}$ & $2.44_{-6.98}^{+8.88}$ \\
	WASP-94A & 6078$_{-13}^{+9}$ & 4.12$_{-0.01}^{+0.01}$ & $1.36^{+0.77}_{-0.67}$ & $1.338^{+0.041}_{-0.047}$ & $1.366^{+0.117}_{-0.061}$ & $15.87^{+10.27}_{-10.09}$ & $14.70_{-10.08}^{+10.27}$ \\
	WASP-69  & 4729$_{-3}^{+2}$ & 4.51$_{-0.01}^{+0.01}$ & $10.10^{+4.78}_{-2.77}$ & $0.814^{+0.034}_{-0.035}$ & $0.781^{+0.049}_{-0.049}$ & $-31.56^{+13.81}_{-13.43}$ & $-33.90_{-13.39}^{+14.27}$ \\
	WASP-52  & 5051$_{-29}^{+24}$ & 4.53$_{-0.01}^{+0.01}$ & $12.38^{+2.54}_{-2.55}$ & $0.850^{+0.021}_{-0.021}$ & $0.862^{+0.043}_{-0.043}$ & $-8.32^{+4.98}_{-5.11}$ & $-2.39_{-10.08}^{+10.27}$ \\
	WASP-47  & 5400$_{-11}^{+12}$ & 4.36$_{-0.01}^{+0.01}$ & $9.83^{+0.54}_{-0.54}$ & $1.024^{+0.014}_{-0.016}$ & $1.332^{+0.077}_{-0.077}$ & $1.77^{+3.85}_{-4.46}$ & $4.11_{-3.96}^{+4.49}$ \\
	WASP-18  & 6312$_{-1}^{+2}$ & 4.27$_{-0.01}^{+0.01}$ & $2.36^{+0.71}_{-0.86}$ & $1.245^{+0.059}_{-0.058}$ & $1.345^{+0.181}_{-0.181}$ & $14.20^{+10.27}_{-10.09}$ & $3.37_{-7.61}^{+10.81}$ \\
	WASP-15  & 6371$_{-13}^{+14}$ & 4.18$_{-0.01}^{+0.01}$ & $1.85^{+0.76}_{-0.76}$ & $1.285^{+0.042}_{-0.042}$ & $1.350^{+0.264}_{-0.132}$ & $18.58^{+10.27}_{-10.09}$ & $22.99_{-10.08}^{+10.27}$ \\
	TOI-824  & 4794$_{-48}^{+22}$ & 3.64$_{-0.07}^{+0.05}$ & $10.86^{+4.02}_{-4.02}$ & $0.751^{+0.029}_{-0.029}$ & $0.713^{+0.036}_{-0.036}$ & $36.16^{+10.27}_{-10.09}$ & $81.15_{-10.08}^{+10.27}$ \\
	TOI-561  & 5565$_{-44}^{+44}$ & 4.52$_{-0.01}^{+0.02}$ & $11.62^{+2.00}_{-2.07}$ & $0.791^{+0.014}_{-0.014}$ & $0.812^{+0.026}_{-0.026}$ & $-6.31^{+7.40}_{-7.68}$ & $2.00_{-10.08}^{+10.27}$ \\
	TOI-1130 & 4477$_{-28}^{+27}$ & 4.98$_{-0.11}^{+0.06}$ & $11.74^{+3.18}_{-3.19}$ & $0.671^{+0.028}_{-0.028}$ & $0.640^{+0.024}_{-0.024}$ & $-3.25^{+13.43}_{-14.77}$ & $-5.40_{-10.37}^{+14.19}$ \\
	TOI-125  & 5216$_{-8}^{+8}$ & 4.50$_{-0.01}^{+0.01}$ & $13.95^{+1.01}_{-0.79}$ & $0.803^{+0.010}_{-0.010}$ & $0.851^{+0.019}_{-0.019}$ & $-0.76^{+3.67}_{-4.30}$ & $11.63_{-10.08}^{+10.27}$ \\
	NGTS-2   & 6587$_{-16}^{+21}$ & 4.10$_{-0.01}^{+0.01}$ & $2.44^{+0.75}_{-0.75}$ & $1.359^{+0.174}_{-0.174}$ & $1.866^{+0.669}_{-0.669}$ & $-8.97^{+8.01}_{-10.02}$ & $-19.20_{-8.28}^{+12.45}$ \\
	LTT 9779 & 5329$_{-1}^{+1}$ & 4.44$_{-0.01}^{+0.01}$ & $10.28^{+3.29}_{-3.29}$ & $0.942^{+0.021}_{-0.022}$ & $1.008^{+0.096}_{-0.096}$ & $-4.58^{+4.39}_{-4.40}$ & $1.12_{-10.08}^{+10.27}$ \\
	Kepler-12 & 6029$_{-10}^{+10}$ & 4.19$_{-0.01}^{+0.02}$ & $2.34^{+0.95}_{-0.96}$ & $1.203^{+0.027}_{-0.026}$ & $1.185^{+0.154}_{-0.154}$ & $11.50^{+10.27}_{-10.09}$ & $10.67_{-10.08}^{+10.27}$ \\
	HD 106315 & 6439$_{-2}^{+3}$ & 4.22$_{-0.01}^{+0.01}$ & $1.22^{+0.98}_{-0.91}$ & $1.243^{+0.041}_{-0.044}$ & $1.238^{+0.093}_{-0.093}$ & $10.15^{+10.27}_{-10.09}$ & $15.88_{-10.08}^{+10.27}$ \\
	HAT-P-30 & 6136$_{-2}^{+4}$ & 4.18$_{-0.01}^{+0.01}$ & $0.63^{+0.65}_{-0.50}$ & $1.236^{+0.016}_{-0.016}$ & $1.185^{+0.037}_{-0.037}$ & $3.07^{+10.27}_{-10.09}$ & $2.32_{-10.08}^{+10.27}$ \\
	GJ 9827  & 4182$_{-1}^{+1}$ & 4.65$_{-0.01}^{+0.01}$ & $10.70^{+4.18}_{-4.18}$ & $0.675^{+0.027}_{-0.026}$ & $0.636^{+0.041}_{-0.041}$ & $11.64^{+15.90}_{-21.18}$ & $-2.86_{-19.80}^{+31.05}$ \\
	GJ 504  & 5931$_{-2}^{+2}$ & 4.23$_{-0.01}^{+0.01}$ & $1.07^{+0.66}_{-0.66}$ & $1.244^{+0.015}_{-0.015}$ & $1.205^{+0.049}_{-0.049}$ & $3.90^{+10.27}_{-10.09}$ & $4.02_{-10.08}^{+10.27}$ \\
	14 Her  & 5177$_{-1}^{+1}$ & 4.42$_{-0.01}^{+0.01}$ & $4.88^{+2.88}_{-2.93}$ & $1.019^{+0.031}_{-0.033}$ & $1.004^{+0.108}_{-0.108}$ & $5.94^{+5.91}_{-7.97}$ & $5.13_{-9.77}^{+11.89}$ \\
	\enddata
	\tablecomments{a. $T_{\rm eff}$ from Gaia DR3 for comparison, HD 3167 does not have DR3 $T_{\rm eff}$, so we adopt $T_{\rm eff}$ from DR2. \\
		b. Age, mass, and radius of the star from isochrone fitting using the q$^2$ package.
		c. Original and GCE-corrected condensation temperature slope ($T_{\rm cond}$) for each star.}
\end{deluxetable*}

\section{Discussion}

\subsection{Host Stellar Abundances and Planetary Properties}

\begin{figure}
	\centering
	\includegraphics[width=0.46\textwidth]{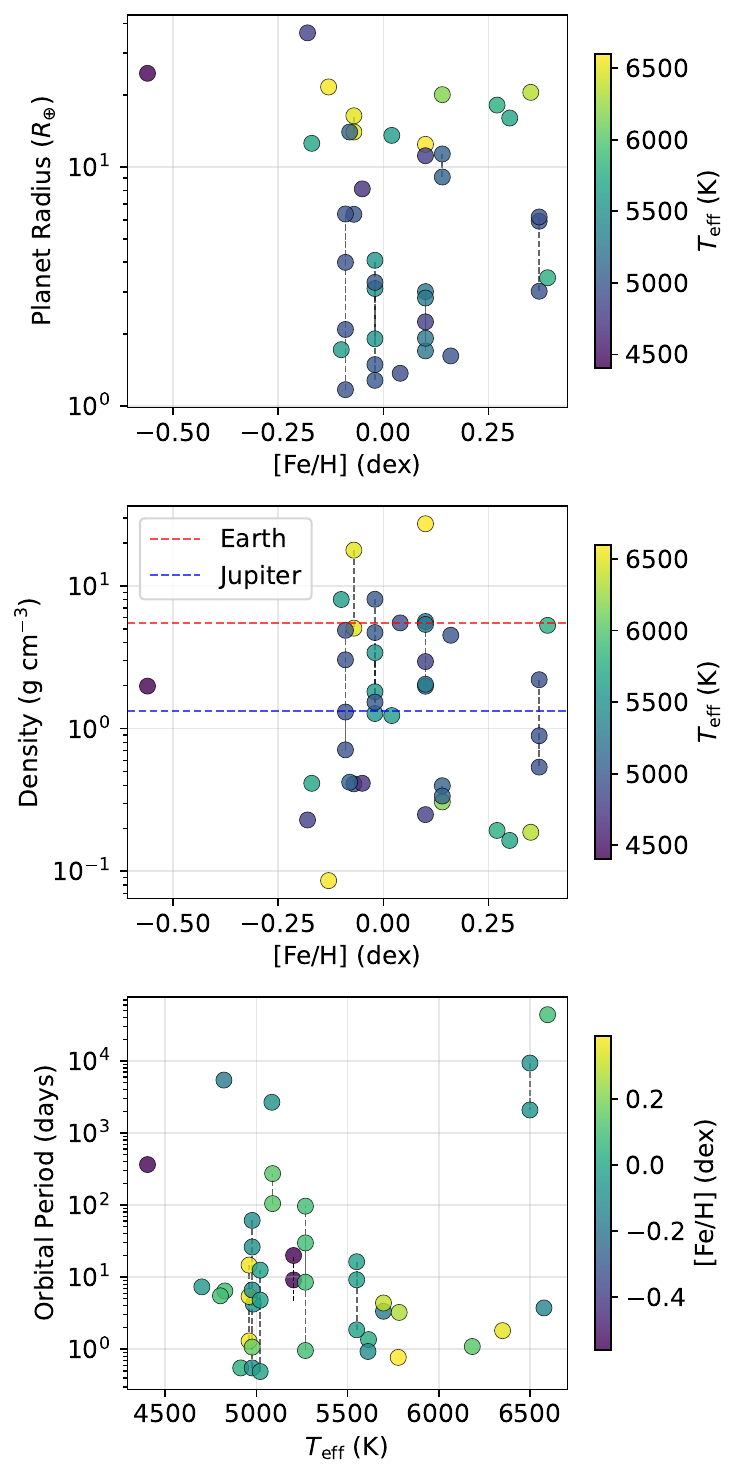}
	\caption{Relations between stellar and planetary properties. In the top two panels, planet radius and density are shown as functions of stellar metallicity ([Fe/H]), with points color-coded by stellar effective temperature ($T_{\rm eff}$). The bottom panel displays orbital period versus $T_{\rm eff}$, color-coded by [Fe/H]. The multi-planet systems are linked by black-dashed lines. The sample is small and selected from specific JWST observing programs targeting interesting planetary systems; therefore the distributions are subject to selection bias, and any apparent trends should not be interpreted as representative of the broader exoplanet population.}
	\label{fig1}
\end{figure}

Elemental abundances of planet-hosting stars provide important context for both stellar and planetary formation. For instance, gas giants are preferentially found around metal-rich stars, consistent with the core-accretion scenario (e.g., \citealt{2005ApJ...622.1102F, 2004A&A...415.1153S, 2010PASP..122..905J}). The JEWELS II sample considered here is not designed to probe population-level trends; instead, we present the stellar and planetary properties to illustrate the diversity of host stars and planetary systems (Figure~\ref{fig1}).

\begin{figure}
	\centering
	\includegraphics[width=0.46\textwidth]{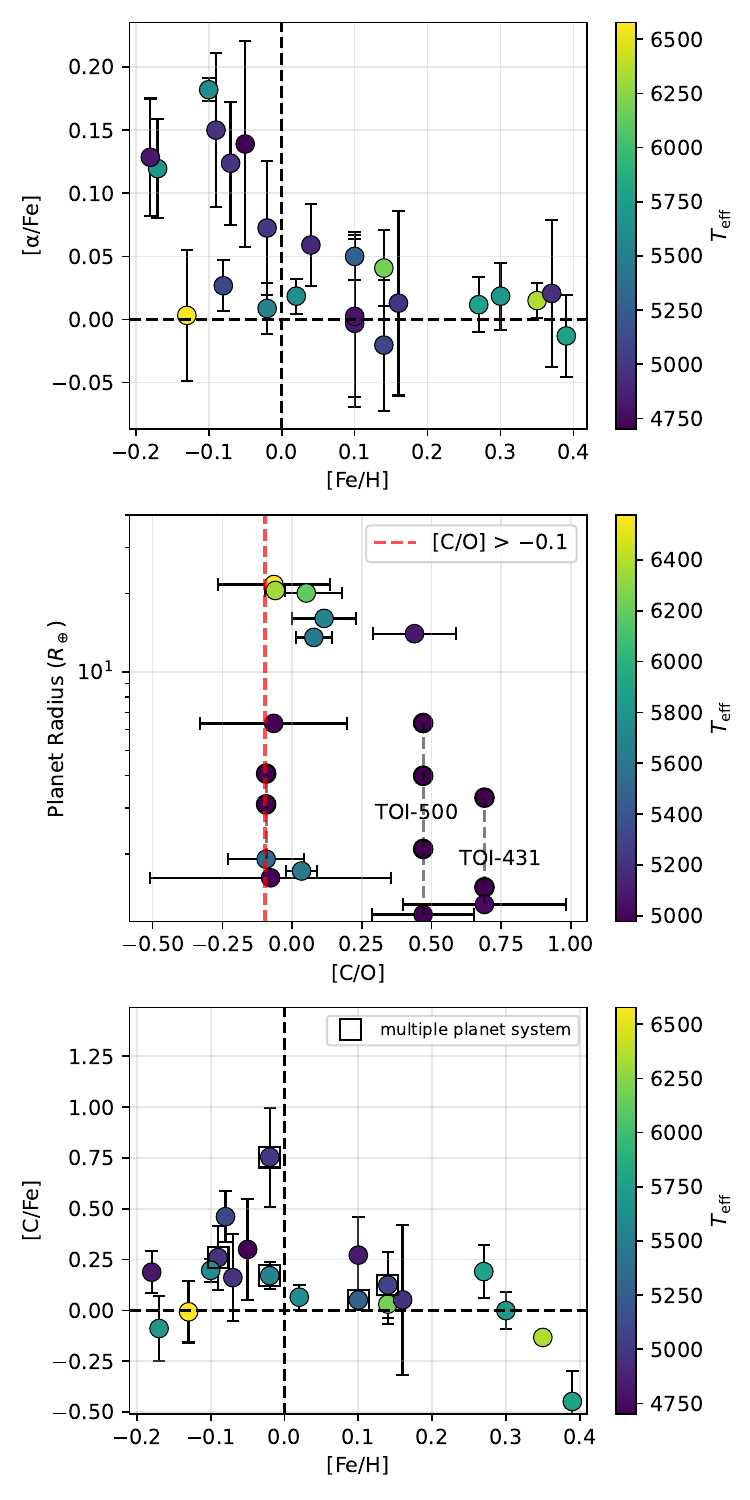}
	\caption{Relations between planetary properties and host stellar abundances. In the top panel, $[\alpha/Fe]$ is computed as the average of individual abundance ratios for $\alpha$-elements (Mg, Si, Ca, Ti) relative to Fe. The middle panel shows planet radius as a function of [C/O]. Both upper panels are color-coded by stellar $T_{\rm eff}$. The bottom panel presents [C/Fe] versus [Fe/H], where several metal-poor but carbon-enhanced stars are identified.}
	\label{fig2}
\end{figure}

We next examine planetary properties in the context of stellar chemical abundances. The top panel of Figure~\ref{fig2} shows the mean differential $\alpha$-element abundance, computed from Mg, Si, Ca, and Ti, relative to iron as a function of [Fe/H]. These elements are primarily produced in core-collapse supernovae and trace the chemical enrichment history of the interstellar medium. Most stars in our sample cluster near solar [$\alpha$/Fe] across the full metallicity range. A small number of slightly metal-poor stars show mild $\alpha$-element enhancement, consistent with the low-metallicity extension of the Galactic disk sequence (e.g., \citealt{2015ApJ...808..132H}), while stars at supersolar metallicity remain close to solar $\alpha$ abundance. We do not identify a distinct population of $\alpha$-rich, metal-rich stars.

The middle panel of Figure~\ref{fig2} shows planet radius as a function of stellar [C/O]. Carbon-to-oxygen ratios influence the dominant molecular species in exoplanet atmospheres, with higher values favoring CO and CO$_2$ over H$_2$O (e.g., \citealt{2012ApJ...758...36M, 2016ApJ...831...20B}). In this sample, planets span a wide range of radii at both low and high [C/O], including several multi-planet systems (e.g. TOI-500, TOI-431) at relatively high [C/O].

The bottom panel of Figure~\ref{fig2} presents [C/Fe] as a function of [Fe/H]. A few stars at subsolar metallicity exhibit modest carbon enhancement. While such variations may affect disk composition and planet formation (e.g., \citealt{2014ApJ...785..125A, 2023A&A...678A..33J, 2024ApJ...976..202P, 2025ApJ...991L..46C}), the enhancements observed here are small and do not indicate a distinct carbon-enhanced population. Alternative origins, such as internal mixing (e.g., \citealt{2008ApJ...679.1541D}) or binary interaction (e.g., \citealt{2015A&A...581A..22A}), may also contribute in individual cases; however, these mechanisms typically produce ultra carbon-rich, metal-poor stars, which is not the case here. Moreover, the observed carbon enhancements are modest and, when accounting for uncertainties, some targets may no longer be classified as carbon-rich.

Overall, the JEWELS II sample spans a broad range of stellar metallicities, $\alpha$-element abundances, and carbon ratios, reflecting diverse chemical environments. Owing to the small and targeted nature of the sample, we do not identify strong correlations with planetary properties, but instead use these data to provide context for individual systems and comparison with larger samples.

\subsection{The $T_{\rm cond}$ slope distribution}

For the 39 targets in JEWELS I \& II, we quantify the condensation temperature ($T_{\rm cond}$) trends by performing linear fits to the measured abundance ratios, [X/Fe], as a function of $T_{\rm cond}$. The slopes and their uncertainties are derived using a Markov Chain Monte Carlo (MCMC) approach implemented with the \texttt{emcee} package, accounting for the observational uncertainties in [X/Fe], following methods similar to those described in \citet{2025ApJ...980..179S, 2025A&A...701A.107S}.

To account for the effects of Galactic chemical evolution (GCE), we correct the elemental abundances using the empirical [X/Fe]–age relations derived in Section~\ref{sec:GCE}. The GCE-corrected abundances are then used to recompute the $T_{\rm cond}$ slopes following the same MCMC procedure. This enables a direct comparison between the original and GCE-corrected $T_{\rm cond}$ trends for each star. The resulting $T_{\rm cond}$ slopes are listed in Table~\ref{tab:age}, with the corresponding linear fits for each star shown in Figures~\ref{fig:Tc_JEWELS1}–\ref{fig:Tc_JEWELS2_GCE}.

\begin{figure*}
	\centering
	\includegraphics[width=1.00\textwidth]{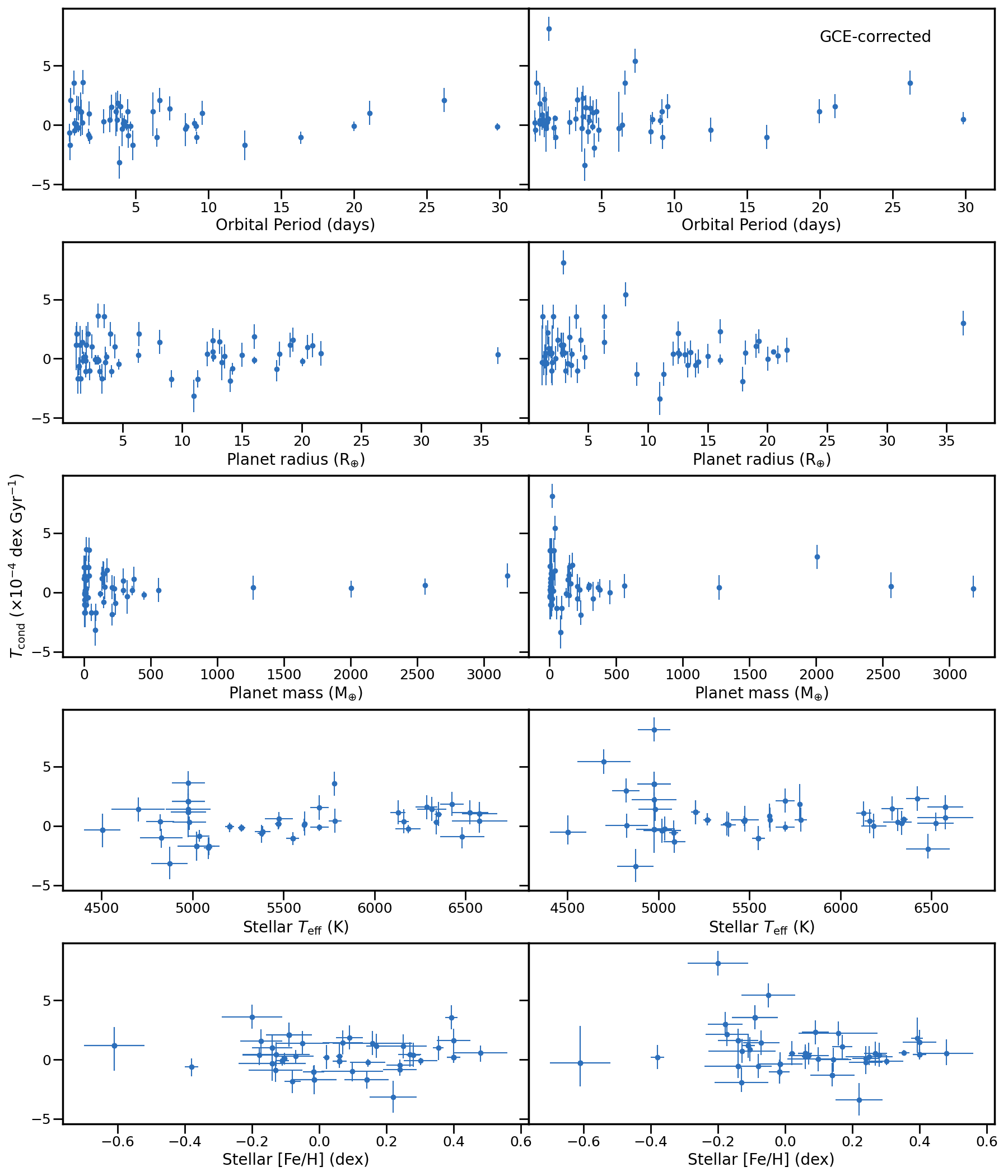}
	\caption{$T_{\rm cond}$ slopes versus stellar and planetary properties. Left panels show slopes from the original abundances, while right panels show slopes after applying GCE corrections.}
	\label{fig:Tc_dist}
\end{figure*}

We find that the $T_{\rm cond}$ trend slopes exhibit substantial star-to-star scatter, both before and after applying GCE corrections. While the GCE adjustment alters the slopes for some stars, it does not systematically reduce the dispersion or drive the slopes toward a common value. Some stars show significant changes after correction, whereas others remain largely unaffected. This non-uniform response suggests that a simple one-dimensional GCE correction cannot fully capture the chemical diversity of the sample, and that the $T_{\rm cond}$ slopes reflect intrinsic differences among the stars. In Figure~\ref{fig:Tc_dist}, we examine the dependence of $T_{\rm cond}$ slopes on planetary properties, including planet mass, radius, and orbital period, and find no correlations, regardless of whether GCE correction is applied. We also do not observe clear trends with stellar $T_{\rm eff}$ or [Fe/H].

These results imply that no single stellar or planetary parameter dominates the observed $T_{\rm cond}$ behavior. Instead, the lack of correlations likely reflects the combined influence of multiple processes. Unlike studies of solar twins (e.g., \citealt{2025ApJ...980..179S, 2025A&A...701A.107S}), our sample spans a broader range of stellar parameters, introducing entangled effects such as GCE, variations in stellar structure, mixing efficiency, and planet formation environment. Together, these factors increase the dispersion in $T_{\rm cond}$ slopes, making it challenging to isolate subtle signatures potentially linked to planet. Nevertheless, individual stars with extreme $T_{\rm cond}$ slopes could be interesting for follow-up studies. Stars with strongly positive slopes may be good candidates to look for planet engulfment signatures, while those with negative slopes could help probe possible refractory element depletion related to planet formation.

\section*{acknowledgements}

This work is supported by the National Key R\&D Program of China under Grant No. 2024YFA1611801, the Science and Technology Commission of Shanghai Municipality under Grant No. 25ZR1402244, and the Shanghai Jiao Tong University Funds Program No. AF4260012. This work is also supported in part by Office of Science and Technology, Shanghai Municipal Government (grant Nos. 24DX1400100, ZJ2023-ZD-001). This work is based, in part, on observations made with the NASA/ESA/CSA James Webb Space Telescope.

Some of the data presented herein were obtained at Keck Observatory, which is a private 501(c)3 non-profit organization operated as a scientific partnership among the California Institute of Technology, the University of California, and the National Aeronautics and Space Administration. The Observatory was made possible by the generous financial support of the W. M. Keck Foundation.This research has made use of the Keck Observatory Archive (KOA), which is operated by the W. M. Keck Observatory and the NASA Exoplanet Science Institute (NExScI), under contract with the National Aeronautics and Space Administration.

This research is also based on data obtained from the ESO Science Archive Facility under request 108.22JQ.001, 089.C-0471(A), 094.A-9012(A), 109.23FU.005, 1102.C-0744(C), 094.A-9010(A), 090.C-0146(A), 1102.C-0249(F), 110.246L.001, 112.25HY.001, 112.25T4.001, 106.21ER.001, 097.C-0280(A), 192.C-0852(A), 1102.C-0249(C), 112.25T4.001, 099.C-0431(A), 0103.C-0360(A), 60.A-9122(B), 60.A-9036(A), 092.A-9029(A), 0102.D-0185(A), 089.D-0202(A).
	
\bibliography{sun26_JWST}{}

\begin{thebibliography}{}
\expandafter\ifx\csname natexlab\endcsname\relax\def\natexlab#1{#1}\fi
\providecommand{\url}[1]{\href{#1}{#1}}
\providecommand{\dodoi}[1]{doi:~\href{http://doi.org/#1}{\nolinkurl{#1}}}
\providecommand{\doeprint}[1]{\href{http://ascl.net/#1}{\nolinkurl{http://ascl.net/#1}}}
\providecommand{\doarXiv}[1]{\href{https://arxiv.org/abs/#1}{\nolinkurl{https://arxiv.org/abs/#1}}}

\bibitem[{{Abate} {et~al.}(2015){Abate}, {Pols}, {Izzard}, \&
  {Karakas}}]{2015A&A...581A..22A}
{Abate}, C., {Pols}, O.~R., {Izzard}, R.~G., \& {Karakas}, A.~I. 2015, \aap,
  581, A22, \dodoi{10.1051/0004-6361/201525876}

\bibitem[{{Adibekyan} {et~al.}(2015){Adibekyan}, {Santos}, {Figueira}, {Dorn},
  {Sousa}, {Delgado-Mena}, {Israelian}, {Hakobyan}, \&
  {Mordasini}}]{2015A&A...581L...2A}
{Adibekyan}, V., {Santos}, N.~C., {Figueira}, P., {et~al.} 2015, \aap, 581, L2,
  \dodoi{10.1051/0004-6361/201527059}

\bibitem[{{Ahrer} {et~al.}(2023){Ahrer}, {Stevenson}, {Mansfield}, {Moran},
  {Brande}, {Morello}, {Murray}, {Nikolov}, {Petit dit de la Roche},
  {Schlawin}, {Wheatley}, {Zieba}, {Batalha}, {Damiano}, {Goyal}, {Lendl},
  {Lothringer}, {Mukherjee}, {Ohno}, {Batalha}, {Battley}, {Bean}, {Beatty},
  {Benneke}, {Berta-Thompson}, {Carter}, {Cubillos}, {Daylan}, {Espinoza},
  {Gao}, {Gibson}, {Gill}, {Harrington}, {Hu}, {Kreidberg}, {Lewis}, {Line},
  {L{\'o}pez-Morales}, {Parmentier}, {Powell}, {Sing}, {Tsai}, {Wakeford},
  {Welbanks}, {Alam}, {Alderson}, {Allen}, {Anderson}, {Barstow}, {Bayliss},
  {Bell}, {Blecic}, {Bryant}, {Burleigh}, {Carone}, {Casewell}, {Changeat},
  {Chubb}, {Crossfield}, {Crouzet}, {Decin}, {D{\'e}sert}, {Feinstein},
  {Flagg}, {Fortney}, {Gizis}, {Heng}, {Iro}, {Kempton}, {Kendrew}, {Kirk},
  {Knutson}, {Komacek}, {Lagage}, {Leconte}, {Lustig-Yaeger}, {MacDonald},
  {Mancini}, {May}, {Mayne}, {Miguel}, {Mikal-Evans}, {Molaverdikhani},
  {Palle}, {Piaulet}, {Rackham}, {Redfield}, {Rogers}, {Roy}, {Rustamkulov},
  {Shkolnik}, {Sotzen}, {Taylor}, {Tremblin}, {Tucker}, {Turner}, {de
  Val-Borro}, {Venot}, \& {Zhang}}]{2023Natur.614..653A}
{Ahrer}, E.-M., {Stevenson}, K.~B., {Mansfield}, M., {et~al.} 2023, \nat, 614,
  653, \dodoi{10.1038/s41586-022-05590-4}

\bibitem[{{Ahrer} {et~al.}(2025){Ahrer}, {Gandhi}, {Alderson}, {Kirk}, {Teske},
  {Booth}, {McDonald}, {Christie}, {Claringbold}, {Nealon}, {Panwar}, {Veras},
  {Wakeford}, {Wheatley}, \& {Zamyatina}}]{2025MNRAS.540.2535A}
{Ahrer}, E.-M., {Gandhi}, S., {Alderson}, L., {et~al.} 2025, \mnras, 540, 2535,
  \dodoi{10.1093/mnras/staf819}

\bibitem[{{Ali-Dib} {et~al.}(2014){Ali-Dib}, {Mousis}, {Petit}, \&
  {Lunine}}]{2014ApJ...785..125A}
{Ali-Dib}, M., {Mousis}, O., {Petit}, J.-M., \& {Lunine}, J.~I. 2014, \apj,
  785, 125, \dodoi{10.1088/0004-637X/785/2/125}

\bibitem[{{Anderson} {et~al.}(2010){Anderson}, {Hellier}, {Gillon}, {Triaud},
  {Smalley}, {Hebb}, {Collier Cameron}, {Maxted}, {Queloz}, {West}, {Bentley},
  {Enoch}, {Horne}, {Lister}, {Mayor}, {Parley}, {Pepe}, {Pollacco},
  {S{\'e}gransan}, {Udry}, \& {Wilson}}]{2010ApJ...709..159A}
{Anderson}, D.~R., {Hellier}, C., {Gillon}, M., {et~al.} 2010, \apj, 709, 159,
  \dodoi{10.1088/0004-637X/709/1/159}

\bibitem[{{Baratella} {et~al.}(2020){Baratella}, {D'Orazi}, {Biazzo},
  {Desidera}, {Gratton}, {Benatti}, {Bignamini}, {Carleo}, {Cecconi}, {Claudi},
  {Cosentino}, {Ghedina}, {Harutyunyan}, {Lanza}, {Malavolta}, {Maldonado},
  {Mallonn}, {Messina}, {Micela}, {Molinari}, {Poretti}, {Scandariato}, \&
  {Sozzetti}}]{2020A&A...640A.123B}
{Baratella}, M., {D'Orazi}, V., {Biazzo}, K., {et~al.} 2020, \aap, 640, A123,
  \dodoi{10.1051/0004-6361/202038511}

\bibitem[{{Barros} {et~al.}(2023){Barros}, {Demangeon}, {Armstrong}, {Delgado
  Mena}, {Acu{\~n}a}, {Fern{\'a}ndez Fern{\'a}ndez}, {Deleuil}, {Collins},
  {Howell}, {Ziegler}, {Adibekyan}, {Sousa}, {Stassun}, {Grieves}, {Lillo-Box},
  {Hellier}, {Wheatley}, {Brice{\~n}o}, {Collins}, {Hawthorn}, {Hoyer},
  {Jenkins}, {Law}, {Mann}, {Matson}, {Mousis}, {Nielsen}, {Osborn}, {Osborn},
  {Paegert}, {Papini}, {Ricker}, {Rudat}, {Santos}, {Seager}, {Stockdale},
  {Str{\o}m}, {Twicken}, {Udry}, {Wang}, {Vanderspek}, \&
  {Winn}}]{2023AA...673A...4B}
{Barros}, S.~C.~C., {Demangeon}, O.~D.~S., {Armstrong}, D.~J., {et~al.} 2023,
  \aap, 673, A4, \dodoi{10.1051/0004-6361/202245741}

\bibitem[{{Bedell} {et~al.}(2018){Bedell}, {Bean}, {Mel{\'e}ndez}, {Spina},
  {Ram{\'\i}rez}, {Asplund}, {Alves-Brito}, {dos Santos}, {Dreizler}, {Yong},
  {Monroe}, \& {Casagrande}}]{2018ApJ...865...68B}
{Bedell}, M., {Bean}, J.~L., {Mel{\'e}ndez}, J., {et~al.} 2018, \apj, 865, 68,
  \dodoi{10.3847/1538-4357/aad908}

\bibitem[{{Bergemann} {et~al.}(2010){Bergemann}, {Pickering}, \&
  {Gehren}}]{2010MNRAS.401.1334B}
{Bergemann}, M., {Pickering}, J.~C., \& {Gehren}, T. 2010, \mnras, 401, 1334,
  \dodoi{10.1111/j.1365-2966.2009.15736.x}

\bibitem[{{Bergemann} {et~al.}(2019){Bergemann}, {Gallagher}, {Eitner},
  {Bautista}, {Collet}, {Yakovleva}, {Mayriedl}, {Plez}, {Carlsson},
  {Leenaarts}, {Belyaev}, \& {Hansen}}]{2019A&A...631A..80B}
{Bergemann}, M., {Gallagher}, A.~J., {Eitner}, P., {et~al.} 2019, \aap, 631,
  A80, \dodoi{10.1051/0004-6361/201935811}

\bibitem[{{Biazzo} {et~al.}(2022){Biazzo}, {D'Orazi}, {Desidera}, {Turrini},
  {Benatti}, {Gratton}, {Magrini}, {Sozzetti}, {Baratella}, {Bonomo}, {Borsa},
  {Claudi}, {Covino}, {Damasso}, {Di Mauro}, {Lanza}, {Maggio}, {Malavolta},
  {Maldonado}, {Marzari}, {Micela}, {Poretti}, {Vitello}, {Affer}, {Bignamini},
  {Carleo}, {Cosentino}, {Fiorenzano}, {Giacobbe}, {Harutyunyan}, {Leto},
  {Mancini}, {Molinari}, {Molinaro}, {Nardiello}, {Nascimbeni}, {Pagano},
  {Pedani}, {Piotto}, {Rainer}, \& {Scandariato}}]{2022A&A...664A.161B}
{Biazzo}, K., {D'Orazi}, V., {Desidera}, S., {et~al.} 2022, \aap, 664, A161,
  \dodoi{10.1051/0004-6361/202243467}

\bibitem[{{Bonomo} {et~al.}(2017){Bonomo}, {Desidera}, {Benatti}, {Borsa},
  {Crespi}, {Damasso}, {Lanza}, {Sozzetti}, {Lodato}, {Marzari}, {Boccato},
  {Claudi}, {Cosentino}, {Covino}, {Gratton}, {Maggio}, {Micela}, {Molinari},
  {Pagano}, {Piotto}, {Poretti}, {Smareglia}, {Affer}, {Biazzo}, {Bignamini},
  {Esposito}, {Giacobbe}, {H{\'e}brard}, {Malavolta}, {Maldonado}, {Mancini},
  {Martinez Fiorenzano}, {Masiero}, {Nascimbeni}, {Pedani}, {Rainer}, \&
  {Scandariato}}]{2017AA...602A.107B}
{Bonomo}, A.~S., {Desidera}, S., {Benatti}, S., {et~al.} 2017, \aap, 602, A107,
  \dodoi{10.1051/0004-6361/201629882}

\bibitem[{{Brewer} \& {Fischer}(2016)}]{2016ApJ...831...20B}
{Brewer}, J.~M., \& {Fischer}, D.~A. 2016, \apj, 831, 20,
  \dodoi{10.3847/0004-637X/831/1/20}

\bibitem[{{Cugno} \& {Grant}(2025)}]{2025ApJ...991L..46C}
{Cugno}, G., \& {Grant}, S.~L. 2025, \apjl, 991, L46,
  \dodoi{10.3847/2041-8213/ae0290}

\bibitem[{{da Silva} {et~al.}(2024){da Silva}, {Danielski}, {Delgado Mena},
  {Magrini}, {Turrini}, {Biazzo}, {Tsantaki}, {Rainer}, {Helminiak}, {Benatti},
  {Adibekyan}, {Sanna}, {Sousa}, {Casali}, \& {Van der
  Swaelmen}}]{2024A&A...688A.193D}
{da Silva}, R., {Danielski}, C., {Delgado Mena}, E., {et~al.} 2024, \aap, 688,
  A193, \dodoi{10.1051/0004-6361/202450604}

\bibitem[{{Dawson} \& {Johnson}(2018)}]{2018ARA&A..56..175D}
{Dawson}, R.~I., \& {Johnson}, J.~A. 2018, \araa, 56, 175,
  \dodoi{10.1146/annurev-astro-081817-051853}

\bibitem[{{Deeg} {et~al.}(2023){Deeg}, {Georgieva}, {Nowak}, {Persson}, {Cale},
  {Murgas}, {Pall{\'e}}, {Godoy-Rivera}, {Dai}, {Ciardi}, {Murphy}, {Beck},
  {Burke}, {Cabrera}, {Carleo}, {Cochran}, {Collins}, {Csizmadia}, {El Mufti},
  {Fridlund}, {Fukui}, {Gandolfi}, {Garc{\'\i}a}, {Guenther}, {Guerra},
  {Grziwa}, {Isaacson}, {Isogai}, {Jenkins}, {K{\'a}bath}, {Korth}, {Lam},
  {Latham}, {Luque}, {Lund}, {Livingston}, {Mathis}, {Mathur}, {Narita},
  {Orell-Miquel}, {Osborne}, {Parviainen}, {Plavchan}, {Redfield}, {Rodriguez},
  {Schwarz}, {Seager}, {Smith}, {Van Eylen}, {Van Zandt}, {Winn}, \&
  {Ziegler}}]{2023AA...677A..12D}
{Deeg}, H.~J., {Georgieva}, I.~Y., {Nowak}, G., {et~al.} 2023, \aap, 677, A12,
  \dodoi{10.1051/0004-6361/202346370}

\bibitem[{{Delgado Mena} {et~al.}(2021){Delgado Mena}, {Adibekyan}, {Santos},
  {Tsantaki}, {Gonz{\'a}lez Hern{\'a}ndez}, {Sousa}, \& {Bertr{\'a}n de
  Lis}}]{2021A&A...655A..99D}
{Delgado Mena}, E., {Adibekyan}, V., {Santos}, N.~C., {et~al.} 2021, \aap, 655,
  A99, \dodoi{10.1051/0004-6361/202141588}

\bibitem[{{Denissenkov} \& {Pinsonneault}(2008)}]{2008ApJ...679.1541D}
{Denissenkov}, P.~A., \& {Pinsonneault}, M. 2008, \apj, 679, 1541,
  \dodoi{10.1086/587681}

\bibitem[{{Dr{\k{a}}{\.z}kowska} \& {Alibert}(2017)}]{2017A&A...608A..92D}
{Dr{\k{a}}{\.z}kowska}, J., \& {Alibert}, Y. 2017, \aap, 608, A92,
  \dodoi{10.1051/0004-6361/201731491}

\bibitem[{{Dyrek} {et~al.}(2024){Dyrek}, {Min}, {Decin}, {Bouwman}, {Crouzet},
  {Molli{\`e}re}, {Lagage}, {Konings}, {Tremblin}, {G{\"u}del}, {Pye},
  {Waters}, {Henning}, {Vandenbussche}, {Ardevol Martinez}, {Argyriou},
  {Ducrot}, {Heinke}, {van Looveren}, {Absil}, {Barrado}, {Baudoz},
  {Boccaletti}, {Cossou}, {Coulais}, {Edwards}, {Gastaud}, {Glasse}, {Glauser},
  {Greene}, {Kendrew}, {Krause}, {Lahuis}, {Mueller}, {Olofsson}, {Patapis},
  {Rouan}, {Royer}, {Scheithauer}, {Waldmann}, {Whiteford}, {Colina}, {van
  Dishoeck}, {{\"O}stlin}, {Ray}, \& {Wright}}]{2024Natur.625...51D}
{Dyrek}, A., {Min}, M., {Decin}, L., {et~al.} 2024, \nat, 625, 51,
  \dodoi{10.1038/s41586-023-06849-0}

\bibitem[{{Edvardsson} {et~al.}(1993){Edvardsson}, {Andersen}, {Gustafsson},
  {Lambert}, {Nissen}, \& {Tomkin}}]{1993AA...275..101E}
{Edvardsson}, B., {Andersen}, J., {Gustafsson}, B., {et~al.} 1993, \aap, 275,
  101

\bibitem[{{Edwards} {et~al.}(2023){Edwards}, {Changeat}, {Tsiaras}, {Yip},
  {Al-Refaie}, {Anisman}, {Bieger}, {Gressier}, {Shibata}, {Skaf}, {Bouwman},
  {Cho}, {Ikoma}, {Venot}, {Waldmann}, {Lagage}, \&
  {Tinetti}}]{2023ApJS..269...31E}
{Edwards}, B., {Changeat}, Q., {Tsiaras}, A., {et~al.} 2023, \apjs, 269, 31,
  \dodoi{10.3847/1538-4365/ac9f1a}

\bibitem[{{Feng} {et~al.}(2019){Feng}, {Anglada-Escud{\'e}}, {Tuomi}, {Jones},
  {Chanam{\'e}}, {Butler}, \& {Janson}}]{2019MNRAS.490.5002F}
{Feng}, F., {Anglada-Escud{\'e}}, G., {Tuomi}, M., {et~al.} 2019, \mnras, 490,
  5002, \dodoi{10.1093/mnras/stz2912}

\bibitem[{{Filomeno} {et~al.}(2024){Filomeno}, {Biazzo}, {Baratella},
  {Benatti}, {D'Orazi}, {Desidera}, {Mancini}, {Messina}, {Polychroni},
  {Turrini}, {Cabona}, {Carleo}, {Damasso}, {Malavolta}, {Mantovan},
  {Nardiello}, {Scandariato}, {Sozzetti}, {Zingales}, {Andreuzzi},
  {Antoniucci}, {Bignamini}, {Bonomo}, {Claudi}, {Cosentino}, {Fiorenzano},
  {Fonte}, {Harutyunyan}, \& {Knapic}}]{2024A&A...690A.370F}
{Filomeno}, S., {Biazzo}, K., {Baratella}, M., {et~al.} 2024, \aap, 690, A370,
  \dodoi{10.1051/0004-6361/202450611}

\bibitem[{{Fischer} \& {Valenti}(2005)}]{2005ApJ...622.1102F}
{Fischer}, D.~A., \& {Valenti}, J. 2005, \apj, 622, 1102,
  \dodoi{10.1086/428383}

\bibitem[{{Gaia Collaboration} {et~al.}(2023){Gaia Collaboration}, {Vallenari},
  {Brown}, {Prusti}, {de Bruijne}, {Arenou}, {Babusiaux}, {Biermann},
  {Creevey}, {Ducourant}, {Evans}, {Eyer}, {Guerra}, {Hutton}, {Jordi},
  {Klioner}, {Lammers}, {Lindegren}, {Luri}, {Mignard}, {Panem}, {Pourbaix},
  {Randich}, {Sartoretti}, {Soubiran}, {Tanga}, {Walton}, {Bailer-Jones},
  {Bastian}, {Drimmel}, {Jansen}, {Katz}, {Lattanzi}, {van Leeuwen}, {Bakker},
  {Cacciari}, {Casta{\~n}eda}, {De Angeli}, {Fabricius}, {Fouesneau},
  {Fr{\'e}mat}, {Galluccio}, {Guerrier}, {Heiter}, {Masana}, {Messineo},
  {Mowlavi}, {Nicolas}, {Nienartowicz}, {Pailler}, {Panuzzo}, {Riclet}, {Roux},
  {Seabroke}, {Sordo}, {Th{\'e}venin}, {Gracia-Abril}, {Portell}, {Teyssier},
  {Altmann}, {Andrae}, {Audard}, {Bellas-Velidis}, {Benson}, {Berthier},
  {Blomme}, {Burgess}, {Busonero}, {Busso}, {C{\'a}novas}, {Carry}, {Cellino},
  {Cheek}, {Clementini}, {Damerdji}, {Davidson}, {de Teodoro}, {Nu{\~n}ez
  Campos}, {Delchambre}, {Dell'Oro}, {Esquej}, {Fern{\'a}ndez-Hern{\'a}ndez},
  {Fraile}, {Garabato}, {Garc{\'\i}a-Lario}, {Gosset}, {Haigron}, {Halbwachs},
  {Hambly}, {Harrison}, {Hern{\'a}ndez}, {Hestroffer}, {Hodgkin}, {Holl},
  {Jan{\ss}en}, {Jevardat de Fombelle}, {Jordan}, {Krone-Martins}, {Lanzafame},
  {L{\"o}ffler}, {Marchal}, {Marrese}, {Moitinho}, {Muinonen}, {Osborne},
  {Pancino}, {Pauwels}, {Recio-Blanco}, {Reyl{\'e}}, {Riello}, {Rimoldini},
  {Roegiers}, {Rybizki}, {Sarro}, {Siopis}, {Smith}, {Sozzetti}, {Utrilla},
  {van Leeuwen}, {Abbas}, {{\'A}brah{\'a}m}, {Abreu Aramburu}, {Aerts},
  {Aguado}, {Ajaj}, {Aldea-Montero}, {Altavilla}, {{\'A}lvarez}, {Alves},
  {Anders}, {Anderson}, {Anglada Varela}, {Antoja}, {Baines}, {Baker},
  {Balaguer-N{\'u}{\~n}ez}, {Balbinot}, {Balog}, {Barache}, {Barbato},
  {Barros}, {Barstow}, {Bartolom{\'e}}, {Bassilana}, {Bauchet}, {Becciani},
  {Bellazzini}, {Berihuete}, {Bernet}, {Bertone}, {Bianchi}, {Binnenfeld},
  {Blanco-Cuaresma}, {Blazere}, {Boch}, {Bombrun}, {Bossini}, {Bouquillon},
  {Bragaglia}, {Bramante}, {Breedt}, {Bressan}, {Brouillet}, {Brugaletta},
  {Bucciarelli}, {Burlacu}, {Butkevich}, {Buzzi}, {Caffau}, {Cancelliere},
  {Cantat-Gaudin}, {Carballo}, {Carlucci}, {Carnerero}, {Carrasco},
  {Casamiquela}, {Castellani}, {Castro-Ginard}, {Chaoul}, {Charlot}, {Chemin},
  {Chiaramida}, {Chiavassa}, {Chornay}, {Comoretto}, {Contursi}, {Cooper},
  {Cornez}, {Cowell}, {Crifo}, {Cropper}, {Crosta}, {Crowley}, {Dafonte},
  {Dapergolas}, {David}, {David}, {de Laverny}, {De Luise}, \& {De
  March}}]{2023A&A...674A...1G}
{Gaia Collaboration}, {Vallenari}, A., {Brown}, A.~G.~A., {et~al.} 2023, \aap,
  674, A1, \dodoi{10.1051/0004-6361/202243940}

\bibitem[{{Gardner} {et~al.}(2023){Gardner}, {Mather}, {Abbott}, {Abell},
  {Abernathy}, {Abney}, {Abraham}, {Abraham}, {Abul-Huda}, {Acton}, {Adams},
  {Adams}, {Adler}, {Adriaensen}, {Aguilar}, {Ahmed}, {Ahmed}, {Ahmed},
  {Albat}, {Albert}, {Alberts}, {Aldridge}, {Allen}, {Allen}, {Altenburg},
  {Altunc}, {Alvarez}, {{\'A}lvarez-M{\'a}rquez}, {Alves de Oliveira},
  {Ambrose}, {Anandakrishnan}, {Andersen}, {Anderson}, {Anderson}, {Anderson},
  {Anderson}, {Aprea}, {Archer}, {Arenberg}, {Argyriou}, {Arribas}, {Artigau},
  {Arvai}, {Atcheson}, {Atkinson}, {Averbukh}, {Aymergen}, {Bacinski},
  {Baggett}, {Bagnasco}, {Baker}, {Balzano}, {Banks}, {Baran}, {Barker},
  {Barrett}, {Barringer}, {Barto}, {Bast}, {Baudoz}, {Baum}, {Beatty},
  {Beaulieu}, {Bechtold}, {Beck}, {Beddard}, {Beichman}, {Bellagama}, {Bely},
  {Berger}, {Bergeron}, {Bernier}, {Bertch}, {Beskow}, {Betz}, {Biagetti},
  {Birkmann}, {Bjorklund}, {Blackwood}, {Blazek}, {Blossfeld}, {Bluth},
  {Boccaletti}, {Boegner}, {Bohlin}, {Boia}, {B{\"o}ker}, {Bonaventura},
  {Bond}, {Bosley}, {Boucarut}, {Bouchet}, {Bouwman}, {Bower}, {Bowers},
  {Bowers}, {Boyce}, {Boyer}, {Boyer}, {Boyer}, {Boyer}, {Bradley}, {Brady},
  {Brandl}, {Brannen}, {Breda}, {Bremmer}, {Brennan}, {Bresnahan}, {Bright},
  {Broiles}, {Bromenschenkel}, {Brooks}, {Brooks}, {Brown}, {Brown}, {Brown},
  {Bruce}, {Bryson}, {Bujanda}, {Bullock}, {Bunker}, {Bureo}, {Burt}, {Bush},
  {Bushouse}, {Bussman}, {Cabaud}, {Cale}, {Calhoon}, {Calvani}, {Canipe},
  {Caputo}, {Cara}, {Carey}, {Case}, {Cesari}, {Cetorelli}, {Chance},
  {Chandler}, {Chaney}, {Chapman}, {Charlot}, {Chayer}, {Cheezum}, {Chen},
  {Chen}, {Cherinka}, {Chichester}, {Chilton}, {Chittiraibalan}, {Clampin},
  {Clark}, {Clark}, {Clark}, {Claybrooks}, {Cleveland}, {Cohen}, {Cohen},
  {Col{\'o}n}, {Coleman}, {Colina}, {Comber}, {Comeau}, {Comer}, {Conde Reis},
  {Connolly}, {Conroy}, {Contos}, {Contreras}, {Cook}, {Cooper}, {Cooper},
  {Correia}, {Correnti}, {Cossou}, {Costanza}, {Coulais}, {Cox}, {Coyle},
  {Cracraft}, {Crew}, {Curtis}, {Cusveller}, {Da Costa Maciel}, {Dailey},
  {Daugeron}, {Davidson}, {Davies}, {Davis}, {Davis}, {Day}, {de Chambure}, {de
  Jong}, {De Marchi}, {Dean}, {Decker}, {Delisa}, {Dell}, \&
  {Dellagatta}}]{2023PASP..135f8001G}
{Gardner}, J.~P., {Mather}, J.~C., {Abbott}, R., {et~al.} 2023, \pasp, 135,
  068001, \dodoi{10.1088/1538-3873/acd1b5}

\bibitem[{{Godoy} {et~al.}(2024){Godoy}, {Choquet}, {Serabyn}, {Danielski},
  {Stolker}, {Charnay}, {Hinkley}, {Lagage}, {Ressler}, {Tremblin}, \&
  {Vigan}}]{2024AA...689A.185G}
{Godoy}, N., {Choquet}, E., {Serabyn}, E., {et~al.} 2024, \aap, 689, A185,
  \dodoi{10.1051/0004-6361/202449951}

\bibitem[{{Gustafsson} {et~al.}(2008){Gustafsson}, {Edvardsson}, {Eriksson},
  {J{\o}rgensen}, {Nordlund}, \& {Plez}}]{2008A&A...486..951G}
{Gustafsson}, B., {Edvardsson}, B., {Eriksson}, K., {et~al.} 2008, \aap, 486,
  951, \dodoi{10.1051/0004-6361:200809724}

\bibitem[{{Hartman} {et~al.}(2011{\natexlab{a}}){Hartman}, {Bakos}, {Kipping},
  {Torres}, {Kov{\'a}cs}, {Noyes}, {Latham}, {Howard}, {Fischer}, {Johnson},
  {Marcy}, {Isaacson}, {Quinn}, {Buchhave}, {B{\'e}ky}, {Sasselov}, {Stefanik},
  {Esquerdo}, {Everett}, {Perumpilly}, {L{\'a}z{\'a}r}, {Papp}, \&
  {S{\'a}ri}}]{2011ApJ...728..138H}
{Hartman}, J.~D., {Bakos}, G.~{\'A}., {Kipping}, D.~M., {et~al.}
  2011{\natexlab{a}}, \apj, 728, 138, \dodoi{10.1088/0004-637X/728/2/138}

\bibitem[{{Hartman} {et~al.}(2011{\natexlab{b}}){Hartman}, {Bakos}, {Sato},
  {Torres}, {Noyes}, {Latham}, {Kov{\'a}cs}, {Fischer}, {Howard}, {Johnson},
  {Marcy}, {Buchhave}, {F{\"u}resz}, {Perumpilly}, {B{\'e}ky}, {Stefanik},
  {Sasselov}, {Esquerdo}, {Everett}, {Csubry}, {L{\'a}z{\'a}r}, {Papp}, \&
  {S{\'a}ri}}]{2011ApJ...726...52H}
{Hartman}, J.~D., {Bakos}, G.~{\'A}., {Sato}, B., {et~al.} 2011{\natexlab{b}},
  \apj, 726, 52, \dodoi{10.1088/0004-637X/726/1/52}

\bibitem[{{Hartman} {et~al.}(2020){Hartman}, {Jord{\'a}n}, {Bayliss}, {Bakos},
  {Bento}, {Bhatti}, {Brahm}, {Csubry}, {Espinoza}, {Henning}, {Mancini},
  {Penev}, {Rabus}, {Sarkis}, {Suc}, {de Val-Borro}, {Zhou}, {Crane},
  {Shectman}, {Teske}, {Wang}, {Butler}, {L{\'a}z{\'a}r}, {Papp}, {S{\'a}ri},
  {Anderson}, {Hellier}, {West}, {Barkaoui}, {Pozuelos}, {Jehin}, {Gillon},
  {Nielsen}, {Lendl}, {Udry}, {Ricker}, {Vanderspek}, {Latham}, {Seager},
  {Winn}, {Christiansen}, {Crossfield}, {Henze}, {Jenkins}, {Smith}, \&
  {Ting}}]{2020AJ....159..173H}
{Hartman}, J.~D., {Jord{\'a}n}, A., {Bayliss}, D., {et~al.} 2020, \aj, 159,
  173, \dodoi{10.3847/1538-3881/ab7821}

\bibitem[{{Hayden} {et~al.}(2015){Hayden}, {Bovy}, {Holtzman}, {Nidever},
  {Bird}, {Weinberg}, {Andrews}, {Majewski}, {Allende Prieto}, {Anders},
  {Beers}, {Bizyaev}, {Chiappini}, {Cunha}, {Frinchaboy},
  {Garc{\'\i}a-Her{\'n}andez}, {Garc{\'\i}a P{\'e}rez}, {Girardi}, {Harding},
  {Hearty}, {Johnson}, {M{\'e}sz{\'a}ros}, {Minchev}, {O'Connell}, {Pan},
  {Robin}, {Schiavon}, {Schneider}, {Schultheis}, {Shetrone}, {Skrutskie},
  {Steinmetz}, {Smith}, {Wilson}, {Zamora}, \&
  {Zasowski}}]{2015ApJ...808..132H}
{Hayden}, M.~R., {Bovy}, J., {Holtzman}, J.~A., {et~al.} 2015, \apj, 808, 132,
  \dodoi{10.1088/0004-637X/808/2/132}

\bibitem[{{Hebb} {et~al.}(2009){Hebb}, {Collier-Cameron}, {Loeillet},
  {Pollacco}, {H{\'e}brard}, {Street}, {Bouchy}, {Stempels}, {Moutou},
  {Simpson}, {Udry}, {Joshi}, {West}, {Skillen}, {Wilson}, {McDonald},
  {Gibson}, {Aigrain}, {Anderson}, {Benn}, {Christian}, {Enoch}, {Haswell},
  {Hellier}, {Horne}, {Irwin}, {Lister}, {Maxted}, {Mayor}, {Norton}, {Parley},
  {Pont}, {Queloz}, {Smalley}, \& {Wheatley}}]{2009ApJ...693.1920H}
{Hebb}, L., {Collier-Cameron}, A., {Loeillet}, B., {et~al.} 2009, \apj, 693,
  1920, \dodoi{10.1088/0004-637X/693/2/1920}

\bibitem[{{Hellier} {et~al.}(2012){Hellier}, {Anderson}, {Collier Cameron},
  {Doyle}, {Fumel}, {Gillon}, {Jehin}, {Lendl}, {Maxted}, {Pepe}, {Pollacco},
  {Queloz}, {S{\'e}gransan}, {Smalley}, {Smith}, {Southworth}, {Triaud},
  {Udry}, \& {West}}]{2012MNRAS.426..739H}
{Hellier}, C., {Anderson}, D.~R., {Collier Cameron}, A., {et~al.} 2012, \mnras,
  426, 739, \dodoi{10.1111/j.1365-2966.2012.21780.x}

\bibitem[{{Hinkley} {et~al.}(2023){Hinkley}, {Lacour}, {Marleau}, {Lagrange},
  {Wang}, {Kammerer}, {Cumming}, {Nowak}, {Rodet}, {Stolker}, {Balmer}, {Ray},
  {Bonnefoy}, {Molli{\`e}re}, {Lazzoni}, {Kennedy}, {Mordasini}, {Abuter},
  {Aigrain}, {Amorim}, {Asensio-Torres}, {Babusiaux}, {Benisty}, {Berger},
  {Beust}, {Blunt}, {Boccaletti}, {Bohn}, {Bonnet}, {Bourdarot}, {Brandner},
  {Cantalloube}, {Caselli}, {Charnay}, {Chauvin}, {Chomez}, {Choquet},
  {Christiaens}, {Cl{\'e}net}, {Coud{\'e} du Foresto}, {Cridland}, {Delorme},
  {Dembet}, {Drescher}, {Duvert}, {Eckart}, {Eisenhauer}, {Feuchtgruber},
  {Galland}, {Garcia}, {Garcia Lopez}, {Gardner}, {Gendron}, {Genzel},
  {Gillessen}, {Girard}, {Grandjean}, {Haubois}, {Hei{\ss}el}, {Henning},
  {Hippler}, {Horrobin}, {Houll{\'e}}, {Hubert}, {Jocou}, {Keppler},
  {Kervella}, {Kreidberg}, {Lapeyr{\`e}re}, {Le Bouquin}, {L{\'e}na}, {Lutz},
  {Maire}, {Mang}, {M{\'e}rand}, {Meunier}, {Monnier}, {Mouillet}, {Nasedkin},
  {Ott}, {Otten}, {Paladini}, {Paumard}, {Perraut}, {Perrin}, {Philipot},
  {Pfuhl}, {Pourr{\'e}}, {Pueyo}, {Rameau}, {Rickman}, {Rubini}, {Rustamkulov},
  {Samland}, {Shangguan}, {Shimizu}, {Sing}, {Straubmeier}, {Sturm}, {Tacconi},
  {van Dishoeck}, {Vigan}, {Vincent}, {Ward-Duong}, {Widmann}, {Wieprecht},
  {Wiezorrek}, {Woillez}, {Yazici}, {Young}, \& {Zicher}}]{2023AA...671L...5H}
{Hinkley}, S., {Lacour}, S., {Marleau}, G.~D., {et~al.} 2023, \aap, 671, L5,
  \dodoi{10.1051/0004-6361/202244727}

\bibitem[{{Hobson} {et~al.}(2023){Hobson}, {Trifonov}, {Henning}, {Jord{\'a}n},
  {Rojas}, {Espinoza}, {Brahm}, {Eberhardt}, {Jones}, {Mekarnia},
  {Kossakowski}, {Schlecker}, {Tala Pinto}, {Torres Miranda}, {Abe},
  {Barkaoui}, {Bendjoya}, {Bouchy}, {Buttu}, {Carleo}, {Collins}, {Col{\'o}n},
  {Crouzet}, {Dragomir}, {Dransfield}, {Gasparetto}, {Goeke}, {Guillot},
  {G{\"u}nther}, {Howard}, {Jenkins}, {Korth}, {Latham}, {Lendl}, {Lissauer},
  {Mann}, {Mireles}, {Ricker}, {Saesen}, {Schwarz}, {Seager}, {Sefako},
  {Shporer}, {Stockdale}, {Suarez}, {Tan}, {J. Triaud}, {Ulmer-Moll},
  {Vanderspek}, {Winn}, {Wohler}, \& {Zhou}}]{2023AJ....166..201H}
{Hobson}, M.~J., {Trifonov}, T., {Henning}, T., {et~al.} 2023, \aj, 166, 201,
  \dodoi{10.3847/1538-3881/acfc1d}

\bibitem[{{Howard} {et~al.}(2025){Howard}, {Sinukoff}, {Blunt}, {Petigura},
  {Crossfield}, {Isaacson}, {Kosiarek}, {Rubenzahl}, {Brewer}, {Fulton},
  {Dressing}, {Hirsch}, {Knutson}, {Livingston}, {Mills}, {Roy}, {Weiss},
  {Benneke}, {Ciardi}, {Christiansen}, {Cochran}, {Crepp}, {Gonzales},
  {Hansen}, {Hardegree-Ullman}, {Howell}, {L{\'e}pine}, {Martinez}, {Rogers},
  {Schlieder}, {Werner}, {Polanski}, {Angelo}, {Beard}, {Behmard}, {Bouma},
  {Brinkman}, {Chontos}, {Dai}, {Dalba}, {Giacalone}, {Grunblatt}, {Hill},
  {Kane}, {Lubin}, {Mayo}, {Mocnik}, {Murphy}, {Rice}, {Rosenthal}, {Tyler},
  {Van Zandt}, \& {Yee}}]{2025ApJS..278...52H}
{Howard}, A.~W., {Sinukoff}, E., {Blunt}, S., {et~al.} 2025, \apjs, 278, 52,
  \dodoi{10.3847/1538-4365/adc5e4}

\bibitem[{{Jiang} {et~al.}(2023){Jiang}, {Wang}, {Ormel}, {Krijt}, \&
  {Dong}}]{2023A&A...678A..33J}
{Jiang}, H., {Wang}, Y., {Ormel}, C.~W., {Krijt}, S., \& {Dong}, R. 2023, \aap,
  678, A33, \dodoi{10.1051/0004-6361/202346637}

\bibitem[{{Johnson} {et~al.}(2016){Johnson}, {Walsh}, {Minton}, {Krot}, \&
  {Levison}}]{2016SciA....2E1658J}
{Johnson}, B.~C., {Walsh}, K.~J., {Minton}, D.~A., {Krot}, A.~N., \& {Levison},
  H.~F. 2016, Science Advances, 2, e1601658, \dodoi{10.1126/sciadv.1601658}

\bibitem[{{Johnson} {et~al.}(2010){Johnson}, {Aller}, {Howard}, \&
  {Crepp}}]{2010PASP..122..905J}
{Johnson}, J.~A., {Aller}, K.~M., {Howard}, A.~W., \& {Crepp}, J.~R. 2010,
  \pasp, 122, 905, \dodoi{10.1086/655775}

\bibitem[{{JWST Transiting Exoplanet Community Early Release Science Team}
  {et~al.}(2023){JWST Transiting Exoplanet Community Early Release Science
  Team}, {Ahrer}, {Alderson}, {Batalha}, {Batalha}, {Bean}, {Beatty}, {Bell},
  {Benneke}, {Berta-Thompson}, {Carter}, {Crossfield}, {Espinoza}, {Feinstein},
  {Fortney}, {Gibson}, {Goyal}, {Kempton}, {Kirk}, {Kreidberg},
  {L{\'o}pez-Morales}, {Line}, {Lothringer}, {Moran}, {Mukherjee}, {Ohno},
  {Parmentier}, {Piaulet}, {Rustamkulov}, {Schlawin}, {Sing}, {Stevenson},
  {Wakeford}, {Allen}, {Birkmann}, {Brande}, {Crouzet}, {Cubillos}, {Damiano},
  {D{\'e}sert}, {Gao}, {Harrington}, {Hu}, {Kendrew}, {Knutson}, {Lagage},
  {Leconte}, {Lendl}, {MacDonald}, {May}, {Miguel}, {Molaverdikhani}, {Moses},
  {Murray}, {Nehring}, {Nikolov}, {Petit dit de la Roche}, {Radica}, {Roy},
  {Stassun}, {Taylor}, {Waalkes}, {Wachiraphan}, {Welbanks}, {Wheatley},
  {Aggarwal}, {Alam}, {Banerjee}, {Barstow}, {Blecic}, {Casewell}, {Changeat},
  {Chubb}, {Col{\'o}n}, {Coulombe}, {Daylan}, {de Val-Borro}, {Decin}, {Dos
  Santos}, {Flagg}, {France}, {Fu}, {Garc{\'\i}a Mu{\~n}oz}, {Gizis},
  {Glidden}, {Grant}, {Heng}, {Henning}, {Hong}, {Inglis}, {Iro}, {Kataria},
  {Komacek}, {Krick}, {Lee}, {Lewis}, {Lillo-Box}, {Lustig-Yaeger}, {Mancini},
  {Mandell}, {Mansfield}, {Marley}, {Mikal-Evans}, {Morello}, {Nixon}, {Ortiz
  Ceballos}, {Piette}, {Powell}, {Rackham}, {Ramos-Rosado}, {Rauscher},
  {Redfield}, {Rogers}, {Roman}, {Roudier}, {Scarsdale}, {Shkolnik},
  {Southworth}, {Spake}, {Steinrueck}, {Tan}, {Teske}, {Tremblin}, {Tsai},
  {Tucker}, {Turner}, {Valenti}, {Venot}, {Waldmann}, {Wallack}, {Zhang}, \&
  {Zieba}}]{2023Natur.614..649J}
{JWST Transiting Exoplanet Community Early Release Science Team}, {Ahrer},
  E.-M., {Alderson}, L., {et~al.} 2023, \nat, 614, 649,
  \dodoi{10.1038/s41586-022-05269-w}

\bibitem[{{Korotin} {et~al.}(2017){Korotin}, {Andrievsky}, {Caffau}, \&
  {Bonifacio}}]{2017ASPC..510..141K}
{Korotin}, S., {Andrievsky}, S., {Caffau}, E., \& {Bonifacio}, P. 2017, in
  Astronomical Society of the Pacific Conference Series, Vol. 510, Stars: From
  Collapse to Collapse, ed. Y.~Y. {Balega}, D.~O. {Kudryavtsev}, I.~I.
  {Romanyuk}, \& I.~A. {Yakunin}, 141

\bibitem[{{Koskinen} {et~al.}(2022){Koskinen}, {Lavvas}, {Huang}, {Bergsten},
  {Fernandes}, \& {Young}}]{2022ApJ...929...52K}
{Koskinen}, T.~T., {Lavvas}, P., {Huang}, C., {et~al.} 2022, \apj, 929, 52,
  \dodoi{10.3847/1538-4357/ac4f45}

\bibitem[{{Kunimoto} {et~al.}(2023){Kunimoto}, {Vanderburg}, {Huang}, {Davis},
  {Affer}, {Cameron}, {Charbonneau}, {Cosentino}, {Damasso}, {Dumusque},
  {Fiorenzano}, {Ghedina}, {Haywood}, {Lienhard}, {L{\'o}pez-Morales}, {Mayor},
  {Pepe}, {Pinamonti}, {Poretti}, {Maldonado}, {Rice}, {Sozzetti}, {Wilson},
  {Udry}, {Baptista}, {Barkaoui}, {Becker}, {Benni}, {Bieryla}, {Bosch-Cabot},
  {Ciardi}, {Collins}, {Collins}, {Evans}, {Dupuy}, {Goliguzova}, {Guerra},
  {Kraus}, {Lissauer}, {Huber}, {Murgas}, {Palle}, {Quinn}, {Safonov},
  {Schwarz}, {Shporer}, {Stassun}, {Jenkins}, {Latham}, {Ricker}, {Seager},
  {Vanderspek}, {Winn}, {Essack}, {Lewis}, \& {Rose}}]{2023AJ....166....7K}
{Kunimoto}, M., {Vanderburg}, A., {Huang}, C.~X., {et~al.} 2023, \aj, 166, 7,
  \dodoi{10.3847/1538-3881/acd537}

\bibitem[{{Lind} {et~al.}(2011){Lind}, {Asplund}, {Barklem}, \&
  {Belyaev}}]{2011A&A...528A.103L}
{Lind}, K., {Asplund}, M., {Barklem}, P.~S., \& {Belyaev}, A.~K. 2011, \aap,
  528, A103, \dodoi{10.1051/0004-6361/201016095}

\bibitem[{{Liu} {et~al.}(2024){Liu}, {Ting}, {Yong}, {Bitsch}, {Karakas},
  {Murphy}, {Joyce}, {Dotter}, \& {Dai}}]{2024Natur.627..501L}
{Liu}, F., {Ting}, Y.-S., {Yong}, D., {et~al.} 2024, \nat, 627, 501,
  \dodoi{10.1038/s41586-024-07091-y}

\bibitem[{{Llop-Sayson} {et~al.}(2021){Llop-Sayson}, {Wang}, {Ruffio}, {Mawet},
  {Blunt}, {Absil}, {Bond}, {Brinkman}, {Bowler}, {Bottom}, {Chontos}, {Dalba},
  {Fulton}, {Giacalone}, {Hill}, {Hirsch}, {Howard}, {Isaacson}, {Karlsson},
  {Lubin}, {Madurowicz}, {Matthews}, {Morris}, {Perrin}, {Ren}, {Rice},
  {Rosenthal}, {Ruane}, {Rubenzahl}, {Sun}, {Wallack}, {Xuan}, \&
  {Ygouf}}]{2021AJ....162..181L}
{Llop-Sayson}, J., {Wang}, J.~J., {Ruffio}, J.-B., {et~al.} 2021, \aj, 162,
  181, \dodoi{10.3847/1538-3881/ac134a}

\bibitem[{{MacDougall} {et~al.}(2023){MacDougall}, {Petigura}, {Gilbert},
  {Angelo}, {Batalha}, {Beard}, {Behmard}, {Blunt}, {Brinkman}, {Chontos},
  {Crossfield}, {Dai}, {Dalba}, {Dressing}, {Fetherolf}, {Fulton}, {Giacalone},
  {Hill}, {Holcomb}, {Howard}, {Huber}, {Isaacson}, {Kane}, {Kosiarek},
  {Lubin}, {Mayo}, {Mo{\v{c}}nik}, {Akana Murphy}, {Pidhorodetska}, {Polanski},
  {Rice}, {Robertson}, {Rosenthal}, {Roy}, {Rubenzahl}, {Scarsdale},
  {Turtelboom}, {Tyler}, {Van Zandt}, {Weiss}, \& {Yee}}]{2023AJ....166...33M}
{MacDougall}, M.~G., {Petigura}, E.~A., {Gilbert}, G.~J., {et~al.} 2023, \aj,
  166, 33, \dodoi{10.3847/1538-3881/acd557}

\bibitem[{{Madhusudhan}(2012)}]{2012ApJ...758...36M}
{Madhusudhan}, N. 2012, \apj, 758, 36, \dodoi{10.1088/0004-637X/758/1/36}

\bibitem[{{Magrini} {et~al.}(2022){Magrini}, {Danielski}, {Bossini}, {Rainer},
  {Turrini}, {Benatti}, {Brucalassi}, {Tsantaki}, {Delgado Mena}, {Sanna},
  {Biazzo}, {Campante}, {Van der Swaelmen}, {Sousa}, {He{\l}miniak}, {Neitzel},
  {Adibekyan}, {Bruno}, \& {Casali}}]{2022A&A...663A.161M}
{Magrini}, L., {Danielski}, C., {Bossini}, D., {et~al.} 2022, \aap, 663, A161,
  \dodoi{10.1051/0004-6361/202243405}

\bibitem[{{Maxted} {et~al.}(2013){Maxted}, {Anderson}, {Collier Cameron},
  {Doyle}, {Fumel}, {Gillon}, {Hellier}, {Jehin}, {Lendl}, {Pepe}, {Pollacco},
  {Queloz}, {S{\'e}gransan}, {Smalley}, {Southworth}, {Smith}, {Triaud},
  {Udry}, \& {West}}]{2013PASP..125...48M}
{Maxted}, P.~F.~L., {Anderson}, D.~R., {Collier Cameron}, A., {et~al.} 2013,
  \pasp, 125, 48, \dodoi{10.1086/669231}

\bibitem[{{McGruder} {et~al.}(2023){McGruder}, {L{\'o}pez-Morales}, {Brahm}, \&
  {Jord{\'a}n}}]{2023ApJ...944L..56M}
{McGruder}, C.~D., {L{\'o}pez-Morales}, M., {Brahm}, R., \& {Jord{\'a}n}, A.
  2023, \apjl, 944, L56, \dodoi{10.3847/2041-8213/acb154}

\bibitem[{{Mel{\'e}ndez} {et~al.}(2009){Mel{\'e}ndez}, {Asplund}, {Gustafsson},
  \& {Yong}}]{2009ApJ...704L..66M}
{Mel{\'e}ndez}, J., {Asplund}, M., {Gustafsson}, B., \& {Yong}, D. 2009, \apjl,
  704, L66, \dodoi{10.1088/0004-637X/704/1/L66}

\bibitem[{{Mesa} {et~al.}(2018){Mesa}, {Baudino}, {Charnay}, {D'Orazi},
  {Desidera}, {Boccaletti}, {Gratton}, {Bonnefoy}, {Delorme}, {Langlois},
  {Vigan}, {Zurlo}, {Maire}, {Janson}, {Antichi}, {Baruffolo}, {Bruno},
  {Cascone}, {Chauvin}, {Claudi}, {De Caprio}, {Fantinel}, {Farisato}, {Feldt},
  {Giro}, {Hagelberg}, {Incorvaia}, {Lagadec}, {Lagrange}, {Lazzoni}, {Lessio},
  {Salasnich}, {Scuderi}, {Sissa}, \& {Turatto}}]{2018AA...612A..92M}
{Mesa}, D., {Baudino}, J.~L., {Charnay}, B., {et~al.} 2018, \aap, 612, A92,
  \dodoi{10.1051/0004-6361/201731649}

\bibitem[{{Miles} {et~al.}(2023){Miles}, {Biller}, {Patapis}, {Worthen},
  {Rickman}, {Hoch}, {Skemer}, {Perrin}, {Whiteford}, {Chen}, {Sargent},
  {Mukherjee}, {Morley}, {Moran}, {Bonnefoy}, {Petrus}, {Carter}, {Choquet},
  {Hinkley}, {Ward-Duong}, {Leisenring}, {Millar-Blanchaer}, {Pueyo}, {Ray},
  {Sallum}, {Stapelfeldt}, {Stone}, {Wang}, {Absil}, {Balmer}, {Boccaletti},
  {Bonavita}, {Booth}, {Bowler}, {Chauvin}, {Christiaens}, {Currie},
  {Danielski}, {Fortney}, {Girard}, {Grady}, {Greenbaum}, {Henning}, {Hines},
  {Janson}, {Kalas}, {Kammerer}, {Kennedy}, {Kenworthy}, {Kervella}, {Lagage},
  {Lew}, {Liu}, {Macintosh}, {Marino}, {Marley}, {Marois}, {Matthews},
  {Matthews}, {Mawet}, {McElwain}, {Metchev}, {Meyer}, {Molliere}, {Pantin},
  {Quirrenbach}, {Rebollido}, {Ren}, {Schneider}, {Vasist}, {Wyatt}, {Zhou},
  {Briesemeister}, {Bryan}, {Calissendorff}, {Cantalloube}, {Cugno}, {De
  Furio}, {Dupuy}, {Factor}, {Faherty}, {Fitzgerald}, {Franson}, {Gonzales},
  {Hood}, {Howe}, {Kraus}, {Kuzuhara}, {Lagrange}, {Lawson}, {Lazzoni}, {Liu},
  {Llop-Sayson}, {Lloyd}, {Martinez}, {Mazoyer}, {Quanz}, {Redai}, {Samland},
  {Schlieder}, {Tamura}, {Tan}, {Uyama}, {Vigan}, {Vos}, {Wagner}, {Wolff},
  {Ygouf}, {Zhang}, {Zhang}, \& {Zhang}}]{2023ApJ...946L...6M}
{Miles}, B.~E., {Biller}, B.~A., {Patapis}, P., {et~al.} 2023, \apjl, 946, L6,
  \dodoi{10.3847/2041-8213/acb04a}

\bibitem[{{Murgas} {et~al.}(2022){Murgas}, {Nowak}, {Masseron}, {Parviainen},
  {Luque}, {Pall{\'e}}, {Korth}, {Carleo}, {Csizmadia}, {Esparza-Borges},
  {Alqasim}, {Cochran}, {Dai}, {Deeg}, {Gandolfi}, {Goffo}, {Kab{\'a}th},
  {Lam}, {Livingston}, {Muresan}, {Osborne}, {Persson}, {Serrano}, {Smith},
  {Van Eylen}, {Orell-Miquel}, {Hinkel}, {Gal{\'a}n}, {Puig-Subir{\`a}},
  {Stangret}, {Fukui}, {Kagetani}, {Narita}, {Ciardi}, {Boyle}, {Ziegler},
  {Brice{\~n}o}, {Law}, {Mann}, {Jenkins}, {Latham}, {Quinn}, {Ricker},
  {Seager}, {Shporer}, {Ting}, {Vanderspek}, \& {Winn}}]{2022AA...668A.158M}
{Murgas}, F., {Nowak}, G., {Masseron}, T., {et~al.} 2022, \aap, 668, A158,
  \dodoi{10.1051/0004-6361/202244459}

\bibitem[{{Nardiello} {et~al.}(2022){Nardiello}, {Malavolta}, {Desidera},
  {Baratella}, {D'Orazi}, {Messina}, {Biazzo}, {Benatti}, {Damasso}, {Rajpaul},
  {Bonomo}, {Capuzzo Dolcetta}, {Mallonn}, {Cale}, {Plavchan}, {El Mufti},
  {Bignamini}, {Borsa}, {Carleo}, {Claudi}, {Covino}, {Lanza}, {Maldonado},
  {Mancini}, {Micela}, {Molinari}, {Pinamonti}, {Piotto}, {Poretti},
  {Scandariato}, {Sozzetti}, {Andreuzzi}, {Boschin}, {Cosentino}, {Fiorenzano},
  {Harutyunyan}, {Knapic}, {Pedani}, {Affer}, {Maggio}, \&
  {Rainer}}]{2022AA...664A.163N}
{Nardiello}, D., {Malavolta}, L., {Desidera}, S., {et~al.} 2022, \aap, 664,
  A163, \dodoi{10.1051/0004-6361/202243743}

\bibitem[{{NASA Exoplanet Archive}(2026{\natexlab{a}})}]{NEA12}
{NASA Exoplanet Archive}. 2026{\natexlab{a}}, Planetary Systems Table,
  NExScI–Caltech/IPAC, \dodoi{10.26133/NEA12}

\bibitem[{{NASA Exoplanet Archive}(2026{\natexlab{b}})}]{NEA13}
---. 2026{\natexlab{b}}, Planetary Systems Composite Table,
  NExScI–Caltech/IPAC, \dodoi{10.26133/NEA13}

\bibitem[{{Newton} {et~al.}(2021){Newton}, {Mann}, {Kraus}, {Livingston},
  {Vanderburg}, {Curtis}, {Thao}, {Hawkins}, {Wood}, {Rizzuto}, {Soubkiou},
  {Tofflemire}, {Zhou}, {Crossfield}, {Pearce}, {Collins}, {Conti}, {Tan},
  {Villeneuva}, {Spencer}, {Dragomir}, {Quinn}, {Jensen}, {Collins},
  {Stockdale}, {Cloutier}, {Hellier}, {Benkhaldoun}, {Ziegler}, {Brice{\~n}o},
  {Law}, {Benneke}, {Christiansen}, {Gorjian}, {Kane}, {Kreidberg}, {Morales},
  {Werner}, {Twicken}, {Levine}, {Ciardi}, {Guerrero}, {Hesse}, {Quintana},
  {Shiao}, {Smith}, {Torres}, {Ricker}, {Vanderspek}, {Seager}, {Winn},
  {Jenkins}, \& {Latham}}]{2021AJ....161...65N}
{Newton}, E.~R., {Mann}, A.~W., {Kraus}, A.~L., {et~al.} 2021, \aj, 161, 65,
  \dodoi{10.3847/1538-3881/abccc6}

\bibitem[{{Osborn} {et~al.}(2021){Osborn}, {Armstrong}, {Cale}, {Brahm},
  {Wittenmyer}, {Dai}, {Crossfield}, {Bryant}, {Adibekyan}, {Cloutier},
  {Collins}, {Delgado Mena}, {Fridlund}, {Hellier}, {Howell}, {King},
  {Lillo-Box}, {Otegi}, {Sousa}, {Stassun}, {Matthews}, {Ziegler}, {Ricker},
  {Vanderspek}, {Latham}, {Seager}, {Winn}, {Jenkins}, {Acton}, {Addison},
  {Anderson}, {Ballard}, {Barrado}, {Barros}, {Batalha}, {Bayliss}, {Barclay},
  {Benneke}, {Berberian}, {Bouchy}, {Bowler}, {Brice{\~n}o}, {Burke},
  {Burleigh}, {Casewell}, {Ciardi}, {Collins}, {Cooke}, {Demangeon},
  {D{\'\i}az}, {Dorn}, {Dragomir}, {Dressing}, {Dumusque}, {Espinoza},
  {Figueira}, {Fulton}, {Furlan}, {Gaidos}, {Geneser}, {Gill}, {Goad},
  {Gonzales}, {Gorjian}, {G{\"u}nther}, {Helled}, {Henderson}, {Henning},
  {Hogan}, {Hojjatpanah}, {Horner}, {Howard}, {Hoyer}, {Huber}, {Isaacson},
  {Jenkins}, {Jensen}, {Jord{\'a}n}, {Kane}, {Kidwell}, {Kielkopf}, {Law},
  {Lendl}, {Lund}, {Matson}, {Mann}, {McCormac}, {Mengel}, {Morales},
  {Nielsen}, {Okumura}, {Osborn}, {Petigura}, {Plavchan}, {Pollacco},
  {Quintana}, {Raynard}, {Robertson}, {Rose}, {Roy}, {Reefe}, {Santerne},
  {Santos}, {Sarkis}, {Schlieder}, {Schwarz}, {Scott}, {Shporer}, {Smith},
  {Stibbard}, {Stockdale}, {Str{\o}m}, {Twicken}, {Tan}, {Tanner}, {Teske},
  {Tilbrook}, {Tinney}, {Udry}, {Villase{\~n}or}, {Vines}, {Wang}, {Weiss},
  {West}, {Wheatley}, {Wright}, {Zhang}, \& {Zohrabi}}]{2021MNRAS.507.2782O}
{Osborn}, A., {Armstrong}, D.~J., {Cale}, B., {et~al.} 2021, \mnras, 507, 2782,
  \dodoi{10.1093/mnras/stab2313}

\bibitem[{{Peng} \& {Valencia}(2024)}]{2024ApJ...976..202P}
{Peng}, B., \& {Valencia}, D. 2024, \apj, 976, 202,
  \dodoi{10.3847/1538-4357/ad6f03}

\bibitem[{{Pollack} {et~al.}(1996){Pollack}, {Hubickyj}, {Bodenheimer},
  {Lissauer}, {Podolak}, \& {Greenzweig}}]{1996Icar..124...62P}
{Pollack}, J.~B., {Hubickyj}, O., {Bodenheimer}, P., {et~al.} 1996, \icarus,
  124, 62, \dodoi{10.1006/icar.1996.0190}

\bibitem[{{Ram{\'\i}rez} {et~al.}(2007){Ram{\'\i}rez}, {Allende Prieto}, \&
  {Lambert}}]{2007AA...465..271R}
{Ram{\'\i}rez}, I., {Allende Prieto}, C., \& {Lambert}, D.~L. 2007, \aap, 465,
  271, \dodoi{10.1051/0004-6361:20066619}

\bibitem[{{Ram{\'\i}rez} {et~al.}(2014{\natexlab{a}}){Ram{\'\i}rez},
  {Mel{\'e}ndez}, {Bean}, {Asplund}, {Bedell}, {Monroe}, {Casagrande},
  {Schirbel}, {Dreizler}, {Teske}, {Tucci Maia}, {Alves-Brito}, \&
  {Baumann}}]{2014A&A...572A..48R}
{Ram{\'\i}rez}, I., {Mel{\'e}ndez}, J., {Bean}, J., {et~al.}
  2014{\natexlab{a}}, \aap, 572, A48, \dodoi{10.1051/0004-6361/201424244}

\bibitem[{{Ram{\'\i}rez} {et~al.}(2014{\natexlab{b}}){Ram{\'\i}rez},
  {Mel{\'e}ndez}, {Bean}, {Asplund}, {Bedell}, {Monroe}, {Casagrande},
  {Schirbel}, {Dreizler}, {Teske}, {Tucci Maia}, {Alves-Brito}, \&
  {Baumann}}]{Ramirez2014}
---. 2014{\natexlab{b}}, \aap, 572, A48, \dodoi{10.1051/0004-6361/201424244}

\bibitem[{{Romaniello}(2022)}]{2022SPIE12186E..0DR}
{Romaniello}, M. 2022, in Society of Photo-Optical Instrumentation Engineers
  (SPIE) Conference Series, Vol. 12186, Observatory Operations: Strategies,
  Processes, and Systems IX, ed. D.~S. {Adler}, R.~L. {Seaman}, \& C.~R.
  {Benn}, 121860D, \dodoi{10.1117/12.2628253}

\bibitem[{{Santos} {et~al.}(2004){Santos}, {Israelian}, \&
  {Mayor}}]{2004A&A...415.1153S}
{Santos}, N.~C., {Israelian}, G., \& {Mayor}, M. 2004, \aap, 415, 1153,
  \dodoi{10.1051/0004-6361:20034469}

\bibitem[{{Schmidt} {et~al.}(2008{\natexlab{a}}){Schmidt}, {Neuh{\"a}user},
  {Seifahrt}, {Vogt}, {Bedalov}, {Helling}, {Witte}, \&
  {Hauschildt}}]{2008AA...491..311S}
{Schmidt}, T.~O.~B., {Neuh{\"a}user}, R., {Seifahrt}, A., {et~al.}
  2008{\natexlab{a}}, \aap, 491, 311, \dodoi{10.1051/0004-6361:20078840}

\bibitem[{{Schmidt} {et~al.}(2008{\natexlab{b}}){Schmidt}, {Neuh{\"a}user},
  {Seifahrt}, {Vogt}, {Bedalov}, {Helling}, {Witte}, \&
  {Hauschildt}}]{2008A&A...491..311S}
---. 2008{\natexlab{b}}, \aap, 491, 311, \dodoi{10.1051/0004-6361:20078840}

\bibitem[{{Serrano} {et~al.}(2022){Serrano}, {Gandolfi}, {Mustill},
  {Barrag{\'a}n}, {Korth}, {Dai}, {Redfield}, {Fridlund}, {Lam}, {D{\'\i}az},
  {Grziwa}, {Collins}, {Livingston}, {Cochran}, {Hellier}, {Bellomo},
  {Trifonov}, {Rodler}, {Alarcon}, {Jenkins}, {Latham}, {Ricker}, {Seager},
  {Vanderspeck}, {Winn}, {Albrecht}, {Collins}, {Csizmadia}, {Daylan}, {Deeg},
  {Esposito}, {Fausnaugh}, {Georgieva}, {Goffo}, {Guenther}, {Hatzes},
  {Howell}, {Jensen}, {Luque}, {Mann}, {Murgas}, {Osborne}, {Palle}, {Persson},
  {Rowden}, {Rudat}, {Smith}, {Twicken}, {Van Eylen}, \&
  {Ziegler}}]{2022NatAs...6..736S}
{Serrano}, L.~M., {Gandolfi}, D., {Mustill}, A.~J., {et~al.} 2022, Nature
  Astronomy, 6, 736, \dodoi{10.1038/s41550-022-01641-y}

\bibitem[{{Stassun} {et~al.}(2017){Stassun}, {Collins}, \&
  {Gaudi}}]{2017AJ....153..136S}
{Stassun}, K.~G., {Collins}, K.~A., \& {Gaudi}, B.~S. 2017, \aj, 153, 136,
  \dodoi{10.3847/1538-3881/aa5df3}

\bibitem[{{Sun}(2026)}]{2026ApJS..282...37S}
{Sun}, Q. 2026, \apjs, 282, 37, \dodoi{10.3847/1538-4365/ae2a29}

\bibitem[{{Sun} {et~al.}(2025{\natexlab{a}}){Sun}, {Ji}, {Wang}, {Lin},
  {Teske}, {Ting}, {Bedell}, \& {Liu}}]{2025A&A...701A.107S}
{Sun}, Q., {Ji}, C., {Wang}, S.~X., {et~al.} 2025{\natexlab{a}}, \aap, 701,
  A107, \dodoi{10.1051/0004-6361/202556272}

\bibitem[{{Sun} {et~al.}(2024){Sun}, {Wang}, {Welbanks}, {Teske}, \&
  {Buchner}}]{2024AJ....167..167S}
{Sun}, Q., {Wang}, S.~X., {Welbanks}, L., {Teske}, J., \& {Buchner}, J. 2024,
  \aj, 167, 167, \dodoi{10.3847/1538-3881/ad298d}

\bibitem[{{Sun} {et~al.}(2025{\natexlab{b}}){Sun}, {Wang}, {Gan}, {Ji}, {Lin},
  {Ting}, {Teske}, {Li}, {Liu}, {Hua}, {Tang}, {Yu}, {Zhang}, {Badenas-Agusti},
  {Vanderburg}, {Ricker}, {Vanderspek}, {Latham}, {Seager}, {Jenkins},
  {Schwarz}, {Guillot}, {Tan}, {Conti}, {Collins}, {Srdoc}, {Stockdale},
  {Suarez}, {Zambelli}, {Radford}, {Barkaoui}, {Evans}, \&
  {Bieryla}}]{2025ApJ...980..179S}
{Sun}, Q., {Wang}, S.~X., {Gan}, T., {et~al.} 2025{\natexlab{b}}, \apj, 980,
  179, \dodoi{10.3847/1538-4357/ad9924}

\bibitem[{{Swastik} {et~al.}(2021){Swastik}, {Banyal}, {Narang}, {Manoj},
  {Sivarani}, {Reddy}, \& {Rajaguru}}]{2021AJ....161..114S}
{Swastik}, C., {Banyal}, R.~K., {Narang}, M., {et~al.} 2021, \aj, 161, 114,
  \dodoi{10.3847/1538-3881/abd802}

\bibitem[{{Tran} {et~al.}(2014){Tran}, {Holt}, {Goodrich}, {Mader}, {Swain},
  {Laity}, {Kong}, {Gelino}, \& {Berriman}}]{2014SPIE.9152E..2IT}
{Tran}, H.~D., {Holt}, J., {Goodrich}, R.~W., {et~al.} 2014, in Society of
  Photo-Optical Instrumentation Engineers (SPIE) Conference Series, Vol. 9152,
  Software and Cyberinfrastructure for Astronomy III, ed. G.~{Chiozzi} \& N.~M.
  {Radziwill}, 91522I, \dodoi{10.1117/12.2054830}

\bibitem[{{Tsantaki} {et~al.}(2025){Tsantaki}, {Magrini}, {Danielski},
  {Bossini}, {Turrini}, {Moedas}, {Folsom}, {Ramler}, {Biazzo}, {Campante},
  {Delgado-Mena}, {da Silva}, {Sousa}, {Benatti}, {Casali}, {He{\l}miniak},
  {Rainer}, \& {Sanna}}]{2025A&A...697A.102T}
{Tsantaki}, M., {Magrini}, L., {Danielski}, C., {et~al.} 2025, \aap, 697, A102,
  \dodoi{10.1051/0004-6361/202453059}

\bibitem[{{Ward-Duong} {et~al.}(2021){Ward-Duong}, {Patience}, {Follette}, {De
  Rosa}, {Rameau}, {Marley}, {Saumon}, {Nielsen}, {Rajan}, {Greenbaum}, {Lee},
  {Wang}, {Czekala}, {Duch{\^e}ne}, {Macintosh}, {Ammons}, {Bailey}, {Barman},
  {Bulger}, {Chen}, {Chilcote}, {Cotten}, {Doyon}, {Esposito}, {Fitzgerald},
  {Gerard}, {Goodsell}, {Graham}, {Hibon}, {Hom}, {Hung}, {Ingraham}, {Kalas},
  {Konopacky}, {Larkin}, {Maire}, {Marchis}, {Marois}, {Metchev},
  {Millar-Blanchaer}, {Oppenheimer}, {Palmer}, {Perrin}, {Poyneer}, {Pueyo},
  {Rantakyr{\"o}}, {Ren}, {Ruffio}, {Savransky}, {Schneider},
  {Sivaramakrishnan}, {Song}, {Soummer}, {Tallis}, {Thomas}, {Wallace},
  {Wiktorowicz}, \& {Wolff}}]{2021AJ....161....5W}
{Ward-Duong}, K., {Patience}, J., {Follette}, K., {et~al.} 2021, \aj, 161, 5,
  \dodoi{10.3847/1538-3881/abc263}

\bibitem[{{Welbanks} {et~al.}(2019){Welbanks}, {Madhusudhan}, {Allard},
  {Hubeny}, {Spiegelman}, \& {Leininger}}]{2019ApJ...887L..20W}
{Welbanks}, L., {Madhusudhan}, N., {Allard}, N.~F., {et~al.} 2019, \apjl, 887,
  L20, \dodoi{10.3847/2041-8213/ab5a89}

\bibitem[{{Wu} {et~al.}(2015){Wu}, {Close}, {Males}, {Barman}, {Morzinski},
  {Follette}, {Bailey}, {Rodigas}, {Hinz}, {Puglisi}, {Xompero}, \&
  {Briguglio}}]{2015ApJ...801....4W}
{Wu}, Y.-L., {Close}, L.~M., {Males}, J.~R., {et~al.} 2015, \apj, 801, 4,
  \dodoi{10.1088/0004-637X/801/1/4}

\bibitem[{{Yi} {et~al.}(2001){Yi}, {Demarque}, {Kim}, {Lee}, {Ree}, {Lejeune},
  \& {Barnes}}]{2001ApJS..136..417Y}
{Yi}, S., {Demarque}, P., {Kim}, Y.-C., {et~al.} 2001, \apjs, 136, 417,
  \dodoi{10.1086/321795}

\end{thebibliography}
\bibliographystyle{aasjournal}
	
\appendix

Table \ref{tab:jwst_host1} summarizes the planets targeted in the JWST Cycle 1 and 2 GTO programs. For each system, we provide the host star’s effective temperature ($T_{\rm eff}$) from the NASA Exoplanet Archive, the source and wavelength range of available archival spectra, the principal investigator (PI), and notes on data quality or special circumstances (e.g., co-added spectra, low S/N, or unavailable public data). Exposure times or S/N are included when known, along with the planned JWST instrument configurations. Systems are organized by spectral type, with FGK stars listed first, followed by A-type and late K/M-type stars. Table \ref{tab:jwst_host2} presents planets from the JWST Cycle 3 program and from Cycles 1–3 DDT programs, using the same set of information. In this study, we focus on FGK stars with spectra that have S/N $>100$ and broad optical wavelength coverage; analysis of M-dwarf hosts will be presented separately.

Several individual cases are worth noting. In Table \ref{tab:jwst_host1}, the host stars of 51~Eri~b, WASP-121~b, GJ~504~b, LTT~9779~b, WASP-52~b, and WASP-69~b are already included in the JWST Cycle 2 GO program, for which a detailed abundance paper has been published (\citealt[][hereafter Sun26]{2026ApJS..282...37S}). Other systems were analyzed in \citet[][hereafter Sun24]{2024AJ....167..167S} or lack public spectra, as indicated in the table, and are therefore not measured here. HD~95086, HR~8799~b–e, and HR~2562 are fast rotators, causing strong line blending that makes most abundance measurements unreliable. For these stars, we adopt stellar parameters from the literature. Potassium (K) can be measured in HD~2562, but other elements are too blended, while HD~95086 does not cover the K line and most lines are blended, so its abundances cannot be determined. From the JWST Cycle 1 and 2 GTO programs, we measure abundances for 19 elements with $Z \le 30$ in three FGK stars, WASP-77~A, WASP-17, and HAT-P-26, and determine the K abundance for HD~2562.

In Table \ref{tab:jwst_host2}, HD~57167 is a fast-rotating spectroscopic binary whose components cannot be separated in the spectrum, and is therefore omitted from our analysis. AF~Lep and V1298~Tau are also rapid rotators, with most spectral lines blended and equivalent widths unreliable. Their public UVES and HARPS spectra do not cover the oxygen (O) triplet or potassium (K) lines. HD~206893 shows similarly blended features, but its spectra include relatively unblended O and K lines, allowing these abundances to be measured. The abundances of other planet-hosting stars are either presented in Sun26 or Sun24, lack public spectra, or do not provide sufficient wavelength coverage for reliable stellar parameters and abundance determination; these details are noted in the table comments. CT~Chamaeleontis (CT~Cha) is a classical T~Tauri star ($\sim$2~Myr) exhibiting subsolar metallicity ([Fe/H] = $-0.56$~dex, \citealt{2021AJ....161..114S}). We find a strong veiling effect in its spectrum, requiring more careful treatment for abundance analysis including metallicity. Observations confirm a brown dwarf/planet companion around CT~Cha \citep{2008A&A...491..311S, 2015ApJ...801....4W}. CT~Cha is a particularly interesting system and will be analyzed in detail in a future study.

Excluding these special cases, we report homogeneous stellar parameters and high-precision abundances for 20 FGK stars for 19 elements with $Z \le 30$. For TOI-4010, TOI-1807, and HAT-P-18, we adopt stellar parameters from the literature to derive abundances for the same 19 elements. Similarly, for HD~206893 and HD~2562 we adopt stellar parameters from the literature. For HD~206893, we report O and K abundances, and for HD~2562 we report the K abundance, as the other elements cannot be measured due to line blending and/or limited wavelength coverage.

\begin{figure*}
	\centering
	\renewcommand\thefigure{A1}
	\includegraphics[width=1.00\textwidth]{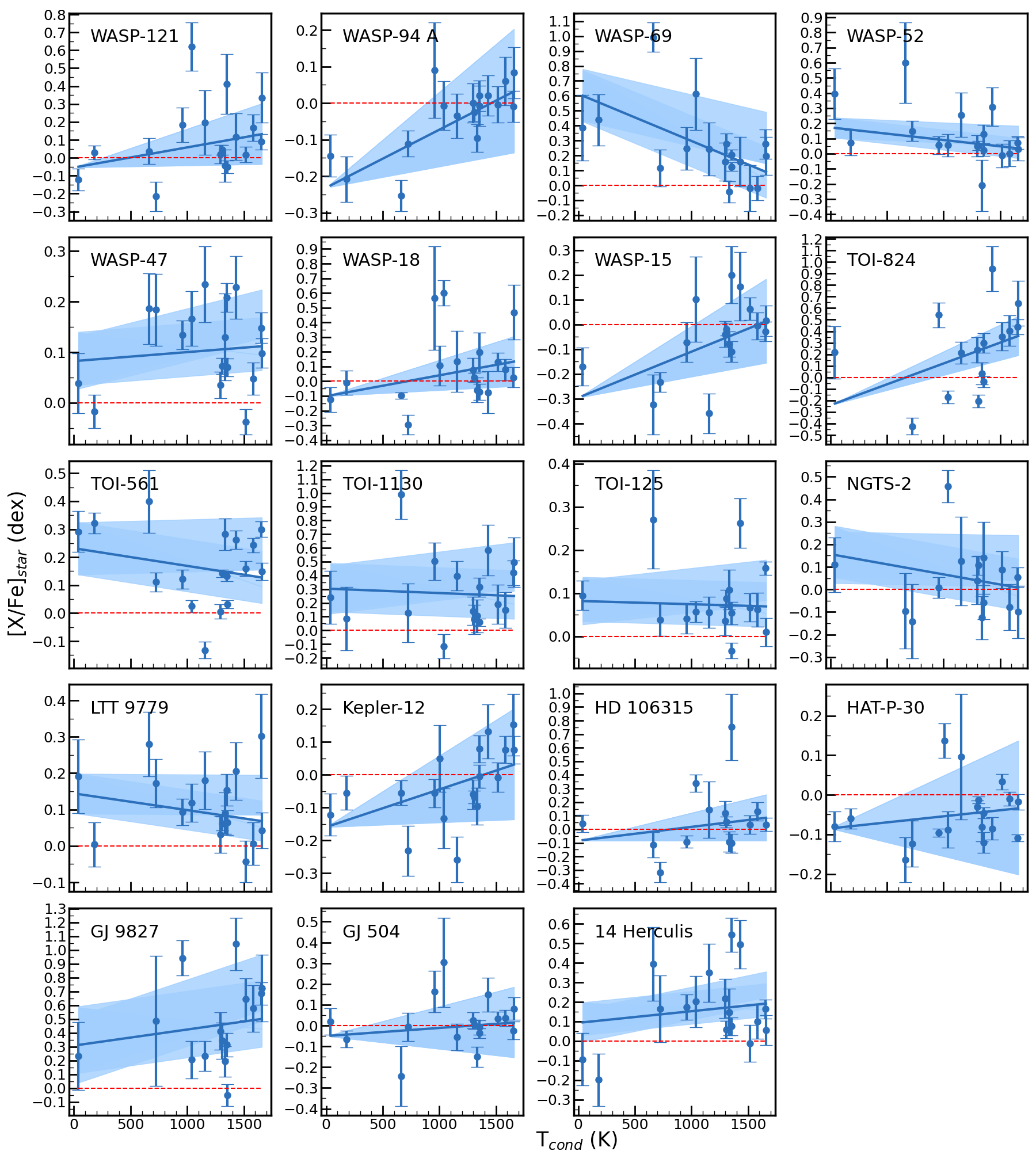}
	\caption{[X/Fe]$_{star}$ versus condensation temperature ($T_{\rm cond}$) for 19 planet-hosting stars in JEWELS I. All abundances are derived on a differential, line-by-line basis relative to the Sun. PH2 is not shown, as we adopt its stellar parameters and abundances from the literature rather than our homogeneous analysis.}
	\label{fig:Tc_JEWELS1}
\end{figure*}

\begin{figure*}
	\centering
	\renewcommand\thefigure{A2}
	\includegraphics[width=1.00\textwidth]{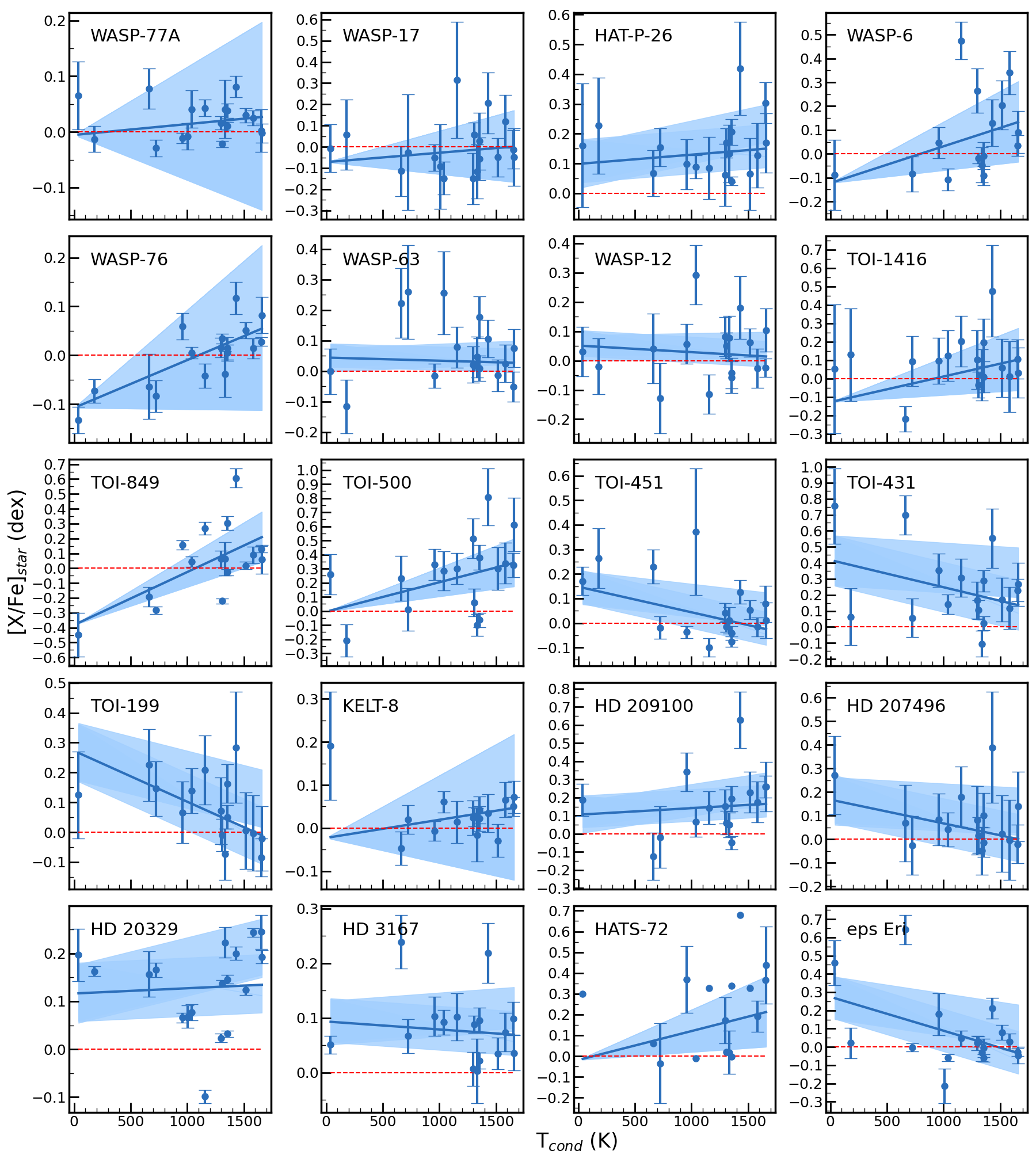}
	\caption{[X/Fe] versus $T_{\rm cond}$ for 20 planet-hosting stars in JEWELS II. Stellar parameters and abundances are derived homogeneously on a differential, line-by-line basis. Five additional stars (TOI-4010, TOI-1807, HAT-P-18, HR~2562, and HD~20689) for which we adopt literature values are not shown.}
	\label{fig:Tc_JEWELS2}
\end{figure*}

\begin{figure*}
	\centering
	\renewcommand\thefigure{A3}
	\includegraphics[width=1.00\textwidth]{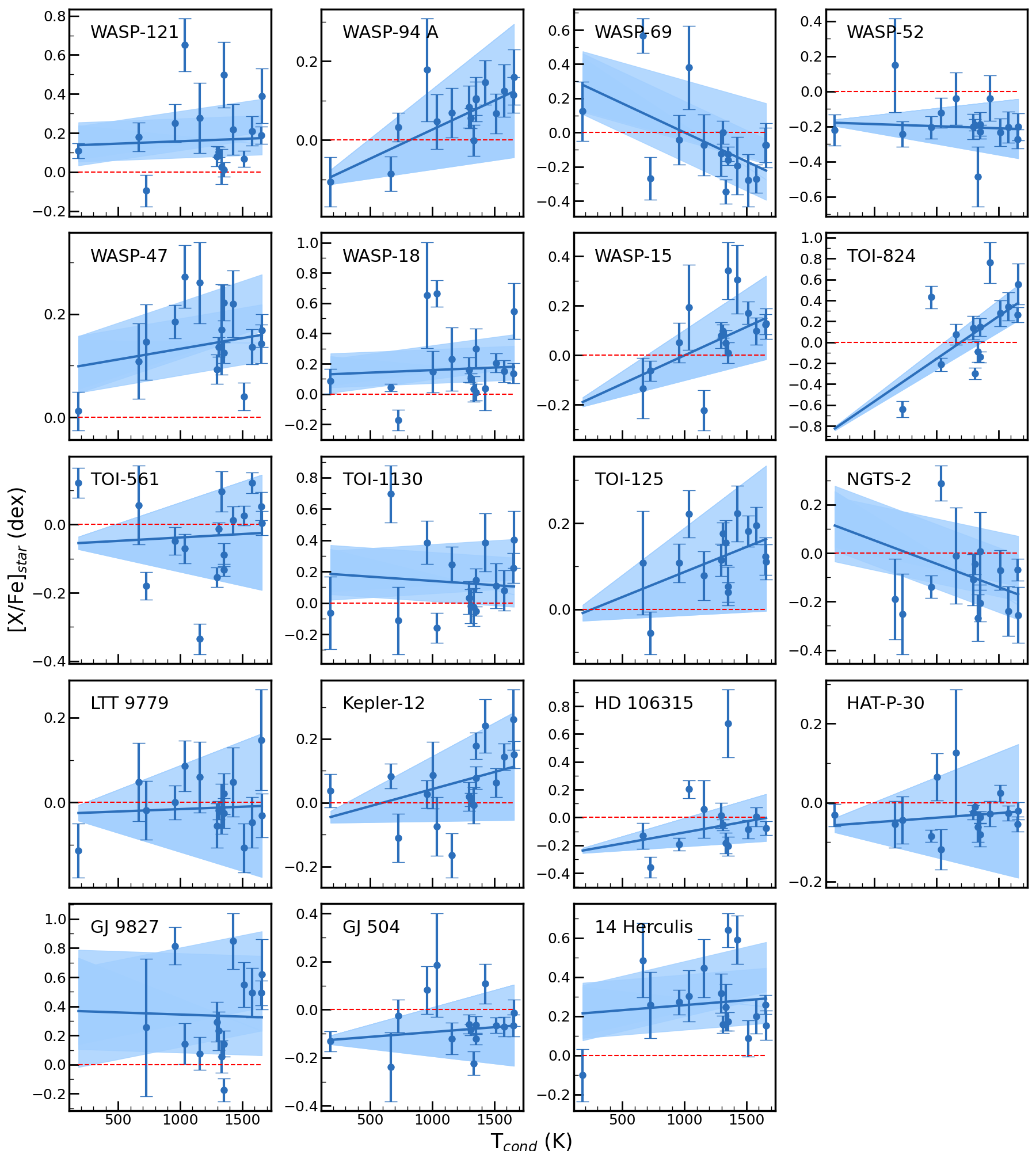}
	\caption{Same as Figure~\ref{fig:Tc_JEWELS1}, but with [X/Fe] corrected for Galactic chemical evolution using the relations in Figure~\ref{fig:GCE}.}
	\label{fig:Tc_JEWELS1_GCE}
\end{figure*}

\begin{figure*}
	\centering
	\renewcommand\thefigure{A4}
	\includegraphics[width=1.00\textwidth]{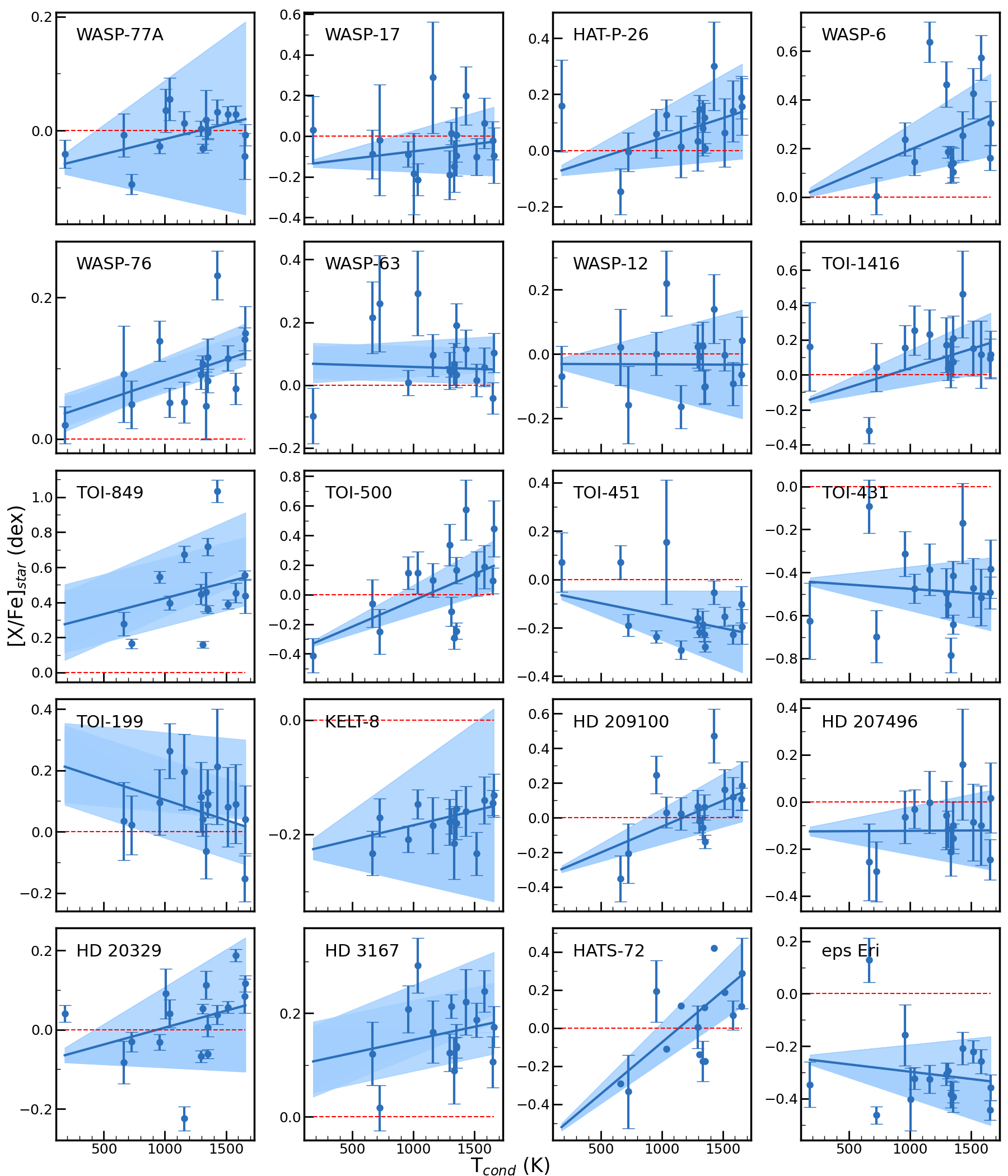}
	\caption{Same as Figure~\ref{fig:Tc_JEWELS2}, but with [X/Fe] corrected for Galactic chemical evolution using the relations in Figure~\ref{fig:GCE}.}
	\label{fig:Tc_JEWELS2_GCE}
\end{figure*}

Table \ref{tab:ew} provides the equivalent width (EW) measurements used in the stellar parameter and abundance analysis. For each line, we list the wavelength, ion, excitation potential, $\log(gf)$, and adopted damping constant. EW measurements from solar spectra obtained with VLT/ESPRESSO, Keck/HIRES, ESO-3.6m/HARPS, ESO-2.2m/FEROS, and VLT/UVES are included for cross-comparison, followed by those for the stellar targets. TOI-4010, TOI-1807, and HAT-P-18 show measurable EWs only in limited spectral regions, as indicated in the table. The EWs of selected lines in HD~206893 and HD~2562 are reported in the main text. A subset of the table is shown here; the complete set of EW measurements is available in machine-readable form online. Table \ref{tab:abund_S} shows synthesized and average S abundances for cool dwarfs in this work and in Sun26. Figures \ref{fig:Tc_JEWELS1}–\ref{fig:Tc_JEWELS2_GCE} show the $T_{\rm cond}$ trends for the 39 stars in JEWELS I \& II, both before and after GCE correction.

\begin{longrotatetable}
	\begin{deluxetable}{p{2.5cm}p{1.0cm}ccp{1.5cm}ccp{2.5cm}}
		\tablecaption{Metadata for planet-hosting stars observed in JWST Cycle 1 and 2 GTO programs. \label{tab:jwst_host1}}
		\renewcommand\thetable{A1}
		\tablehead{
			\colhead{Planet Name$^a$} & \colhead{$T_{\mathrm{eff}}^a$} & \colhead{Archive$^b$} &
			\colhead{Wavelength$^b$} & \colhead{PI$^b$} & \colhead{Note$^b$} & \colhead{Exp/SN$^b$} & \colhead{Instrument(s)$^c$}
		}
		\startdata
		\hline
		\multicolumn{8}{c}{F, G, K-type stars} \\ \hline \hline
		HD 95086 b & 7750 & HARPS & 3782-6913 & LAGRANGE, A.-M. & & 1200 & NIRCam Coronagraph, NIRISS AMI, MIRI Coronagraph \\ \hline
		51 Eri b & 7422 &  &  &  & covered in cycle 2 GO & 1241, 1412 & MIRI Coronagraph, NIRCam Coronagraph \\ \hline
		HR 8799 b–e & 7339 & HARPS & 3781-6912 & CHAUVIN, G. &  & 1200 & NIRSpec IFU, NIRCam Coronagraph, MIRI Coronagraph, NIRISS AMI \\ \hline
		WASP-121 b & 6776 &  &  &  & covered in cycle 2 GO & 1201 & NIRISS SOSS \\ \hline
		HR 2562 b & 6597 & FEROS & 3528-9217 & BAYO, AMELIA &  & 305 (blue), 1241 (red) & MIRI Coronagraph \\ \hline
		WASP-17 b & 6550 & FEROS & 3527-9216 & FAEDI, FRANCESCA & co-add four exposures & 126 (blue), 1353 (red) & NIRISS SOSS, NIRSpec G395H, MIRI LRS \\ \hline
		HD 149026 b & 6147 &  &  &  & no public spectra available & 1274 & NIRCam F322W2 grism, NIRCam F444W grism \\ \hline
		HD 209458 b & 6071 &  &  &  & published in Sun24 & 1274 & NIRCam F322W2 grism, NIRCam F444W grism \\ \hline
		GJ 504 b & 6000 &  &  &  & covered in cycle 2 GO & 1277, 2778 & MIRI Coronagraph, NIRSpec IFU \\ \hline
		HAT-P-1 b & 5975 &  &  &  & published in Sun24 & 1201 & NIRISS SOSS \\ \hline
		WASP-127 b & 5750 &  &  &  & published in Sun24 & 1201 & NIRISS SOSS \\ \hline
		WASP-77 A b & 5605 & ESPRESSO & 3772-7900 & WANG, WEI &  & 280 (blue), 1274 (red) & NIRSpec G395H \\ \hline
		WASP-19 b & 5460 &  &  &  & published in Sun24, S/N $<$ 100 & 1274 & NIRSpec PRISM \\ \hline
		LTT 9779 b & 5445 &  &  &  & covered in cycle 2 GO & 1201 & NIRISS SOSS \\ \hline
		HAT-P-26 b & 5079 & ESPRESSO & 3772-7900 & LAFARGA MAGRO, MARINA & & 264, 1312 & MIRI LRS, NIRISS SOSS, NIRSpec G395H \\ \hline
		WASP-52 b & 5000 &  &  &  & covered in cycle 2 GO & 1201, 1224 & NIRISS SOSS, NIRSpec PRISM \\ \hline
		HD 189733 b & 4969 &  &  &  & published in Sun24 & 1185, 1274 & NIRCam F444W grism, NIRCam F322W2 grism \\ \hline
		WASP-69 b & 4700 &  &  &  & covered in cycle 2 GO & 1177, 1185 & MIRI LRS, NIRCam F444W grism, NIRCam F322W2 grism \\ \hline
		HAT-P-12 b & 4650 &  &  &  & published in Sun24, though S/N$<$100 & 1281 & NIRISS SOSS, NIRSpec G395M, MIRI LRS \\ \hline
		WASP-107 b & 4425 &  &  &  & published in Sun24, S/N$<$100 & 1185, 1201, 1224, 1280 & NIRCam F444W grism, NIRCam F322W2 grism, NIRISS SOSS, NIRSpec G395H, MIRI LRS \\ \hline
		WASP-43 b & 4400 &  &  &  & published in Sun24, S/N$<$100 & 1224 & NIRSpec G395H \\ \hline \hline
		\multicolumn{8}{c}{A-type; late-K/M-type} \\ \hline \hline
		bet Pic b & 8500 &  &  &  &  & 1241 & MIRI Coronagraph \\ \hline
		2MASS J22362452+4751425 b & 4033 &  &  &  &  & 1188 & NIRSpec IFU, MIRI LRS \\ \hline
		WD 0806-661 b & white dwarf &  &  &  &  & 1276 & NIRSpec IFU, NIRCam Imaging, MIRI LRS, MIRI Imaging \\ \hline
		GU Psc b & 1000-1100 &  &  &  &  & 1188 & NIRSpec IFU, MIRI LRS \\ \hline
		WASP-80 b & 4145 &  &  &  &  & 1177, 1185, 1201 & MIRI LRS, NIRCam F444W grism, NIRCam F322W2 grism, NIRISS SOSS \\ \hline
		PDS 70 b & 3972 &  &  &  &  & 1179, 1242 & NIRCam Imaging, NIRISS AMI \\ \hline
		PDS 70 c & 3972 &  &  &  &  & 1179, 1242 & NIRCam Imaging, NIRISS AMI \\ \hline
		GJ 3470 b & 3725 &  &  &  & published in Sun24, S/N$<$100 & 1185, 1201 & NIRCam F444W grism, NIRCam F322W2 grism, NIRSpec G395H \\ \hline
		L231-32 d & 3506 &  &  &  &  & 2759 & NIRISS SOSS \\ \hline
		GJ 357 b & 3505 &  &  &  & covered in cycle 2 GO & 1201 & NIRISS SOSS \\ \hline
		Ross 458 c & 3484 &  &  &  &  & 1277, 1292 & NIRSpec IFU, MIRI MRS, NIRSpec FS \\ \hline
		GJ 436 b & 3416 &  &  &  & & 1177, 1185 & MIRI LRS, NIRCam F444W grism, NIRCam F322W2 grism \\ \hline
		L 98-59 c & 3412 &  &  &  &  & 1201 & NIRISS SOSS \\ \hline
		L 98-59 d & 3412 &  &  &  &  & 1201, 1224 & NIRISS SOSS, NIRSpec G395H \\ \hline
		GJ 1132 b & 3270 &  &  &  &  & 1274 & MIRI LRS \\ \hline
		GJ 1214 b & 3101 &  &  &  &  & 1185 & NIRSpec G395H \\ \hline
		LP 791-18 c & 2960 &  &  &  &  & 1201 & NIRSpec PRISM \\ \hline
		TWA 27 b & 2640 &  &  &  &  & 1270 & NIRSpec IFU, MIRI MRS \\ \hline
		TRAPPIST-1 b,f,e,d & 2566 &  &  &  & covered in cycle 2 GO & 1177, 1201, 1279, 1331 & MIRI F1500W, NIRSpec PRISM, NIRISS SOSS, MIRI F1280W \\ \hline
		HD 106906 b & 1820 &  &  &  &  & 1277 & NIRSpec IFU, MIRI Coronagraph, MIRI LRS \\ \hline
		kap And b & 1700 &  &  &  &  & 1241 & MIRI Coronagraph \\ \hline
		GJ 758 b & 700 &  &  &  &  & 1413 & MIRI Coronagraph \\ \hline
		Delorme 1 AB b & young, hot &  &  &  &  & 2778 & MIRI MRS \\ \hline
		\enddata
		\tablecomments{a. Planets proposed for observation in the JWST Cycle 1--2 GTO programs; host-star effective temperatures ($T_{\rm eff}$) are adopted from the NASA Exoplanet Archive. \\
			b. Information on available ground-based stellar spectra, including the instrument, wavelength coverage, principal investigator of the public data, relevant notes on data quality or processing, and either total exposure time (for HIRES) or representative signal-to-noise ratio (Exp./S/N). \\
			c. JWST instruments proposed for each system. The sample includes both transiting planets observed in time-series spectroscopy and directly imaged planets observed via high-contrast imaging or IFU modes.}
	\end{deluxetable}
\end{longrotatetable}

\begin{longrotatetable}
	\begin{deluxetable}{p{2.5cm}p{1.0cm}ccp{1.5cm}ccp{2.5cm}}
		\tablecaption{Metadata for planet-hosting stars observed in JWST Cycle 3 GO and GTO programs, as well as in DDT programs from all cycles. \label{tab:jwst_host2}}
		\renewcommand\thetable{A2}
		\tablehead{
			\colhead{Planet Name} & \colhead{$T_{\mathrm{eff}}$} & \colhead{Archive} &
			\colhead{Wavelength} & \colhead{PI} & \colhead{Note} & \colhead{Exp/SN} & \colhead{Instrument(s)}
		}
		\startdata
		\hline
		\multicolumn{8}{c}{F, G, K-type stars} \\ \hline \hline
		R Canis Majoris (HD 57167) & 7300, 4350 & UVES & 3023-10430 & PARANAL OBSERVATORY, ESO & co-add all 4 exposures & $>$100 & MIRI MRS \\ \hline
		HR 8799 (b–e) & 7339 &  &  &  & covered in cycles 1–2 GTO &  & MIRI MRS, NIRCam Coronagraph \\ \hline
		KELT-7 b & 6768 &  &  &  & covered in cycle 2 GO &  & NIRISS SOSS \\ \hline
		NGTS-2 b & 6527 &  &  &  & covered in cycle 2 GO &  & NIRISS SOSS \\ \hline
		HD 206893 b & 6500 & FEROS & 3527–9216 & CARSON, JOSEPH &  & 377 & NIRSpec IFU \\ \hline
		WASP-76 b & 6366 & ESPRESSO & 3772–7900 & PEPE, FRANCESCO &  & 319 & NIRSpec G395H, NIRISS SOSS, MIRI LRS \\ \hline
		HAT-P-30 b & 6304 &  &  &  & covered in cycle 2 GO &  & NIRISS SOSS \\ \hline
		WASP-12 b & 6265 & FEROS & 3527–9216 & SOUSA, SERGIO & co-add 10 exposures & $\sim$100 & NIRSpec PRISM \\ \hline
		AF Lep b & 6130 & HARPS & 3783–6913 & RIGLIACO, ELISABETTA & & 181 & NIRSpec IFU, NIRCam Coronagraph \\ \hline
		WASP-94 A b & 6072 &  &  &  & covered in cycle 2 GO &  & NIRISS SOSS \\ \hline
		HAT-P-1 b & 5975 &  &  &  & published in Sun24 &  & NIRSpec G395H \\ \hline
		WASP-103 b & 5843 &  &  &  & FEROS spectrum, S/N too low &  & NIRSpec PRISM \\ \hline
		WASP-63 b & 5715 & FEROS & 3527–9216 & MANCINI, LUIGI & co-add all 5 exposures & 110 & NIRSpec G395H \\ \hline
		KELT-8 b & 5703 & UVES & 4726–6835 & SOUSA, SERGIO & & 141 & MIRI LRS \\ \hline
		V1298 Tau (c–e) & 5700 & UVES & 4726–6835 & SOUSA, SERGIO & same as above & 96 & NIRSpec G395H, NIRISS SOSS \\ \hline
		HD 20329 b & 5596 & ESPRESSO & 3772–7900 & SOARES, BÁRBARA & & 280 & MIRI LRS \\ \hline
		TOI-451 (c–d) & 5550 & ESPRESSO & 3772–7900 & MALLORQUÍN DÍAZ, MANUEL & co-add 5 exposures & 170 & NIRISS SOSS, NIRSpec G395H \\ \hline
		WASP-19 b & 5460 &  &  &  & published in Sun24, S/N $<$ 100 &  & NIRSpec PRISM \\ \hline
		WASP-6 b & 5380 & HARPS & 3781–6912 & CAMERON, A.; EHRENREICH, D.; OSHAGH, M. & coadd 39 low S/N spectra & 105 & NIRSpec PRISM \\ \hline
		TOI-561 b & 5342 &  &  &  & covered in cycle 2 GO &  & MIRI LRS \\ \hline
		WASP-39 b & 5327 &  &  &  & published in Sun24 &  & MIRI LRS, NIRSpec G395H \\ \hline
		HD 3167 b & 5261 & UVES & 3282–4563, 4726–6835 & GUENTHER, EIKE &  & 476, 155 & MIRI LRS \\ \hline
		TOI-849 b & 5257 & HARPS & 3782–6913 & ARMSTRONG, DAVID & co-add 17 exposures & 106 & NIRSpec G395H \\ \hline
		TOI-199 b & 5255 & HARPS & 3782–6913 & BRAHM, RAFAEL & co-add 6 exposures & 131 & NIRSpec G395M \\ \hline
		TOI-2076 (b–d) & 5192 &  &  &  & spectrum not found &  & NIRISS SOSS, NIRSpec G395H \\ \hline
		TOI-2525 (b–c) & 5096 &  &  &  & S/N too low &  & NIRSpec PRISM \\ \hline
		eps Eri b & 5020 & FEROS & 3527–9216 & 2P2 TEAM 22 &  & 249 & NIRSpec IFU \\ \hline
		TOI-4010 (b–d) & 4960 & HIRES & 3360–8100 & Dressing & Selective wavelength exclusion & 128s & NIRSpec PRISM \\ \hline
		TOI-1807 b & 4914 & HIRES & 3360–8100 & Chontos & Selective wavelength exclusion & 15s & MIRI LRS \\ \hline
		TOI-1416 b & 4884 & HIRES & 3360–8100 & Howard & Selective wavelength exclusion & 16s & MIRI LRS \\ \hline
		Kepler-167 e & 4884 & HIRES & 3360–8100 & A. Kraus & S/N too low, drop this & 876 & NIRISS SOSS, NIRSpec PRISM \\ \hline
		TOI-431 b & 4850 & ESPRESSO & 3772–7900 & SOARES, BÁRBARA &  & 329 & MIRI LRS \\ \hline
		HD 207496 b & 4819 & HARPS & 3782–6912 & ARMSTRONG, DAVID &  & 127 & NIRCam F332W2 grism, NIRCam F444W grism \\ \hline
		HAT-P-18 b & 4790 & HIRES & 3360–8100 & Bakos & Selective wavelength exclusion & 1320s & NIRSpec G395M, MIRI LRS \\ \hline
		HAT-P-11 b & 4780 &  &  &  & published in Sun24, S/N $<$ 100 &  & NIRISS SOSS \\ \hline
		WASP-69 b & 4700 &  &  &  & covered in cycles 2 GO &  & NIRISS SOSS \\ \hline
		HATS-72 b & 4656 & ESPRESSO & 3772–7900 & JORDAN, ANDRES & co-add 7 exposures & 143 & NIRSpec PRISM \\ \hline
		HAT-P-12 b & 4650 &  &  &  & published in Sun24, S/N$<$100 &  & MIRI LRS \\ \hline
		eps Ind A b (HD 209100) & 4760 & HARPS & 3782–6912 & UDRY, S. &  & 353 & MIRI Coronagraph \\ \hline
		TOI-500 b & 4440 & ESPRESSO & 3772–7900 & ADIBEKYAN, VARDAN &  & 115 & MIRI LRS \\ \hline
		CT Cha b & 4403 & FEROS & 3528–9217 & GREDEL, ROLAND & co-add 17 spectra & 148 & NIRSpec G140H, NIRSpec G235H, NIRSpec G395H \\ \hline
		GJ 9827 b & 4236 &  &  &  & covered in cycle 2 GO &  & MIRI LRS \\ \hline \hline
		\multicolumn{8}{c}{A-type; late K/M-type stars} \\ \hline \hline
		HD 97048 b & 10889 &  &  &  &  & 4543 & NIRCam Coronagraph \\ \hline
		HIP 75056 A b & 8610 &  &  &  &  & 1902 & NIRCam Image \\ \hline
		WASP-189 b & 8000 &  &  &  &  & 3279 & NIRISS SOSS \\ \hline
		bet Pic b & 7890 &  &  &  &  & 4758 & NIRCam Coronagraph \\ \hline
		WASP-80 b & 4145 &  &  &  & observed in cycle1+2 GTO & 5924 & NIRISS SOSS \\ \hline
		TOI-1075 b & 3875 &  &  &  &  & 4818 & MIRI LRS \\ \hline
		SR 12 c & 3829 &  &  &  &  & 6086 & NIRSpec FS, MIRI MRS \\ \hline
		GJ 3470 b & 3725 &  &  &  & published in Sun24, though S/N$<$100 & 4536 & NIRSpec G359H \\ \hline
		AU Mic b & 3678 &  &  &  &  & 5311 & NIRCam F322W2 grism, NIRCam F444W grism \\ \hline
		TOI-4481 b & 3600 &  &  &  &  & 4931 & MIRI LRS \\ \hline
		TOI-700 d,e & 3459 &  &  &  &  & 6193 & NIRISS SOSS \\ \hline
		COCONUTS-2 b & 3406 &  &  &  &  & 6463 & MIRI MRS \\ \hline
		TOI-1442 b & 3330 &  &  &  &  & 4818 & MIRI LRS \\ \hline
		GJ 486 b & 3317 &  &  &  &  & 5866 & NIRISS SOSS \\ \hline
		PSR J2322-2650 b & 3300 &  &  &  &  & 5263 & NIRSpec PRISM, NIRSpec G235H \\ \hline
		TOI-4336 b & 3298 &  &  &  &  & 4711 & NIRISS SOSS, NIRSpec G395H \\ \hline
		TOI-3884 b & 3180 &  &  &  &  & 5799, 5863 & NIRISS SOSS, NIRSpec G395H, NIRSpec G395M \\ \hline
		CHXR 73 b & 3099 &  &  &  &  & 6361 & NIRSpec G140H, NIRSpec G235H \\ \hline
		LHS 1140 b & 3096 &  &  &  &  & 6543 & NIRISS SOSS \\ \hline
		LP 791-18 d & 2960 &  &  &  &  & 6457 & MIRI F1500W \\ \hline
		TRAPPIST-1 b,c,e & 2566 &  &  &  &  & 5191, 6456 & MIRI Imaging, NIRSpec PRISM, NIRISS SOSS \\ \hline
		\enddata
		\tablecomments{This table summarizes targets from JWST Cycle 3 GO and GTO programs, together with all cycle 1--3 DDT observations, and associated ground-based spectroscopic data used for exoplanet characterization. The column definitions are identical to those in Table~\ref{tab:jwst_host1}. Both transiting and directly imaged planets are included.}
	\end{deluxetable}
\end{longrotatetable}

\begin{longrotatetable}
	\begin{deluxetable*}{lcccccccccccccc}
		\centering
		\renewcommand\thetable{A3}
		\tablecaption{Equivalent Widths and Atomic Data for 23 Planet-Hosting Stars \label{tab:ew}}
		\tablehead{
			\colhead{Wavelength$^a$} & \colhead{Ion$^a$} & \colhead{ExPot$^a$} & \colhead{log($gf$)$^a$} & \colhead{Damping C$^a$} 
			& \colhead{Solar$^b$} & \colhead{WASP-77A} & \colhead{WASP-17} & \colhead{HAT-P-26} & \colhead{WASP-6} & \colhead{WASP-76} & \colhead{TOI-849} & \colhead{TOI-500} & \colhead{HD\,3167} & \colhead{eps Eri$^b$} \\
			\colhead{\AA} & \colhead{} & \colhead{eV} & \colhead{} & \colhead{} 
			& \colhead{m\AA} & \colhead{m\AA} & \colhead{m\AA} & \colhead{m\AA} & \colhead{m\AA} & \colhead{m\AA} & \colhead{m\AA} & \colhead{m\AA} & \colhead{m\AA} & \colhead{m\AA}
		}
		\startdata
		5044.211 & 26.0 & 2.8512 & -2.058 & 2.71E-31 & 73.3 & 82.9 & 45.4 & 106.4 & 77.0 & 77.6 & 91.7 & 133.3 & 75.7 & 116.8 \\
		5054.642 & 26.0 & 3.6400 & -1.921 & 4.68E-32 & 39.3 & 46.4 & 15.8 & 51.1 & 40.0 & 41.0 & 60.8 & 63.0 & 41.0 & 62.2 \\
		5127.359 & 26.0 & 0.9150 & -3.307 & 1.84E-32 & 96.5 & 106.3 & 66.9 & 129.2 & 105.4 & 98.5 & 126.0 & -- & 100.3 & 123.1 \\
		5127.679 & 26.0 & 0.0520 & -6.125 & 1.20E-32 & 19.7 & 26.8 & -- & 59.5 & 30.2 & -- & 50.7 & 97.4 & 24.2 & 44.8 \\
		5198.711 & 26.0 & 2.2230 & -2.135 & 4.61E-32 & 97.2 & 112.1 & 80.4 & 159.6 & 104.9 & 101.9 & 128.8 & -- & 100.8 & -- \\
		5225.525 & 26.0 & 0.1101 & -4.789 & 1.23E-32 & 72.4 & 86.1 & 21.0 & 101.3 & 84.4 & 67.5 & 99.5 & 130.6 & 96.0 & 113.2 \\
		5242.491 & 26.0 & 3.6340 & -0.967 & 4.95E-32 & 85.9 & 93.2 & 68.3 & 102.0 & 18.2 & 94.7 & 108.1 & 113.4 & 99.5 & 107.3 \\
		5250.208 & 26.0 & 0.1212 & -4.938 & 1.23E-32 & 67.3 & 74.7 & 15.4 & 91.1 & 77.0 & 58.6 & 87.1 & 125.2 & 86.2 & 93.0 \\
		5295.312 & 26.0 & 4.4150 & -1.490 & 6.54E-31 & 28.8 & 33.8 & 10.6 & 39.3 & 29.4 & 33.6 & 45.2 & 36.7 & 40.1 & 37.9 \\
		5373.709 & 26.0 & 4.4730 & -0.770 & 7.04E-31 & 62.6 & 69.8 & 49.5 & 78.2 & 64.2 & 70.7 & 77.1 & 120.0 & 77.7 & 81.5 \\
		5386.334 & 26.0 & 4.1540 & -1.740 & 5.27E-31 & 32.4 & 37.9 & 18.1 & 46.1 & 34.6 & 35.9 & 50.5 & 45.0 & 45.5 & 42.5 \\
		5466.396 & 26.0 & 4.3710 & -0.565 & 4.40E-31 & 77.9 & 83.8 & 102.6 & 96.1 & 78.5 & 86.8 & 90.4 & 129.9 & 95.1 & 101.5 \\
		... & ... & ... & ... & ... & ... & ... & ... & ...& ... & ... & ... & ... & ... & ... \\
		5197.577 & 26.1 & 3.2306 & -2.22 & 8.69E-33 & 79.2 & 74.8 & 115.8 & 52.6 & 58.5 & 116.1 & 66.9 & 34.9 & 61.8 & 52.8 \\
		5234.625 & 26.1 & 3.2215 & -2.18 & 8.69E-33 & 82.3 & 80.8 & 112.2 & 56.4 & 64.3 & 120.2 & 72.8 & 28.3 & 75.9 & 59.3 \\
		5264.812 & 26.1 & 3.2304 & -3.13 & 9.43E-33 & 46.3 & 42.2 & 51.0 & 26.5 & 31.1 & 74.9 & 41.1 & -- & 40.9 & 20.7 \\
		5414.073 & 26.1 & 3.2215 & -3.58 & 9.30E-33 & 27.9 & 25.9 & 73.3 & 14.4 & 16.0 & 52.5 & 23.4 & -- & 23.9 & 12.5 \\
		5425.257 & 26.1 & 3.1996 & -3.22 & 8.45E-33 & 41.9 & 39.8 & 55.9 & 27.9 & 28.2 & 69.4 & 33.8 & 36.2 & 35.2 & 26.4 \\
		6247.557 & 26.1 & 3.8918 & -2.3 & 9.43E-33 & 52.3 & 46.3 & 88.6 & 27.5 & 34.9 & 94.7 & 37.8 & 11.4 & 44.4 & 27.2 \\
		6456.383 & 26.1 & 3.9036 & -2.05 & 9.30E-33 & 62.9 & 59.6 & 80.2 & 35.5 & 45.2 & 97.1 & 52.6 & 16.7 & 55.4 & 36.3 \\
		... & ... & ... & ... & ... & ... & ... & ... & ...& ... & ... & ... & ... & ... & ... \\
		7771.944 & 8.0 & 9.146 & 0.37 & 8.41E-32 & 68.2 & 54.2 & 164.6 & 29.1 & -- & 137.2 & -- & 10.2 & 63.4 & 22.3 \\
		7774.166 & 8.0 & 9.146 & 0.22 & 8.41E-32 & 60.2 & 49.0 & 135.5 & 22.6 & -- & 127.4 & -- & -- & 55.6 & 24.3 \\
		7775.388 & 8.0 & 9.146 & 0.001 & 8.41E-32 & 47.1 & 40.2 & 126.1 & 17.9 & -- & 105.9 & -- & -- & 43.4 & 14.8 \\
		... & ... & ... & ... & ... & ... & ... & ... & ...& ... & ... & ... & ... & ... & ... \\
		6154.225 & 11.0 & 2.1023 & -1.547 & 2.8 & 38.6 & 42.9 & 16.3 & 65.5 & 34.2 & 51.8 & 66.5 & 104.3 & 62.8 & 59.2 \\
		6160.747 & 11.0 & 2.1044 & -1.246 & 2.8 & 56.7 & 63.1 & 25.8 & 88.6 & 56.9 & 66.6 & 92.2 & 117.2 & 79.6 & 78.0 \\
		... & ... & ... & ... & ... & ... & ... & ... & ...& ... & ... & ... & ... & ... & ... \\
		6696.018 & 13.0 & 3.143 & -1.481 & 2.8 & 37.7 & 47.6 & 17.9 & 74.9 & 39.2 & 46.2 & 65.5 & 93.8 & 61.6 & 55.1 \\
		6698.667 & 13.0 & 3.143 & -1.782 & 2.8 & 20.2 & 25.9 & 16.8 & 53.5 & 21.8 & 26.1 & 50.0 & 68.7 & 39.7 & 31.7 \\
		... & ... & ... & ... & ... & ... & ... & ... & ...& ... & ... & ... & ... & ... & ... \\
		5517.54 & 14.0 & 5.08 & -2.496 & 2.8 & 13.0 & 13.4 & -- & 16.4 & 14.3 & 20.3 & 19.1 & 18.4 & 14.8 & 10.2 \\
		6125.021 & 14.0 & 5.614 & -1.5 & 2.8 & 31.3 & 32.2 & 24.6 & 29.7 & 22.1 & 49.3 & 36.0 & 15.5 & 35.4 & 24.2 \\
	   ... & ... & ... & ... & ... & ... & ... & ... & ...& ... & ... & ... & ... & ... & ... \\
		6743.54 & 16.0 & 7.866 & -0.6 & 2.8 & 10.6 & -- & 14.6 & -- & -- & 27.5 & -- & -- & -- & -- \\
		6757.153 & 16.0 & 7.87 & -0.15 & 2.8 & 17.1 & 15.2 & 29.2 & -- & -- & 54.5 & 17.2 & -- & 13.8 & -- \\
		... & ... & ... & ... & ... & ... & ... & ... & ...& ... & ... & ... & ... & ... & ... \\
		5260.387 & 20.0 & 2.521 & -1.719 & 7.27E-32 & 32.2 & 40.4 & -- & 61.5 & 36.7 & 41.6 & 52.0 & 78.9 & 50.1 & 56.3 \\
		6166.439 & 20.0 & 2.521 & -1.142 & 5.95E-31 & 70.2 & 78.0 & 50.4 & 109.1 & 77.4 & 77.0 & 92.2 & 149.5 & 94.9 & 106.2 \\
		... & ... & ... & ... & ... & ... & ... & ... & ...& ... & ... & ... & ... & ... & ... \\
		5670.85 & 23.0 & 1.08 & -0.42 & 3.58E-32 & 19.0 & 28.9 & -- & 82.1 & 33.7 & 17.1 & 62.1 & 122.2 & 57.0 & 63.1 \\
		6081.44 & 23.0 & 1.051 & -0.578 & 3.89E-32 & 14.4 & 21.7 & -- & 59.3 & 23.2 & 13.3 & 49.1 & 92.4 & 38.4 & 47.5 \\
		... & ... & ... & ... & ... & ... & ... & ... & ...& ... & ... & ... & ... & ... & ... \\
		5247.566 & 24.0 & 0.96 & -1.59 & 3.92E-32 & 82.5 & 92.3 & 41.0 & 122.8 & 96.1 & 80.7 & 114.0 & -- & 86.2 & 121.3 \\
		5300.744 & 24.0 & 0.982 & -2.13 & 3.92E-32 & 58.1 & 67.6 & 14.4 & 96.7 & 73.1 & 51.4 & 88.2 & 133.5 & 85.8 & 95.8 \\
		... & ... & ... & ... & ... & ... & ... & ... & ...& ... & ... & ... & ... & ... & ... \\
		5483.352 & 27.0 & 1.7104 & -1.49 & 2.89E-32 & 51.6 & 65.2 & 22.9 & 96.5 & 56.5 & 41.5 & 91.3 & 100.3 & 87.3 & 80.5 \\
		5647.23 & 27.0 & 2.28 & -1.56 & 4.14E-32 & 14.3 & 18.6 & -- & 35.8 & 17.3 & 12.3 & 36.9 & 38.6 & 30.0 & 25.5 \\
	    ... & ... & ... & ... & ... & ... & ... & ... & ...& ... & ... & ... & ... & ... & ... \\
		4953.208 & 28.0 & 3.7397 & -0.66 & 3.25E-31 & 55.9 & 61.2 & 29.1 & 68.5 & 50.9 & 63.1 & 73.1 & 67.5 & 68.6 & 67.7 \\
		6108.116 & 28.0 & 1.676 & -2.44 & 2.48E-32 & 64.9 & 71.6 & 20.7 & 85.9 & 65.3 & 63.9 & 89.2 & 84.6 & 84.1 & 79.4 \\
		6643.63 & 28.0 & 1.676 & -2.1 & 2.14E-32 & 93.8 & 103.6 & 48.9 & 118.0 & 94.7 & 94.4 & 126.1 & 127.5 & 114.2 & 113.1 \\
		6767.772 & 28.0 & 1.826 & -2.17 & 2.8 & 78.8 & 85.2 & 38.4 & 94.3 & 83.4 & 84.1 & 99.8 & 108.7 & 94.6 & 98.0 \\
		... & ... & ... & ... & ... & ... & ... & ... & ...& ... & ... & ... & ... & ... & ... \\
		5218.197 & 29.0 & 3.816 & 0.476 & 2.8 & 52.1 & 59.9 & -- & 63.7 & 43.7 & 60.8 & 72.2 & 90.5 & 63.6 & 56.4 \\
		4722.159 & 30.0 & 4.03 & -0.38 & 2.8 & 72.7 & 73.4 & 40.5 & 68.8 & 59.4 & 86.4 & 74.0 & 65.1 & 70.2 & 68.1 \\
		4810.534 & 30.0 & 4.08 & -0.16 & 2.8 & 71.9 & 70.8 & 64.5 & 64.9 & 60.2 & 88.8 & 74.3 & 48.4 & 74.7 & 63.6 \\
		\enddata
		\tablecomments{a. The first five columns list the wavelength in angstroms (\AA), the ion identifier, the excitation potential, the logarithm of the oscillator strength times the statistical weight (log $gf$), and the damping constant. The integer part of the ion identifier corresponds to the atomic number, while the fractional part denotes the ionization state, where ``.0" indicates neutral species and ``.1" indicates singly ionized species. For example, ``26.0" corresponds to Fe I, ``26.1" to Fe II, and ``6.0" to C I. \\
		b. The remaining columns list the equivalent width measurements (in m\AA) for the Sun (from ESPRESSO, HIRES, HARPS, FEROS, and UVES; only the ESPRESSO column is shown here) and for the 23 target stars (only a few are shown here). \\
		(The full table is available in machine-readable form online.)}
	\end{deluxetable*}
\end{longrotatetable}
	
\begin{splitdeluxetable*}{lcccccccccccBccccccccccccc}
	\centering
	\renewcommand\thetable{A4}
	\tablecaption{Sulfur abundances from S~I line synthesis
	\label{tab:abund_S}}
	\tablehead{
		\colhead{Wavelength} &
		\multicolumn{4}{c}{Sun$^a$} &
		\multicolumn{2}{c}{HAT-P-26$^b$} &
		\multicolumn{2}{c}{HATS-72$^b$} &
		\multicolumn{2}{c}{HD~207496$^b$} &
		\multicolumn{2}{c}{HD~209100$^b$} &
		\multicolumn{2}{c}{TOI-500$^b$} &
		\multicolumn{2}{c}{TOI-1416$^b$} &
		\multicolumn{2}{c}{TOI-1130$^b$} &
		\multicolumn{2}{c}{TOI-824$^b$} &
		\multicolumn{2}{c}{WASP-69$^b$} &
		\multicolumn{2}{c}{GJ 9827$^b$} \\
		\colhead{(\AA)} &
		\colhead{ESPRESSO} & \colhead{HARPS} & \colhead{HIRES} & \colhead{UVES} &
		\colhead{$A({\rm S})$} & \colhead{[S/H]} &
		\colhead{$A({\rm S})$} & \colhead{[S/H]} &
		\colhead{$A({\rm S})$} & \colhead{[S/H]} &
		\colhead{$A({\rm S})$} & \colhead{[S/H]} &
		\colhead{$A({\rm S})$} & \colhead{[S/H]} &
		\colhead{$A({\rm S})$} & \colhead{[S/H]} &
		\colhead{$A({\rm S})$} & \colhead{[S/H]} &
		\colhead{$A({\rm S})$} & \colhead{[S/H]} &
		\colhead{$A({\rm S})$} & \colhead{[S/H]} &
		\colhead{$A({\rm S})$} & \colhead{[S/H]}
	}
	\startdata
	6046.00 & 7.72 & 7.75 & 7.71 & 7.72 & 7.88 & 0.16 & 7.70 & $-0.02$ & 7.95 & 0.20 & 7.70 & $-0.05$ & 7.79 & 0.07 & 7.70 & $-0.01$ & \nodata & \nodata & \nodata & \nodata & 7.92 & 0.20 & \nodata & \nodata \\
	6052.66 & 7.31 & 7.20 & 7.27 & 7.27 & 7.25 & $-0.06$ & \nodata & \nodata & 7.45 & 0.25 & 7.43 & 0.23 & \nodata & \nodata & \nodata & \nodata & \nodata & \nodata & 7.69 & 0.49 & \nodata & \nodata & \nodata & \nodata \\
	6743.54 & 7.12 & 7.12 & 7.04 & \nodata & 7.00 & $-0.12$ & 7.18 & 0.06 & 7.45 & 0.33 & 6.93 & $-0.19$ & 7.57 & 0.45 & 7.15 & 0.11 & \nodata & \nodata & \nodata & \nodata & \nodata & \nodata & \nodata & \nodata \\
	6757.15 & 7.18 & 7.27 & 7.19 & \nodata & 7.13 & $-0.05$ & 7.48 & 0.30 & 7.60 & 0.33 & 6.93 & $-0.34$ & 7.37 & 0.19 & 7.27 & 0.08 & \nodata & \nodata & \nodata & \nodata & \nodata & \nodata & \nodata & \nodata \\
	\hline
	Mean [S/H]$^c$       &      &      &      &      &      & $-0.004$ &      & 0.011 &      & 0.166 &      & $-0.302$ &      & 0.142 &      & $-0.062$ &      & \nodata &      & 0.49 &      & 0.20 &      & \nodata \\
	$\sigma_\mu^c$ &      &      &      &      &      & 0.061 &      & 0.096 &      & 0.032 &      & 0.121 &      & 0.097 &      & 0.036 &      & \nodata &      & 0.20 &      & 0.20 &      & \nodata \\
	Total error$^c$ &      &      &      &      &      & 0.077 &      & 0.135 &      & 0.161 &      & 0.130 &      & 0.159 &      & 0.070 &      & \nodata &      & 0.206 &      & 0.222 &      & \nodata \\
	\enddata
	\tablecomments{
		a. Solar reference abundances are listed for each instrument. \\
		b. Abundances are derived from spectral synthesis of individual S~I lines using the {\it synth} task in IRAF. The [S/H] values are calculated by subtracting the solar abundance measured with the corresponding instrument. TOI-1130, TOI-824, WASP-69, GJ 9827 are cool dwarfs in Sun26. \\
		c. Mean abundances are computed by averaging individual line abundances in linear space. $\sigma_\mu$ denotes the standard error of the mean, and the total error is obtained by adding $\sigma_\mu$ and the uncertainties propagated from the stellar atmospheric parameters in quadrature.
	}
\end{splitdeluxetable*}

\end{document}